**Development of Embedded Speed Control System for DC Servo Motor using Wireless Communication**

# By

Santosh Mohan Rajkumar (13-1-3-015)

Sayan Chakraborty (13–1–3-007)

A thesis submitted in partial  fulfillment  for  the

degree of Bachelor of Technology in Electrical  Engineering

Under The Guidance of

Dr. Rajeeb Dey

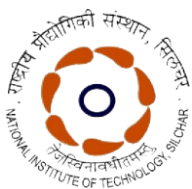

Department of Electrical  Engineering

**National Institute of Technology,  Silchar**

December 2019

# Declaration

We, Santosh Mohan Rajkumar (13-1-3-015), and Sayan Chakraborty (13-1-3-007) declare that this thesis titled, "**Development of Embedded Speed Control System for DC Servo Motor using Wireless Communication**" and the work presented in it are our own. We confirm that:

- This work was done wholly or mainly while in candidature for B.Tech. degree at National Institute of Technology, Silchar.

- Any part of this thesis has not previously been submitted for a degree or any other qualification at this institute or any other institutes.

- Where we have consulted the published work of others, this is always clearly attributed.

- Where we have quoted from the work of others, the source is always given. With the exception of such quotations, this thesis is entirely our own work.

- We have acknowledged all main sources of help.

Signature:

______________________________________________

Date:

______________________________________________

*"Peace comes from within, don't seek it without"*

The Tathagata

# *Abstract*

With the advancement of computer technology and digital systems, wireless control systems or Wireless Networked Control Systems(NCS) are becoming increasingly popular among the scientific community as well as the industry due to their flexibility, convenience & ease of operation. In this experiment, a closed-loop discrete time system for speed control of a permanent magnet DC motor with discrete PI controller is implemented in embedded platform. The design & analysis of the system is based on the mathematical model of the DC motor obtained by system identification technique. After that the closed loop system is distributed through a wireless network created by means of Bluetooth without any change in the discrete controller. The network connects the controller on one side with the sensor, actuator & the plant on the other side. Then the performance of the closed loop system is observed with the wireless network in two configurations : with point-to-point connection between two nodes and a network structure with two intermediate nodes among the controller side node & the plant side node. It has been observed that the performance degrades with only the PI controller a little bit in point to point configuration and the performance severely degrades in intermediate node configuration. To tackle this issue time delay is measured in the WNCS and then a digital smith predictor structure is implemented to obtain better performance. It has been observed that in point to point configuration the time delay remains almost constant and in intermediate node configuration the time delay is varying. Hence the digital smith predictor fails to perform reasonably in intermediate node configuration. An online time delay measurement & estimation procedure has been implemented and proposed in this research. We have implemented an adaptive digital smith predictor using the online time delay measurement and identification. The implemented smith predictor provides good results for both intermediate node and point to point connection. An stability analysis of the DC motor system with variation of delay has been discussed.

# *Acknowledgements*

First and foremost, we feel it as a great privilege in expressing our deepest and most sincere gratitude to our supervisor Dr. Rajeeb Dey, for providing us an opportunity to work under him and for his excellent guidance. His kindness, friendly accessibility and attention to details have been a great inspiration to us. Our heartiest thanks to the supervisor for the support, motivation and the patience he has shown to us. His suggestions and advice have been always constructive. We hope that we will be able to pass on the research values and knowledge that he has given given to us.

We are also very much thankful to Mr. Nalini Prasad Mohanty and Mr. Anirudh Nath (PhD scholars) for their support, guidance and help. Their friendly behaviour and kind cooperation made the environment productive.

We are also grateful to Prof. Binoy Krishna Roy for his instrumental suggestions for continuous improvement in research. He is the back bone of the Control System Research Community of NIT Silchar. We are proud to have such an experienced and wise professor in our department.

We would not have contemplated this road without our parents who support us, care for us and give motivations.

We would also like to express our heartiest gratitute to Mr. Prasanta Roy who taught us the foundations of Control Engineering. He is a great teacher.

All the faculty members and scholars of the EE department of NIT Silchar, helped us in many ways to accomplish this project.. . .

# Contents

*Contents*        ix

# List of Figures

xi

# List of Tables

# Chapter 1

# Introduction

DC motor speed controllers are widely used for motion control of robotics, industrial control and automation systems [8]. Industrial process control has requirements of adjusting motor speed over a wide range with good resolution and reproducibility [1]. Conventional analog speed control methods have certain drawbacks, including nonlinearity in the analog speed transducer and difficulty in accurate transmission of the analog signal [1]. Also in analog methods signal manipulation suffers from errors occurring due to temperature, component aging, and external disturbances [1]. In a digital speed control scheme, there is no nonlinearity associated with speed transducer and the digital signal of speed can be transmitted to long distances without sacrificing accuracy [1]. Also in a digital speed control system, the control signal is not adversely effected by temperature variations, component aging, or noise [1].

For the purpose of speed control of a DC motor, a variable-voltage DC power source (DC Chopper) is needed [1]. Pulse Width Modulation (PWM) technique is used for generating variable voltage for speed control purpose [1]. In home appliances the permanent magnet DC motor has replaced the ac universal motor due to improve speed or drive performance [9]. In this research, an embedded speed control method for a permanent magnet DC motor has been implemented in Arduino Due micro controller board based on the Atmel SAM3X8E ARM Cortex-M3 CPU. The wired speed control is done with the help of single Arduino Due micro controller board as shown in figure2.17. The wired speed control scheme can also be implemented with two Arduino Due micro controller boards serially connected through wires as in figure1.2.Then the controller side and plant side of the embedded control system as in figure1.2are connected with an wireless network (Bluetooth) in order to develop a Wireless Networked Control System (WNCS). The motive of this research is to observe the performance of the closed loop system with wireless network and then develop necessary techniques or algorithms for performance improvement if required.

The advancement of semiconductor technology and electronic devices, the embedded micro controllers are very much inexpensive and can provide reasonably well speed of performance.

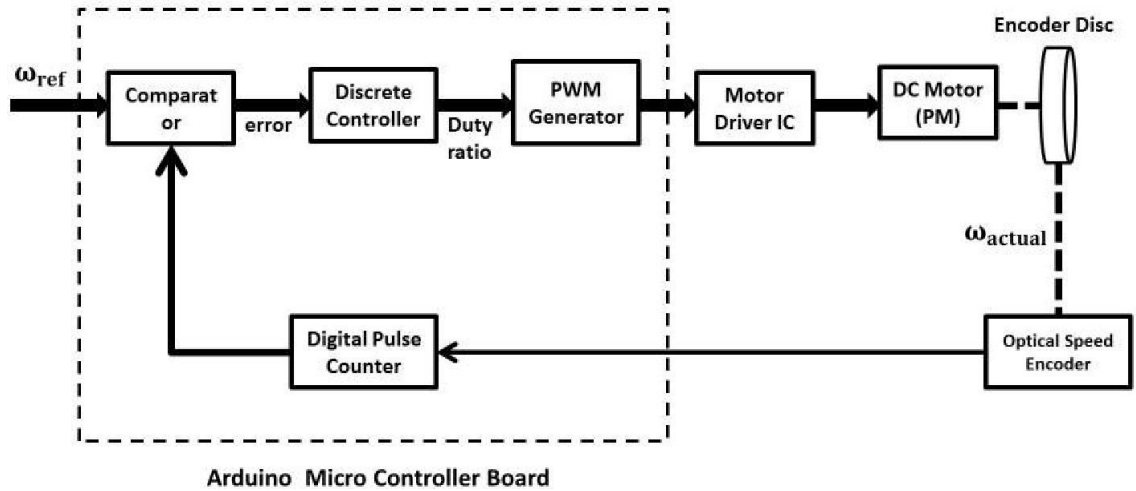

Figure 1.1: The functional block diagram of the basic embedded DC motor speed control system with single micro controller board

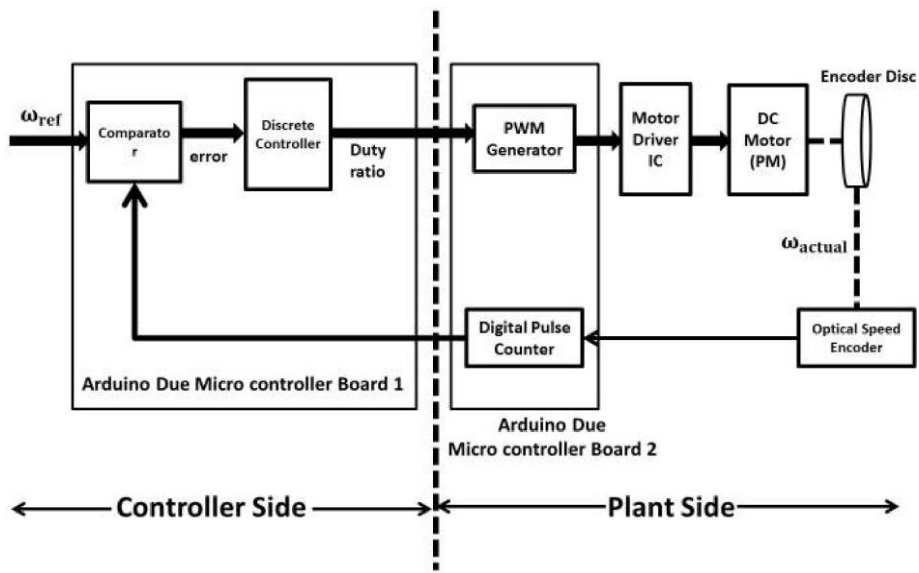

Figure 1.2: The functional block diagram of the embedded DC motor speed control system with two micro controller board

Wireless communication is the transfer of information or power between two or more points that are not connected by an electrical conductor.The use of a wireless network enables enterprises to avoid the costly process of introducing cables into buildings or as a connection between different equipment locations. So, it has gained widespread popularity with time and its use keeps growing. Wireless Networked Control System (WNCS) is a distributed control system with sensor, actuator and controller communication supported by a wireless network. WNCS can also be defined as a spatially distributed control system with sensor, actuator and controller communication supported by a wireless network [19]. WNCSs allow exchange of information among distributed sensors, controllers actuators over the wireless network to achieve certain control objectives [19].

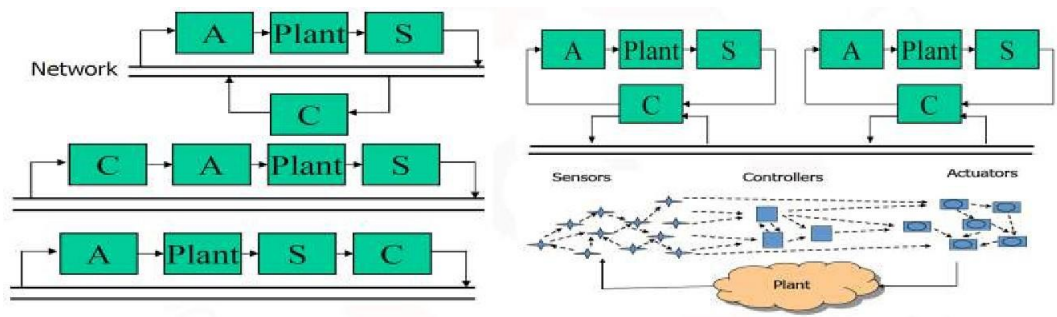

Figure 1.3: Wireless networked control systems

The benefits of wireless networked control are [19]:-

1. It is more flexible. Some of the aspects of flexibility are:-

2. Sensors and actuators can be replaced easily.

3. Less restrictive maneuvers and control actions.

4. Powerful control over distributed computations.

5. Reduction in installation and maintenance costs.

6. Requirement of less cabling.

7. Possibility of efficient monitoring and diagnosis.

Wireless technology has some barriers in control and automation. Some of the key barriers of wireless technology are [19]:

- **Complexity:** Systems designers and programmers need suitable abstractions to hide the complexity from wireless devices and communication.

- **Reliability:** Systems should have robust and predictable behaviour despite characteristics of wireless networks.

- Communication of sensor and actuator data impose uncertainty, disturbances and constraints on control system.

- Communication imperfections in control loops include:

  – Time delay and jitter.

  – Bandwidth limitations.

  – Data loss and bit errors.

  – Outages and disconnection.

  – Topology variations of different networks.

- **Security:** Wireless technology is vulnerable to digital hacking or attacks.

There are several approaches to control in WNCS, they are,

- *Network-aware control:* Modify control algorithms to cope with communication imperfections, for e.g. : predictive controller. In this procedure, in order to adjust control algorithms, one needs to estimate the network states, like,

  – Time Delay

  – Data Loss

  – Bandwidth

- *Control-aware networking:*  Control of communication resources

The parameter of a wireless network which is of most concern is ***time-delay*** . Time delay in wireless communication arises due to the communication medium and additional functionality required for physical signal coding and communication processing.

Time delay in a communication network is expressed as,

$$T_{delay} = T_{device} + T_{network} \qquad (1.1)$$

$$\implies T_{delay} = T_{pre} + T_{wait} + T_{prop} + T_{post} \qquad (1.2)$$

Time delay in a communication network can be of different types as given below [19],

- Varying or fixed.

- known or unknown.

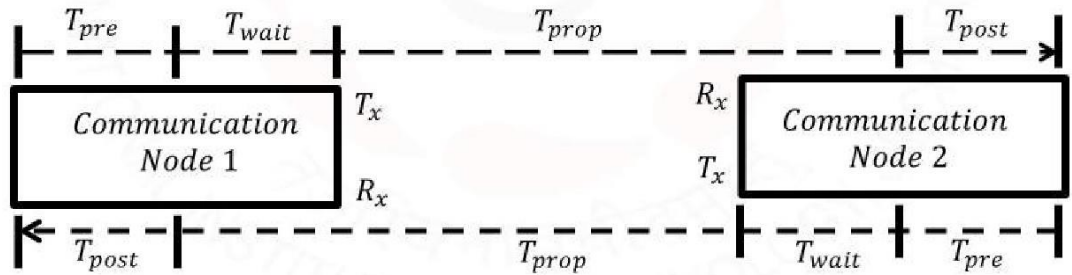

Figure 1.4: Break up of Communication Network Time-delay Into Different Constituents

# 1.1 Motivations

DC motor drives are widely used in applications requiring adjustable speed control, frequent starting, good speed regulation, braking and reversing. Some important applications are paper mills, rolling mills, mine winders, hoists , printing presses, machine tools, traction, textile mills, excavators and cranes. DC motors are widely used as servomotors for tracking and positioning. For industrial applications development of high performance motor drives are very much essential. These applications may demand high speed control accuracy and good dynamic response. Some other examples of DC motor speed control applications are washers, dryers and other household appliances. Applications of DC motor speed control are also found in the automotive area for uses such as fuel pumps, electronic steering, and engine controls.DC motor drives are less costly and less complex compared to AC motor drives. Moreover, use of embedded systems for speed control purpose of DC motor brings the cost to much lower value. Development of very low cost embedded speed control system can also serve as a good laboratory experiment to understand and learn control engineering practically.

Motivating applications of Wireless Networked Control Systems (WNCSs) are as follows,

- **Wireless Process Control Industrial Automation:** Communication cabling is subject to heavy wear and tear in industrial controlled processes & robots and therefore requires frequent maintenance. This implies increased maintenance cost. Replacing cables or wires with wireless network saves cost

- **Medical Applications:** For example, Glucose-Insulin monitoring & control of Diabetic Patient.

- **Practical Fault Tolerant System:** WNCSs are tolerant to sensor controller failure. If a sensor or controller fails, the system can use that of any other connected system within the wireless network.

## 1.2 Problem Statement

In this thesis, design of a closed-loop DC motor (permanent magnet) servo system for speed control will be discussed. Then implementation of the servo system using an embedded system platform (Arduino micro controller board) will be explained. The controller will be a discrete PI controller. Then the servo system will be distributed into two parts : controller side (containing reference input functionality, discrete controller and comparator) and plant side containing (actuator, sensor and the DC motor plant), with both the parts connected through a Bluetooth wireless network. The behaviour of the wireless network in the control loop will be examined. Then the performance of the servo system with the wireless network will be observed. After that measurement, estimation and approximation of the network parameters effecting the performance of the servo system will be discussed in detail. Based on the estimated wireless network parameters (e.g. time delay) , predictive control strategies will be developed for better performance of the servo system. Then the implementation of predictive control strategies in embedded platform will be discussed.

## 1.3 Literature Review

DC motor speed control problem is an age old problem that scientists and researchers are dealing with. There are lots of papers available in literature on DC motor speed control using various analog techniques and power electronic methods. Maloney, T.J. and Alvarado, F.L. (1976) [1] first discussed a digital method for DC motor speed control where digital tachometer was used to remove the nonlinearity associated with analog transducers. A. K. Lin and W. W. Koepsel (1977) [2] first discussed a microprocessor (Intel 8080) based speed control strategy using DC chopper to achieve reasonably low steady state error. J. B. Plant and S. J. Jorna (1980) [3] published an SCR based dc motor drive control where a microprocessor performed both the control law computation and logical functions of SCR. State space methodology was used for the control purpose and stability, error analysis of the controller was provided. A. H. M. S. Ula, J. W. Steadman and J. M. Wu (1988) [4] explained a micro controller based speed control for industrial size DC motor using thyristors. J. Nicolai and T. Castagnet (1993) [5] discussed a permanent magnet DC motor speed control using micro controller generated PWM signal. Software flexibility for modifying drive parameters like maximum power, time constant etc. were also explained in that paper. T. Castagnet and J Nicolai (1994) [9] discussed a micro controller based brush DC motor speed control through direct voltage compensation and motor power limitation. Y. S. E. Ali, S. B. M. Noor, S. M. Uashi and M. K. Hassan (2003) [6] explained a micro controller based speed control and over current protection scheme for a DC motor. These are the key publications found in literature on embedded/ micro controller/ microprocessor based DC motor servo systems.

There are a large number of publications on networked control systems, but the number of research in wireless networked control systems are less in number [11]. J. Eker, A. Cervinnad A. Hrjel (2001) [10] discussed a distributed control method using Bluetooth network and provided solution for two specific problems occurring while using Bluetooth in a control loop: long random delays and bit errors. N. J. Ploplys, P. A. Kawka and A.G. Alleyne (2004) [11] developed a method for near real-time control over a wireless network and implemented for a practical set up. It has been found that The use of event-driven control with clock driven sensing and actuation gives a natural synchronization of the control loop without clock synchronization of the individual nodes [11]. Y. Jianyong, Y. Shimin and W. Haiqing (2004) [11] published survey on the performance analysis of Networked Control Systems (NCSs) where main aspects around performance analysis of NCSs were: network-induced delays, sampling period, jitter, data pocket dropout, network scheduling and stability. C.H. Chen, C.L. Lin and T.S. Hwang (2007) [14] showed that stability of the closed-loop system is guaranteed up to some maximum upper bound of the round trip delay. J. De Boeij, M. Haazen and P. Smulders (2009) [17] presented a new approach for wireless motion control with a new protocol and wireless system that reduce the closed loop transmission delay to less than 300 microseconds. The system was verified in a real control system, and measurements showed the performance, which was more than ten times better than existing techniques. A. Hernandez (2010) [13] discussed use of IEEE 802.15.4 wireless protocol in control applications. An inverted pendulum process is introduced to show the benefits in wireless process control by using the IEEE 802.15.4 (as wireless sensor network) in a real-time control loop process. The extensive experimental results show that packets losses and delays are the essential factors to guarantee the stability of the system. C.L. Lai and P.L. Hsu (2010) [15] discussed a networked remote control system (wired) and online measurement of the round-trip time (RTT) between the application layers of the server and the client. M. Pajic, S. Sundaram, G. Pappas and R. Mangharam (2011) [16] presented a distributed scheme used for control over wireless networks, where where the network itself, with no centralized node, acts as the controller. C. Suryendu, S. Ghosh and B. Subudhi (2017) [18] discussed a gradient descent method based delay estimator for use with a variable gain control strategy for wireless networked control systems. The delay estimator was developed in such a way that its boundedness is ensured. The performance of the estimator with variable gain controller was evaluated on a temperature control plant with network in the feedback loop. From the literature survey on WNCS/NCS applications, we have found that most of the works are done using PCs, high performance computers and hence cost of implementation goes high. We could not find any literature to discuss WNCS implementation using embedded hardware platform.

For compensation of time delays occurring in control systems, predictive control strategies need to be adopted for compensation of time delays. Smith predictor is one of the most popular scheme for time delay compensation. In 1957 O.J. Smith first proposed the smith predictor scheme to compensate time delays occurring in process control applications. Since then it has become one of the most popular dead time compensation scheme and many papers have been published. D. E. Lupfer, M. W. Oglesby (1962) discussed an implementation of smith predictor in automatic process control. Alevisakis & Seborg (1973) extended smith predictor to multivariable systems with single delay and that to multivariable systems with multiple delays by Ogunnaike & Ray (1979). C. Meyer, .D.E. Seborg & R.K. Wood (1976) gave a comparison of smith predictor and conventional PI control scheme having dead time both in forward and feedback path in simulation.JR Schleck & D Hanesian (1978) evaluated linear smith predictor structure for dead time dominating process. A. C. Ioannides, G. J. Rogers & V. Latham (1978) discussed stability limits of smith predictor controller for simple systems when process-model mismatch occurs. C. Meyer, .D.E. Seborg & R.K. Wood (1979) also provided a comparison between analytical predictor and smith predictor scheme. Z. Palmor (1980) also investigated the stability properties of smith time delay compensator scheme with practical conditions of stability. A. Terry Bahill (1983) published a tutorial on simple smith predictor scheme with adaptive nature to process parameter variation. Wellons and Edgar (1985) discussed a Generalized Analytical Predictor (GAP) for dead time systems and its relationship with smith predictor, IMC strategy and discrete analytical predictor in simulation. Z. J. Palmor and D. V. Powers (1985) proposed a modified smith predictor with simultaneous feedforward & feedback actions with a control on degree of dead time compensation.S. K. P. Wong & D. E. Seborg (1986) gave a theoretical analysis of smith predictor and analytical predictors.

Until 1980, smith predictor schemes developed were mostly analog & hence were not used in the industry due complexity of analog technique. Vogel, E. F. and T. F. Edgar, in 1980, proposed a digital control scheme for dead time compensation. In the same year, they also proposed adaptive control strategies for variable dead time compensation in simulation level. C. C. Hang, K. W. Lim & T. T. Tay (1986) also proposed an adaptive digital scheme of smith predictor in simulation. An adaptive digital smith predictor with dual rate sampling, i.e. smaller sampling time for control & larger sampling time for on-line parameter estimation, have been discussed by C. C. Hang, K. W. Lim and B. W. Chong (1989). Many researchers discussed modified smith predictor for system with integrator and long dead time and some of them are K.J. Astrom and Hang Lim (1994), M.R. Matausek and A. D. Micric (1996), J. E. Normey-Rico and E.F. Camacho (1999). V Massimiliano (2003) gave a improved smith predictor structure through automatic dead time computation in practical. Chen Peng, Dong Yue & Ji Sun (2004) proposed smith predictor structure in Networked Control System (NCS) employing on-line time delay identification and provided simulation results. Chun-Hsiung Chen, Chun-Liang Lin, and Thong-Shing Hwang (2007) discussed a robust smith predictor to compensate varying round trip delay time in NCS and the maximum upper bound of the round trip delay providing guaranteed stability in simulation. Chien-Liang Lai and Pau-Lo Hsu (2010) discussed a method for on line measurement of round trip delay time in NCS and based on that round trip delay time an adaptive smith predictor scheme was proposed for a servo system in practical. Rui Wang, Guo-Ping Liu, Wei Wang, David Rees, and Yunbo B. Zhao (2010) proposed an improved predictive controller design strategy for NCS to compensate for the varying time delay and data dropout in both the forward and backward channels to achieve the desired control performance. V Bobl, P Chalupa and P Dostl (2010) proposed a digital adaptive smith predictor based on pole assignment or polynomial design in simulation. V Bobl, P Chalupa, P Dostl and M Kubalk (2014) again proposed a digital smith predictor for control of unstable and integrating time-delay processes.

From the literature review, it has been found that most of the publications discussed simulation study of various smith predictor strategies for time delayed systems. Very few of them mentioned about practical implementation (e.g. V Massimiliano (2003), Chien-Liang Lai and Pau-Lo Hsu (2010)). Although all of the practical implementations found in literature were done using sophisticated computing systems and we could hardly find any paper that discusses real time embedded system application. There are a handful amount of work on smith predictor strategy for NCS, but we could not find any publication that provides practical implementation for Wireless Networked Control System in embedded platform.

## 1.4 Thesis Outline

The thesis is divided into 7 chapters as follows,

- **Chapter 1:** It discusses an introduction to embedded speed control of DC motor and Wireless Networked Control Systems (WNCSs). Also motivations for work, problem statement and literature review has been presented.

- **Chapter 2:** This chapter provides DC motor plant description, speed control methodology, modeling and system identification of the plant, discretization of the system and design of discrete PI controller. Then practical implementation of the servo system is discussed.

- **Chapter 3:** This chapter deals with details on Bluetooth communication technology, behaviour of the Bluetooth network to be used in control loop, development and implementation of WNCS for the DC motor servo system.

- **Chapter 4:** This chapter discusses time delay measurement, estimation and approximation in the WNCS. Practical embedded implementation for time delay estimation as RTT (Round Trip Time) has been discussed. Then time delay compensation using digital Smith Predictor in Arduino embedded platform for fixed and varying time delays in the wireless network is presented.

- **Chapter 5:** This chapter gives an stability analysis of the DC motor plant with time delay.

- **Chapter 6:** This chapter discusses modifications from existing literature adopted in the practical WNCS set up and contributions or proposal of the research.

- **Chapter 7:** This chapter provides conclusions and future works.

## 1.5 Chapter Summary

In this chapter basic idea about the embedded platform based speed control system for a PMDC motor has been provided. We have also discussed introductory ideas about WNCSs. Then motivations and problem definition are provided. Literature survey related to the research work presented in the thesis has also been provided.

# Chapter 2

# Plant Modeling, Controller Design and The Practical DC Motor Speed Control System

There are various types of speed control methods used for DC motor's speed control. In our experiment, the DC motor is of permanent magnet type i.e.separately excited. Various speed control methods used for DC separately excited motors are[5]:

- Armature Voltage Control Method

- Armature Resistance Control Method

- Field Control Method

In our case, we are going with armature voltage control due to the following reasons:

- Field control method is not application as the DC motor used is of permanent magnet type.

- Rotor resistance control method suffers from extra power loss.

- Armature voltage control is preferred because of high efficiency, good transient response and good speed regulation.

# 2.1 Armature Voltage Control For Speed Regulation

It is the method by which the speed of a DC motor is controlled by giving variable input armature voltage to the motor. The motor we use here is a permanent magnet separately excited dc motor therefore the field flux $\varphi$ is constant.

Therefore, back emf of motor $E = K\omega$, where $\omega$ is the speed of the motor K is back emf constant.

Now, for V being terminal voltage, Ia being the armature current, R being the armature resistance, the back emf of the motor is given as[5]:-

$$E = K\omega = V - I_a R \qquad (2.1)$$

$$\implies \omega = \frac{V - I_a R}{K} \qquad (2.2)$$

Therefore, from equation 2.2, we can see that by varying the voltage V, we can vary the speed of the  motor.

There are various methods by which armature voltage control can be done.

When the supply is ac, we can use [5]:

- Ward-Leonard schemes
- Transformer with taps and an uncontrolled bridge rectifier
- Static Ward-Leonard scheme or controlled rectifiers

As we are using a dc supply,the pulse width is varied using Pulse Width Modulation (PWM) technique. It is a chopper control strategy to produce variable voltage from a fixed voltage.

## 2.1.1 Chopper Control of Separately Excited dc Motors

The waveform of a PWM (Pulse Width Modulation) chopper device used for DC motor speed control is shown in figure 2.1 The frequency of PWM scheme should be high to ensure continuous conduction[5].

The waveform is given as:-

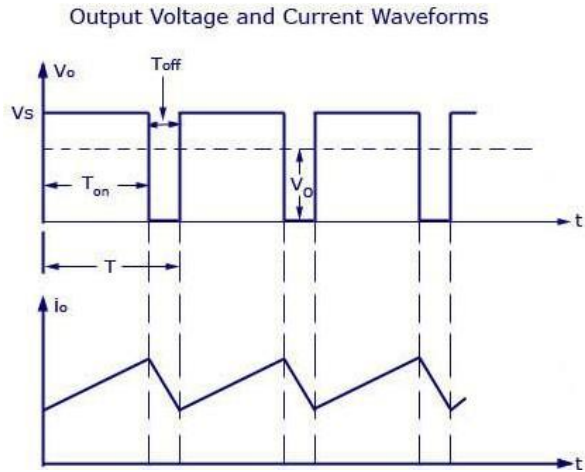

Figure 2.1: Waveforms for PWM chopper control of separately excited dc motor

Ratio of duty interval $t_on$ to chopper period T is called duty ratio or duty cycle ($\delta$). Thus [5]

$$\delta = \frac{Duty interval}{T} = \frac{t_{on}}{t}$$                                    (2.3)

$$V_a = \delta V$$                                    (2.4)

Now as [5],

$$I_a = \frac{\delta V - E}{R_a}$$                                    (2.5)

From equations 2.2 and 2.5

$$\omega_m = \frac{\delta V}{K} - \frac{R_a T}{K^2}$$                                    (2.6)

In this way the speed of a DC motor with permanent magnet poles can be controlled.

## 2.2 Hardware Components Used

### 2.2.1 Arduino Micro Controller  Board

Arduino is an open source hardware and software project first introduced in 2005 aiming to provide an accessible way for novices and professionals to create devices that interact with their environment using sensors and actuators. Common examples of such devices include simple robots, thermostats, and motion detectors.

Arduino is based on micro controller board designs, which use inputs and outputs in the same way  an ordinary computer does.  Inputs capture information from the user or the environment while outputs do something with the information that has been captured. An input could be digital or analog, and could come from the environment or a user. Outputs can control and turn on and off devices such as motors or other computers. These systems provide sets of digital and analog input/output (I/O) pins that can interface to various expansion boards (termed shields) and other circuits. The boards feature serial communication interfaces, including Universal Serial Bus (USB) on some models, for loading programs from personal computers.

### 2.2.1.1 Ardunio UNO

This is the entry level board from Arduino family.

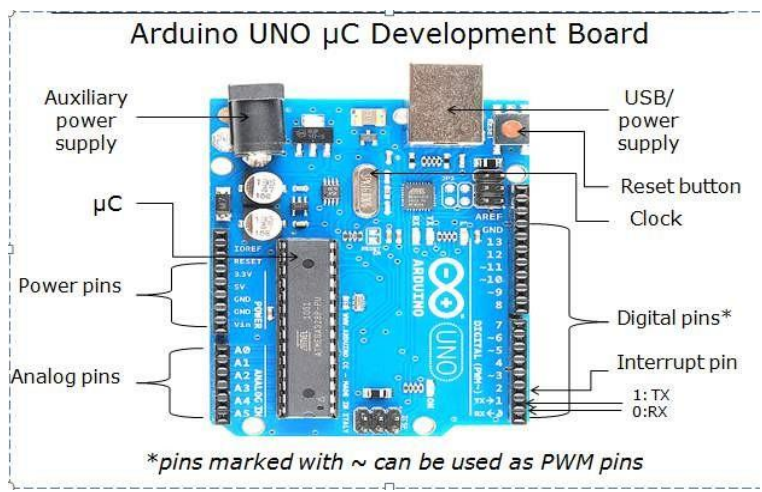

Figure 2.2: Arduino UNO board

The specifications of Arduino UNO board are given below,

- The micro controller is ATmega328P (8 bit, AVR)

- 14Digital  input/output  pins

- 6  Analog  input/output pins

- 6/14 Digital I/O pins can be use as PWM pins

- The clock is 16 MHz quartz crystal

- A USB connection/power jack  (5V)

- A auxiliary power jack  (5V)

- Flash Memory 32 KB

- SRAM 2 KB

- EEPROM 1 KB

- One Serial Transmitter (Tx)

- One Serial Receiver (Rx)

### 2.2.1.2 Arduino Due

This is one of the most advanced Arduino boards.

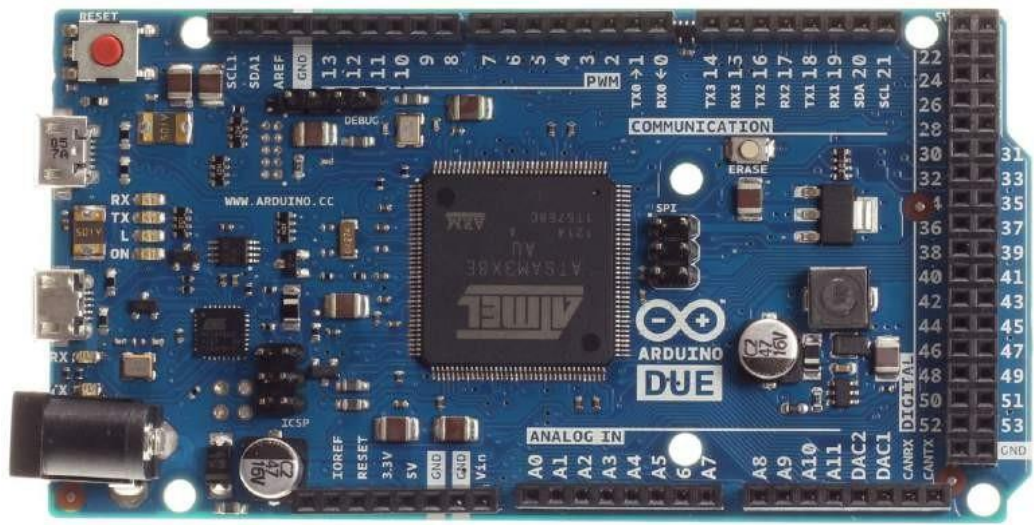

Figure 2.3: Arduino Due Board

The specifications of Arduino Due board are given below,

- Atmel SAM3X8E ARM Cortex-M3 CPU (32-bit)

- 54 Digital input/output pins

- 12 PWM pins among the 54 digital I/O pins.

- 12 analog input pins.

- The clock is 84 MHz quartz crystal.

- A USB connection and power jack (5V).

- 2 DAC pins.

- Flash Memory 512 KB.

- SRAM 96 KB

- EEPROM 1 KB

- 4 UART transmitter-receiver pin pairs.

## 2.2.2 The Bluetooth module

Bluetooth is a wireless technology standard for exchanging data over short distances using short-wavelength UHF radio waves. The model we are using is $HC-05$ Bluetooth Serial Port Protocol (SPP) Module.

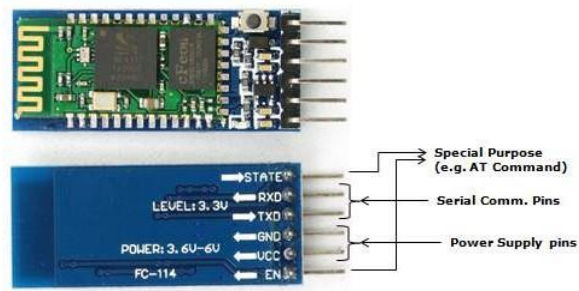

Figure 2.4: The Bluetooth Module

Specifications:

- Bluetooth V2.0+EDR (Enhanced Data Rate)

- 3 Mbps Data Rate

- 2.4GHz radio transceiver and baseband

- Low Power 1.8V Operation ,1.8 to 3.6V I/O

- Supported baud rate: 9600,19200,38400,57600,115200,230400,460800

- Data bits:8, Stop bit :1, No Parity.

## 2.2.3 The Motor Driver IC

Motor drivers act as current and/or voltage amplifiers since they take a low-current and/or low-voltage control signal and provide a higher-current signal. This higher current signal is used to drive the motors. The Motor Drive IC used is L-293D.

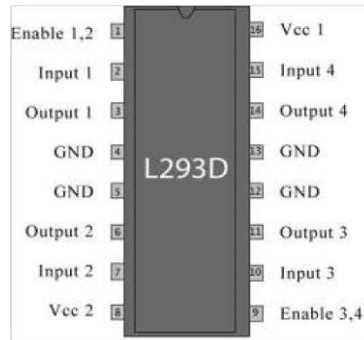

Figure 2.5: The Motor driver IC

### 2.2.4  The Optical Speed Encoder

• Model : HC-020k Module with Opto-coupler unit.

• Slotted Encoder Disc (20 slots)

• Measurement frequency: 100 KHz

• Module Working Voltage: 4.5-5.5V

• Encoder resolution: 20 lines

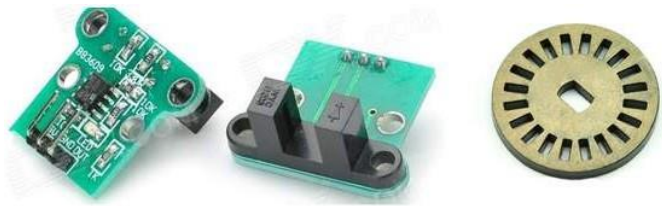

Figure 2.6: The Speed Encoder With Slotted Disc

## 2.3 Speed Measurement by Optical Encoder

The optical speed encoder consists of two parts:

• An optical encoder or optocoupler unit.

• A slotted encoder disc.

## 2.3.1 Optocoupler

An Optocoupler, also known as an Opto-isolator or Photo-coupler, is an electronic component that transfers electrical signals between two electrical circuits with the help of light.  The main purpose of opto-coupler is to provide isolation between two electrical circuits.

An opto-isolator connects input and output sides with a beam of light modulated by input current. It transforms useful input signal into light, sends it across the dielectric channel, captures light on the output side and transforms it back into electric signal.

A basic optoisolator consists of a light-emitting diode (LED), IRED (infrared-emitting diode) or laser diode for signal transmission and a photosensor (or phototransistor) for signal reception. Using an optocoupler, when an electrical current is applied to the LED, infrared light is produced and passes through the material inside the optoisolator. The beam travels across a transparent gap and is picked up by the receiver, which converts the modulated light or IR back into an electrical signal. In the absence of light, the input and output circuits are electrically isolated from each other.

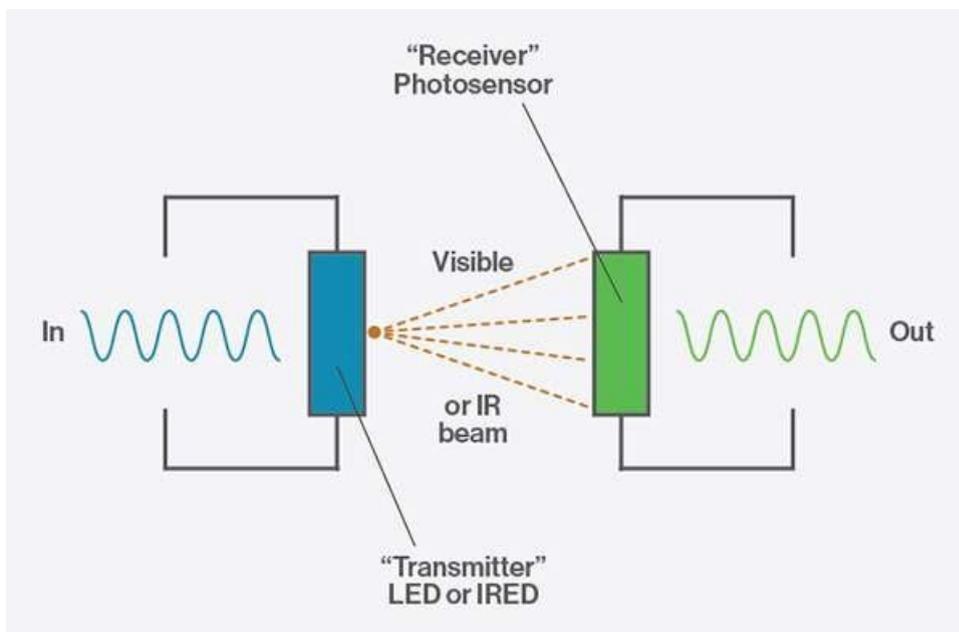

Figure 2.7: A basic optocoupler unit

Electronic equipment, as well as signal and power transmission lines, are subject to voltage surges from radio frequency transmissions, lightning strikes and spikes in the power supply. To avoid disruptions, optoisolators offer a safe interface between high-voltage components and low-voltage  devices.

A circuit breaker performs the same function in many electrical circuits but if fault persists, then the circuit will trip which disrupts the continuity of flow. The optocoupler will simply block the transient due to fault and passes the normal signal.

### 2.3.2 Speed Encoder Disc

An encoder disc is a device attached to the shaft of a servo motor to provide feedback for precisely control of the electical current supplied to the motor and in turn, control its motion.

Optical, the most common type of speed encoders use a (usually) metal disk (the code wheel) with uniformly spaced slots to alternately allow and block light from an LED(s) or other light source reaching a photo-transistor(s). As the motor shaft turns, the slots pass the LED - photo-transistor pairs creating a series of electrical ON/OFF states that can be used to compute the position, speed and acceleration of the motor and then be used to determine the correct current to command to cause the motor to follow the desired motion profile.

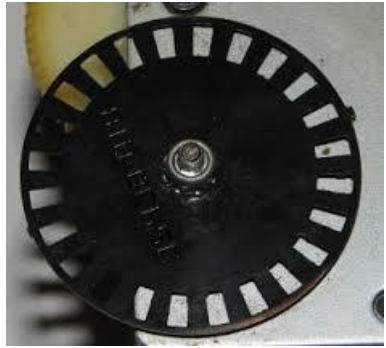

Figure 2.8: Slots of a speed encoder

### 2.3.3 Calculations for Speed Measurement

Here, no of slots of the speed encoder=20

Let the no of transitions per 20 ms be= $x$

Then, no of revolutions per ms is= $(x/20)$

Thus, speed in rps= $(x/(20 * 20)) * 100 = x * 2.5$

# 2.4 System Identification

System Identification is the process of building or describing a relationship between the input & output of a system or process from observed or measured data [1]. System identification is a statistical methods in control system engineering to build mathematical and analytical models of dynamic systems from the already inferred data .It is an experimental or empirical Modeling technique [1].

Model is the relationship among cause-effect signals and it provides a description of the system. Models serve as good mathematical substitutes for the system. To infer a model, input and output signals from the system are recorded and analysed [1].

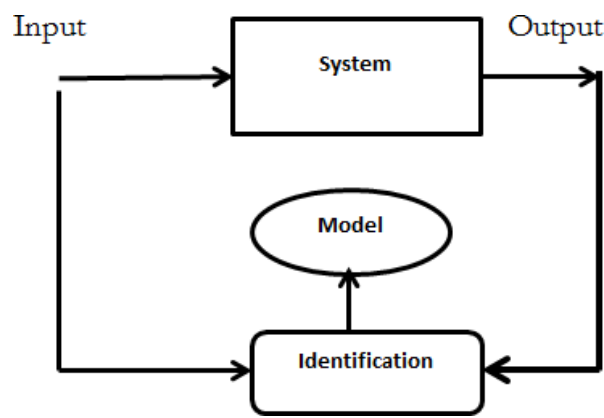

Figure 2.9:  System Identification & the Mode [1]

## 2.4.1 Purpose of System Identification

The main purposes of using system identification are as follows [1]:

- To obtain a mathematical model for controller  design.

- To explain or understand observed phenomena

- To forecast events in future.

- To obtain a model of signal in filter design.

## 2.4.2 Different Approaches For System Identification

There are two types of approach for system identification [1]:

1. **Non-parametric Approach :** Non-parametric approach gives basic idea about the system is useful for validation. A good estimate of time delay is provided & some insights to order of the process along with other variable information such as time constants, presence of integrators, inverse response is found. This approach aims at determining a time or frequency response directly without first selecting a possible set of models. A structure &the order of the parametric model is found using the information gathered from non-parametric model. Three most common non-parametric models are:

   - Finite Impulse Response (FIR) Model.
   - Step Response Model
   - Frequency Response Model

2. **Parametric Approach :** Assumptions on model structure is required in the parametric approach. It becomes more complicated than non-parametric approach as the search for the best model within the candidate set becomes a problem of determining the model parameters. The various types of parametric models are:

   - Auto-Regressive Exogenous (ARX).
   - Auto-Regressive Moving Average (ARMAX)
   - Output Error (OE)
   - Box-Jenkins Model(BJ)
   - State Space Model.

### 2.4.3 Systematic Procedure for Identification

The construction of model from input-output data involves the following basic entities [1]:

1. **The Data :** There may be two possibilities: Specially designed identification experiment, where user may determine which signals to measure & when to measure them & may also choose the input signals or the user may not have the possibility to affect the experiment, but must use data from the normal operation of the system.

2. **A Set of Candidate Models :** By specifying within which the collection of models we are going to look for a suitable one, a set of candidate models is obtained. Then a model with some unknown physical parameters is developed from basic physical laws other well established relationships. In other cases two types of linear models can be employed:

- **Black Box Model:** Models set with adjustable parameters with no physical interpretation.

- **Grey Box Model:** Model set with adjustable parameters with physical interpretation

3. **A Rule by Which Candidate Models Can be Assessed Using the Data:** Guided by data, the best model is determined. The assessment of model quality is typically based on how the model performs when they attempt to reproduce the measured data.

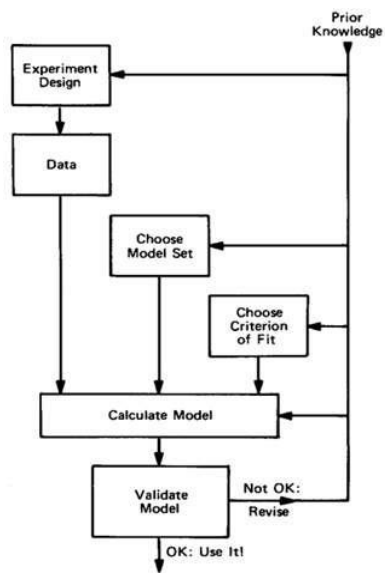

Figure 2.10: System Identification Loop [1]

## 2.4.4 The Toolbox Used For System Identification

We are using the **MATLAB System Identification Toolbox** for identification & modeling of the DC motor plant. MATLAB functions, Simulink blocks,  and an app for constructing mathematical models of dynamic systems from measured input-output data are provided by the System Identification  Toolbox.

The toolbox lets us create and use models of dynamic systems not easily modeled from first principles or specifications. We can use time-domain and frequency-domain input-output data to identify continuous-time and discrete-time transfer functions, process models, and state-space models. The toolbox also provides algorithms for embedded online parameter estimation. The app can be started using the MATLAB command **ident**.

The toolbox provides identification techniques such as maximum likelihood, prediction-error minimization (PEM), and subspace system identification. To represent nonlinear system dynamics, you can estimate Hammerstein-Weiner models and nonlinear ARX models with wavelet network, tree-partition, and sigmoid network nonlinearities. The toolbox performs grey-box system identification for estimating parameters of a user-defined model.

The principal contributor of the toolbox is **Prof. Lennart Ljung** from **the Department of Automatic Control, Linkoping University**.

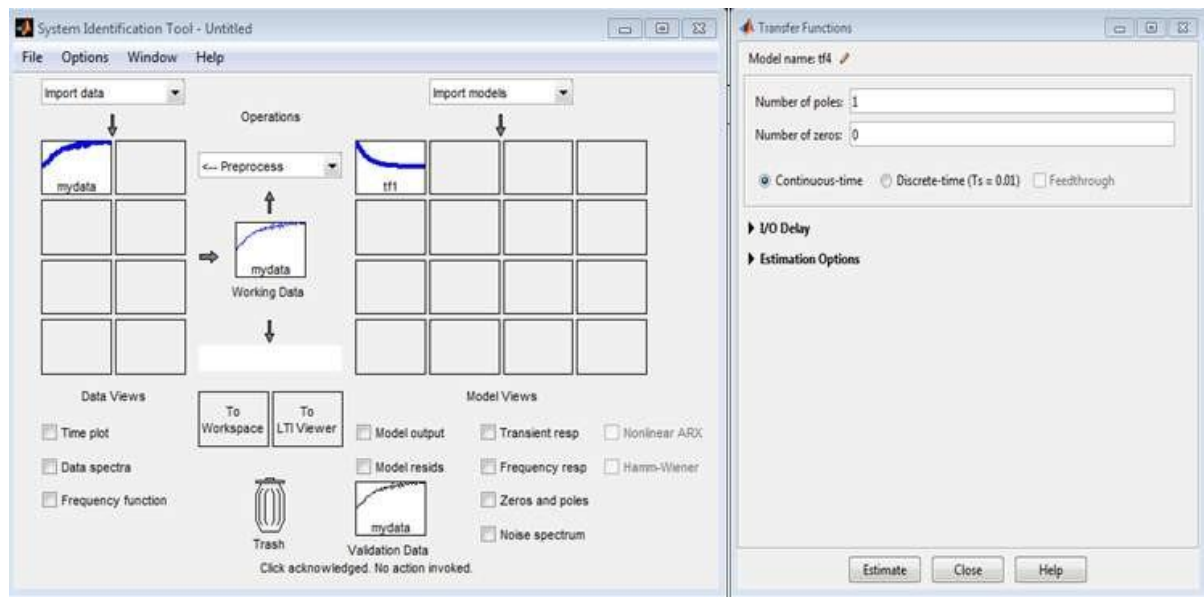

Figure 2.11: System Identification Toolbox App User Interface

## 2.4.5 The Data Samples Used For Model Identification

For identification & modeling of the DC Motor system the input-output data used are as follows:

- Input :Voltage (Range: 0-12.5 Volt) (It is the PWM voltage output from Motor Driver IC) (Randomly given with potentiometer)

- Output: Speed in RPS (obtained from Digital Speed Encoder as a number)

We are using two sets of independent practical input-output data for the identification process:

- Data Set for Model Identification : 2500 samples (Graph shown in  figure 2.12)

- Data Set for cross validation of identified model : 998 samples (Graph shown in figure 2.13).

The sampling time for each data set is selected as 20*ms*.

It is important to note the the input-output data sets are normalized to values (0-1) during identification process and the normalization is done using formula,

$$x_{norm} = \frac{x - x_m in}{x_m ax - x_m in} \tag{2.7}$$

Where, x is any data point in the input-output data set.

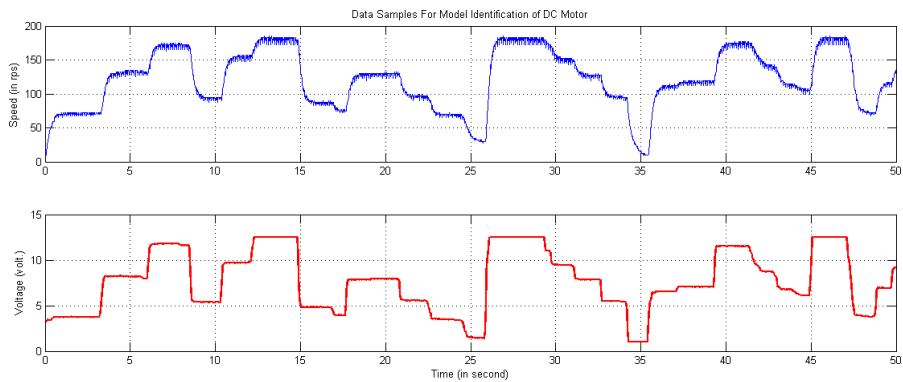

Figure 2.12: Data Set for Model Identification

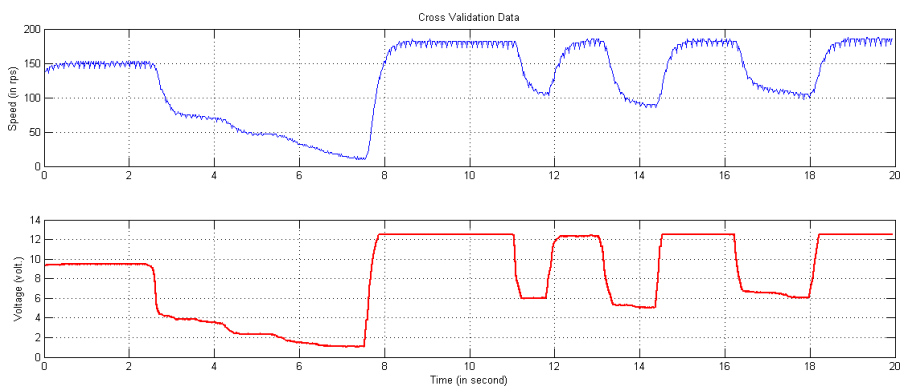

Figure 2.13: Data Set for Cross Validation of identified Model

## 2.4.6 Linear Least-Squared Error Minimization for  System Identification

It is a technique used for minimizing errors during System Identification. It is also called Prediction Error Minimization  [1].

For simplicity let us take a simple ARX(221) (Auto Regressive Exogeneous Input with $n_a = 2$, $n_b = 2n_k = 1$) model (equation-error structure) consisting of 2 input & 2 output parameters as [1],

$$A(q)y(t) = B(q)u(t) + e(t) \tag{2.8}$$

Where, $y(t)$ is the output signal, $u(t)$ is the input signal & $e(t)$ is the error signal or white noise [1].

Here,

$$A(q) = 1 + a_1q^{-1} + a_2q^{-}2 \tag{2.9}$$

$$B(q) = b_1q^{-1} + b_2q^{-}2 \tag{2.10}$$

Where, $q^{-1}$ is the delay or shift operator i.e. $q^{-l}x(t) = x(t - l)$

The transfer function of the plant when mean squared error is zero or minimum is given as [1],

$$G(q) = \frac{B(q)}{A(q)} \tag{2.11}$$

It has been assumed that $B(q)$ & $A(q)$ has no factor in common.

Now,from equation 2.8 we can write the equation of the system as [1],

$$y(t) = -a_1y(t-1) - a_2y(t-2) + b_1u(t-1) + b_2u(t-2) + e(t) \tag{2.12}$$

Now, from equation 2.12, we can define a linear *regressor* matrix as [1],

$$\Phi = \begin{bmatrix} -y(t-1) & -y(t-2) & u(t-1) & u(t-2) \end{bmatrix}$$

Also, we define parameter matrix as [1],

$$\theta = \begin{bmatrix} a_1 & a_2 & b_1 & b_2 \end{bmatrix}^T$$

Now from equation 2.12 we get,

$$e(t) = y(t) - \Phi\theta \tag{2.13}$$

Now, we have to minimize $e^2(t)$ by using the *regressor* matrix, therefore we define a objective function $J(\theta)$ as [1],

$$J(\theta) = e^2(t) = [y(t) - \Phi\theta]^T \tag{2.14}$$

Now, gradient of $J$ is given as,

$$\frac{\partial J(\theta)}{\partial \theta} = -2\Phi^T y + 2\Phi^T \Phi \theta \qquad (2.15)$$

The least squared error will be minimum when the gradient of $J$ is zero, which gives,

$$\theta = (\Phi^T \Phi)^{-1} \Phi^T y(t) \qquad (2.16)$$

Now this parameters of equation 2.16 obtained by least square estimation give us the transfer function of the plant. Here, the estimation is done in terms of the parameters in the continuous time system which are sampled to match the observed data.

## 2.4.7 Nonlinear Least-Squared Error Minimization for System Identification

This technique is used by the MATLAB System Identification Toolbox for identification of the continuous time model of a plant based on input-output data. Nonlinear least squares regression extends linear least squares regression for use with a much larger and more general class of functions. A nonlinear regression model can incorporate almost any function that can be written in a closed form. There are very few limitations on the way parameters can be used in the functional part of a nonlinear regression model which is not the case in linear regression. However, conceptually the unknown parameters in the function are estimated in the same way as it is in linear least squares regression [2].

A non-linear model is any model of the form [2]:

$$y = f(x; \beta) + s \qquad (2.17)$$

Where:

1. the functional part of the model is not linear with respect to the unknown parameters $\beta_0$, $\beta_1$, ...

2. the method of least squares is used to estimate the values of the unknown parameters.

   However, it is often much easier to work with models that meet two additional criteria because of the way in which the unknown parameters of the function are usually estimated. Those are:

3. the function is smooth with respect to the unknown parameters.

4. there is a unique solution for the least squares criterion that is used to obtain the parameter estimates.

These last two criteria are not essential parts of the definition of a nonlinear least squares model, but hold practical importance [2].

The biggest advantage of nonlinear least squares regression over many other techniques is the broad range of functions that can be fit. Many scientific and engineering processes can be described well using linear models, or other relatively simple types of models, but still many other processes that occur are inherently nonlinear. For example, the strengthening of concrete as it cures is a nonlinear process. Research on concrete strength shows that the increase in strength is quick at first and then levels off, or it approaches an asymptote in mathematical terms, over time. Processes that asymptote are not very well described by linear models because for all linear functions the function value cant increase or decrease at a declining rate as the explanatory variables go to the extremes. On the other hand, many types of nonlinear models describe the asymptotic behavior of a process well. Similarly, other features of physical processes can often be expressed more easily using nonlinear models than with simpler model types [2].

Being a least squares procedure, nonlinear least squares has similar advantages (and disadvantages) as that of the linear least squares regression over other methods. One common advantage is efficient use of data. With relatively small data sets, nonlinear regression can produce good estimates of the unknown parameters in the model. Another advantage that both linear and nonlinear least squares share is a fairly well-developed theory for computing confidence, prediction and calibration intervals to answer scientific and engineering questions. In most cases the probabilistic interpretation of the intervals produced by nonlinear regression are only approximately correct, but these intervals still work very well in practice [2].

The major cost of moving from simpler modeling techniques like linear least squares to nonlinear least squares regression is the need to use iterative optimization procedures to compute the parameter estimates. With functions that are linear in the parameters, the least squares estimates of the parameters can always be obtained analytically, while for nonlinear models it is generally not the case. For the iterative procedures, the user is required to provide starting values for the unknown parameters before the software can start the optimization. The starting values must be reasonably close to the as yet unknown parameter estimates or the optimization procedure may not converge. If the starting values are not close enough to the unknown parameters, then the software might converge to a local minimum rather than the global minimum which defines the least squares estimates [2].

Common disadvantages with the linear least squares procedure includes a strong sensitivity to outliers. Similar to a linear least squares analysis, the presence of one or two outliers in the data can seriously affect the results of a nonlinear analysis. In addition, unfortunately model validation tools are fewer for the detection of outliers in nonlinear regression than there are for linear regression [2]

## 2.4.8 Mathematical Approach to Modeling of DC Motor

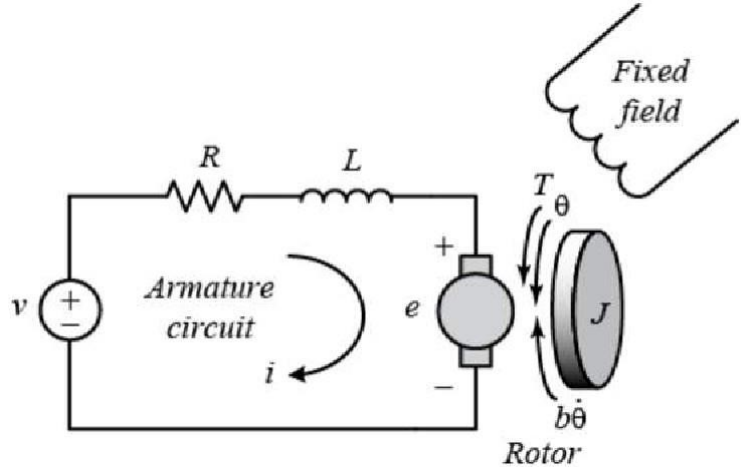

Figure 2.14: The Circuit Representation of A Permanent Magnet DC Motor [4]

Where, $v$ is the terminal voltage of the motor, $J$ is the rotational inertia of of the motor, $b$ is the viscous friction constant, $e$ is the back emf, $i$ is the armature current & $\theta$ is the angular rotation of motor shaft.

Therefore, $\frac{d\theta}{dt} = \omega$ (speed of the motor in rps)

Here, the motor is permanent magnet type, the field flux $\varphi$ is constant.

The motor torque $T$ is proportional to only the armature current $i$ by a constant factor called motor torque constant $K_t$ as [4],

$$T = K_t i \qquad (2.18)$$

The back emf, $e$, is proportional to by a constant factor $K_e$ as [4],

$$e = K_e \omega \qquad (2.19)$$

Now, based on Newton's 2nd law of motion & Kirchhoff's voltage law, the governing equations of a DC motor can be written as [4],

$$J \frac{d\omega}{dt} + b\omega = Ki \qquad (2.20)$$

$$L\frac{di}{dt} + iR = V - K\omega \tag{2.21}$$

Here, $K = K_t = K_e$ in SI unit.

Now assuming initially relaxed condition, the transfer function of the motor is obtained by taking Laplace Transform of the equations (1.12) & (1.13) as,

$$G(s) = \frac{\omega(s)}{v(s)} = \frac{K}{(Js + B)(Ls + R) + K^2} \tag{2.22}$$

Therefore equation (1.14) represents the standard 2nd order continuous domain model of a permanent magnet DC motor.

For the motor used in our experiment, we are devoid of the values $J$, $b$, $L$ of the motor, hence we have decided to go with the system identification approach.

### 2.4.9 Result of System Identification of the DC Motor

Using the MATLAB System Identification toolbox, we tried to approximate both the 1st order & 2nd order standard transfer function of the DC motor using the data set for model identification.

The 1st order model comes out to be,

$$G(s) = \frac{4.159}{s + 3.888} \tag{2.23}$$

The 2nd order model comes out to be,

$$G^j(s) = \frac{10.84}{s^2 + 338.6s + 155.5} \tag{2.24}$$

### 2.4.10 Selection of Model Representing the Actual System

The inputs of the cross validation data are applied to the identified models $G(s)$ & $G^j(s)$ and the responses are obtained. Both the identified models $G(s)$ & $G^j(s)$ give similar fitting to the cross validation data set.

The Percentage fitting of identified model response $y_{id}$ to actual plant response y, for the cross validation data set, is given by [3],

$$percentage fit = 100 * (1 - \frac{norm(y_{id} - y)}{norm(y - mean(y))}) \tag{2.25}$$

Now out of the two identified models $G(s)$ & $G^j(s)$, the model $G(s)$ is selected as the actual representation of the system based on the following criteria,

- Both $G(s)$ & $G^j(s)$ provides similar fitting to the cross validation data, but the order of G'(s) is more than G(s). Hence, from complexity point of view G(s) is better representation of the DC motor plant.

- Also if we observe the root locus plot, then one of the poles of $G^j(s)$ lies very far from the origin on s-plane. Hence, $G^j(s)$ almost behaves like a first order system.

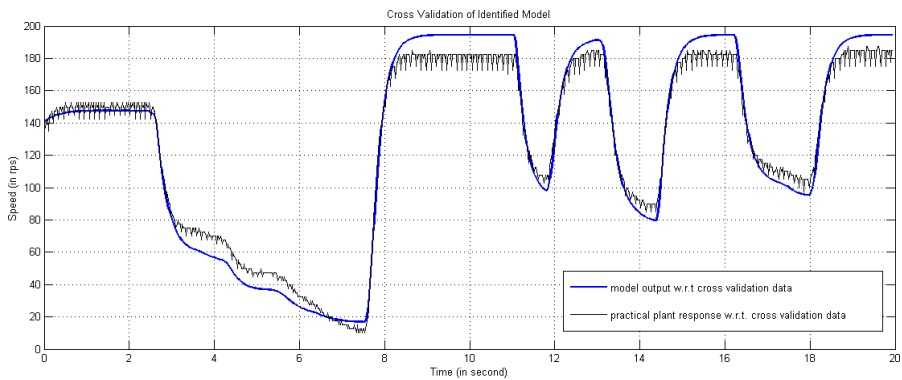

Figure 2.15: The Response of $tt(s)$ to the Cross Validation Data

The percentage fitting given by $G(s)$ to the cross validation data is **83.75 %**.

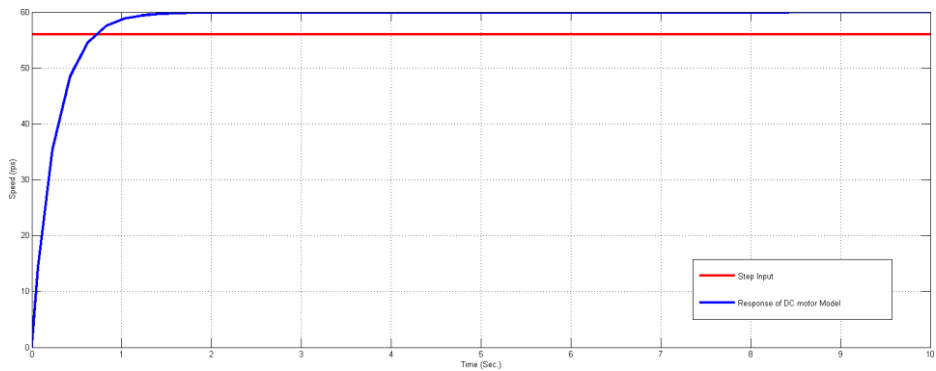

Figure 2.16: Step Response of Identified Model $tt(s)$

Therefore, from the system identification experiment, we get the transfer function of the DC motor as,

$$G(s) = \frac{4.159}{s + 3.888} \qquad (2.26)$$

The step response of the identified model is shown in figure 2.16 We can also represent the transfer function of the DC motor as,

$$G(s) = \frac{1.0697}{\tau s + 1} \tag{2.27}$$

Where, $\tau = 0.2577$ s, is the time constant of the system.

## 2.5 Sampling Time Decision For Discretization of the Plant

Selection of proper sampling time is a matter of utmost importance for a discrete data control system due to the following reasons,

- A very low sampling time leads to unnecessary computational overhead.

- To explain or understand observed phenomena

- A very high sampling time may lead to loss of significant information about the system dynamics.

- A sufficiently high sampling time may lead to deterioration of performance & even to instability in a discrete data control system.

Therefore we need to keep the points above in mind while discretizing the continuous time transfer function & while implementing the closed loop system. Sampling time should be such the chance of data loss is minimized at the same time leaving enough time for the micro controller process the data within two sampling instants.

Usually selection of sampling time depends greatly on the measurement & instrumentation system. In our experiment the speed encoder has a good resolution when the sampling time is $20ms$ i.e the encoder can give an acceptable speed measurement at a minimum interval of $20ms$ between two consecutive samples. Also, for execution of a single statement, the Arduino micro controller requires a time of the order of $10^-6$ second. Therefore, the selection of sampling time as $T_s = 20ms$ leaves the micro controller with enough time for processing of data and also does not lead to loss of significant information. The DC motor identified transfer function has a time constant of $\tau = 0.2577 second$ which is almost 13 times of $20ms$. Therefore we have selected sampling time as $20ms$ for discretization of the transfer function of the DC motor as well as for the discrete system design.

# 2.6 Discretization of the Closed Loop System

The functional block diagram of the closed loop system is shown below.

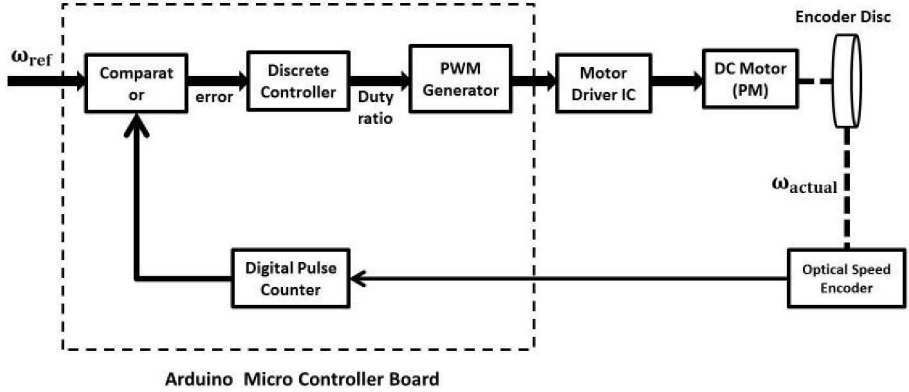

Figure 2.17: The functional block diagram of the basic embedded DC motor speed control system with single micro controller board

For simplicity in realization of the control theory block diagram, from figure 2.17, the PWM generator block of the micro controller board is considered as the Zero Order Hold Circuit (ZOH) block as it converts the digitally calculated duty ratio by the dicrete controller into output DC analog PWM voltage and it holds a value of the duty ratio until next sample comes. The comparator can simply be represented as a summer block and the driver IC (actuator) along with the DC motor plant has been simply represented as DC motor block. The encoder disc along with the speed optical encoder and the pulse counter is considered as the sensor block. The sensor block output is the reading of speed of the DC motor plant and hence the feedback loop is considered to be unity feedback.

## 2.6.1 The Closed Loop Pulse Transfer Function

Here,

$G_p(s)$ = Transfer Function of DC Motor Plant

$G_h(s)$ = Transfer Function of ZOH circuit (D/A Converter)

$G(s) = G_p(s) * G_h(s)$

$G_c^*(s)$ = Transfer Function of Discetre Controller

$H(s)$ = Transfer Function of speed sensor = 1 (taken as unity)

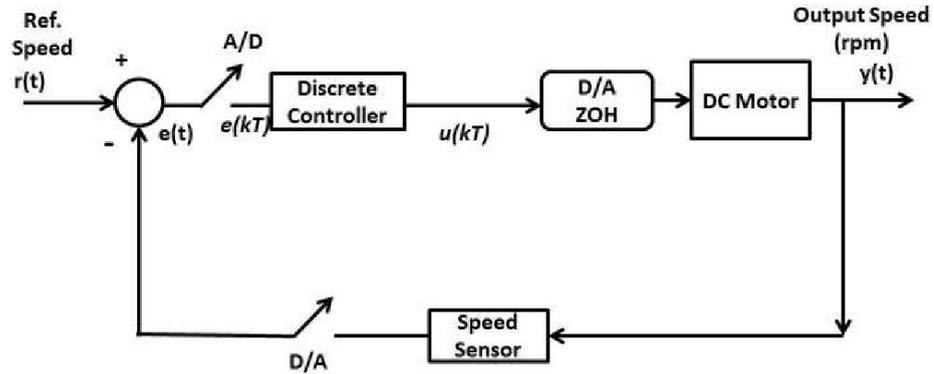

Figure 2.18: The simplified block diagram representation of the DC motor speed control system

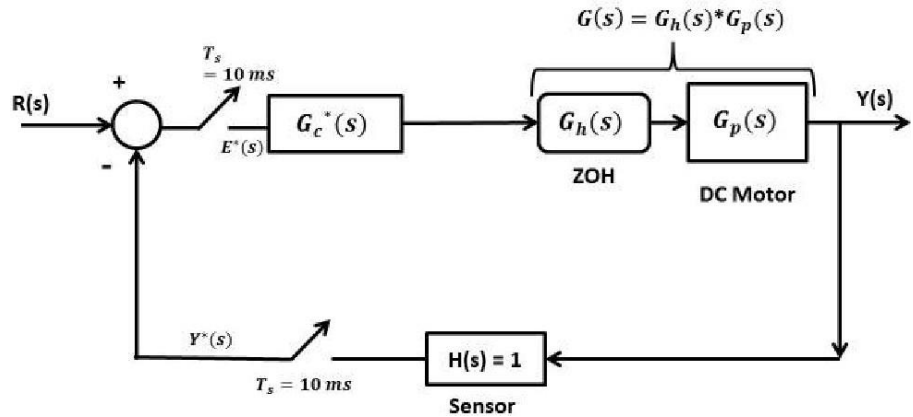

Figure 2.19: The simplified s-domain block diagram representation of the DC motor speed control system

From figure 2.19,

$$E^*(s) = R(s) - Y^*(s) \tag{2.28}$$

$$Y(s) = E^*(s)G_c^*(s)G(s) \tag{2.29}$$

Now, equation 2.28 ⇒

$$E^*(s) = R(s) - E^*(s)G_c^*(s)G(s) \tag{2.30}$$

By taking starred Laplace transform on equation 2.30, we obtain,

$$E^*(s) = R^*(s) - E^*(s)G_c^*(s)G^*(s) \tag{2.31}$$

$$\Rightarrow E^*(s) = \frac{R^*(s)}{1 + G_c^*(s)G^*(s)} \tag{2.32}$$

Now, taking starred Laplace transform on  equation2.29 $\Rightarrow$

$$Y^*(s) = E^*(s)G_c^*(s)G^*(s) \tag{2.33}$$

Using2.32

$$Y^*(s) = \frac{R^*(s)}{1 + G_c^*(s)G^*(s)}(s)G_c^*(s)G^*(s) \tag{2.34}$$

$$\Rightarrow \frac{Y^*(s)}{R^*(s)} = \frac{G_c^*(s)G^*(s)}{1 + G^*(s)G^*(s)} \tag{2.35}$$

The transfer function of the closed loop system is,

$$C^*(s) = \frac{Y^*(s)}{R^*(s)} = \frac{G_c^*(s)G^*(s)}{1 + G_c^*(s)G^*(s)} \tag{2.36}$$

Therefore, the pulse transfer function of the closed loop system is,

$$C(z) = \frac{Y(z)}{R(z)} = \frac{G_c(z)G(z)}{1 + G_c(z)G(z)} \tag{2.37}$$

Here,

$$G_p(s) = \frac{4.159}{s + 3.888} \tag{2.38}$$

$$G_h(s) = \frac{1 - e^{-sT_s}}{s} \tag{2.39}$$

$T_s$ = sampling time = 20 ms

Therefore,

$$G(s) = G_h(s) * G_h(s) = \frac{1 - e^{-sT_s}}{s}\frac{4.159}{s + 3.888} \tag{2.40}$$

Now, taking z transform,

$$G(z) = (1 - z^{-1})\mathsf{Z}[\frac{4.159}{s + 3.888}] \tag{2.41}$$

$$\Rightarrow G(z) = (1 - z^{-1})\mathsf{Z}[\frac{1.0697}{s} - \frac{1.0697}{s + 3.888}]$$

(2.42)

$$\Rightarrow G(z) = (1 - z^{-1})[\frac{1.0697z}{z - 1} - \frac{14.03z}{z - 0.92}] \tag{2.43}$$

$$\Rightarrow G(z) = \frac{0.0831}{z - 0.92} \tag{2.44}$$

### 2.6.2 Pulse Transfer Function of Dicrete PI Controller

The PI control action for analog controllers is given by [6],

$$u(t) = K_c[e(t) + \frac{1}{T_i}\int_0^t e(t)] \tag{2.45}$$

Where,

$u(t)$ = output of the  controllers

$e(t)$ = error input to the controllers

$K_c$ = proportional gain of controllers

$T_i$ = integral time or reset time of controllers

To obtain the pulse transfer function, we discretize equation 2.45 with integral term approximated by trapezoidal summation. We get,

$$u(kT) = K_c[e(kT) + \frac{T}{T_i}[\frac{e(0) + e(T)}{2} + \frac{e(T) + e(T)}{2T} + ......... + \frac{e((k-1)T) + e(kT)}{2}]] \tag{2.46}$$

$$\Rightarrow u(kT) = K_c[e(kT) + \frac{T}{T_i}[\sum_{h=1}^{k}\frac{e((h-1)T) + e(hT)}{2}]] \tag{2.47}$$

Where,

T = sampling interval

We take,

$$\frac{e((h-1)T) + e(hT)}{2} = f(hT), f(0) = 0 \tag{2.48}$$

Taking Z-transform,

$$Z[\frac{e((h-1)T) + e(hT)}{2}] = \frac{1 + z^{-1}}{2}E(z) = F(z) \tag{2.49}$$

Now,

$$Z[\sum_{h=1}^{\Sigma} f(hT)] = Z[\sum_{h=0}^{\Sigma} f(hT)] - f(0) = \frac{1}{1 - z^{-1}}F(z) = \frac{1 + z^{-1}}{2(1 - z^{-1})}E(z) \tag{2.50}$$

Taking Z-transform on equation 2.47 using equation 2.50,

$$U(z) = K_c[1 + \frac{T}{2T_i}\frac{1 + z^{-1}}{1 - z^{-1}}]E(z) \tag{2.51}$$

$$\Rightarrow U(z) = K_c[1 - \frac{T}{2T_i} + \frac{T}{T_i}\frac{1}{1 - z^{-1}}]E(z) \tag{2.52}$$

$$\Rightarrow U(z) = [(K_c - K_c\frac{T}{2T_i}) + K_c\frac{T}{T_i}\frac{1}{1 - z^{-1}}]E(z) \tag{2.53}$$

Now, we take,

$$K_P = K_c - K_c\frac{T}{2T_i}, K_I = \frac{K_cT}{T_i} \tag{2.54}$$

Therefore, pulse transfer function of the PI controller becomes,

$$G_c(z) = \frac{U(z)}{E(z)} = K_P + K_I\frac{1}{1 - z^{-1}} \tag{2.55}$$

We can modify the pulse transfer function of PI Controller as,

$$G_c(z) = (K_P + K_I)\frac{z - \frac{K_P}{K_P + K_I}}{z - 1} \tag{2.56}$$

$$\Rightarrow G_c(z) = K\frac{z - \frac{K_P}{K}}{z - 1} \tag{2.57}$$

Where, $K = K_P + K_I$

### 2.6.3 Tuning of the Discrete PI  Controller

The transfer function of the PI controller implemented is given as,

$$G_c(z) = \frac{U(z)}{E(z)} = K_P + K_I \, \frac{1}{1-z^{-1}} \tag{2.58}$$

#### 2.6.3.1 Tuning Discrete PI Controller in MATLAB PID Tuner

Control System Toolbox of MATLAB has PID Tuner to perform automatic, interactive tuning of PID controllers for plants represented by LTI models.  We have used MATLAB PID controller tuner for tuning of the discrete PI controller for the DC motor speed control system. Configurable options availabble in the MATLAB PID Controller block include:

- Controller type (PID, PI, PD, P, or  I)

- Controller form (Parallel or  Ideal)

- Time domain (continuous or  discrete)

- Initial conditions and reset  trigger

- Output saturation limits and built-in anti-windup mechanism

- Signal tracking for bumpless control transfer and multiloop control

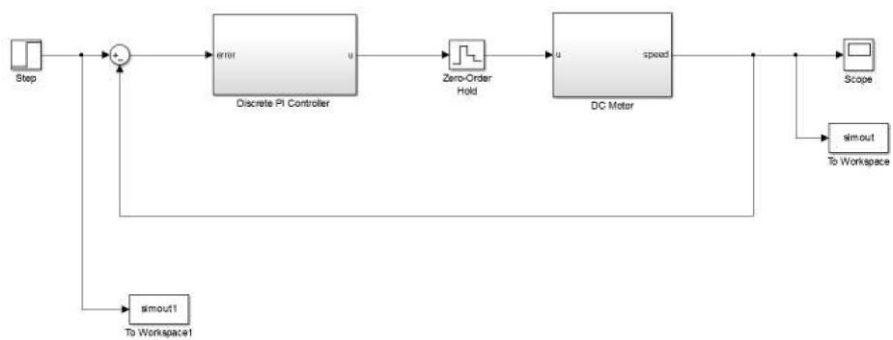

Figure 2.20: The SIMULINK block diagram of the closed loop system used in tuning

The default parallel form of PI controller is selected in the PID tuner for tuning purpose.The Backward Euler approximation of the integrator part which uses the Backward Rectangular (right-hand) approximation. An advantage of the Backward Euler method is that discretizing a stable continuous-time system using this method always yields a stable discrete-time result. We have also put a output saturation of the discrete PI controller to a value 255 which is the Arduino micro controller's highest value for PWM duty ration (100 %).

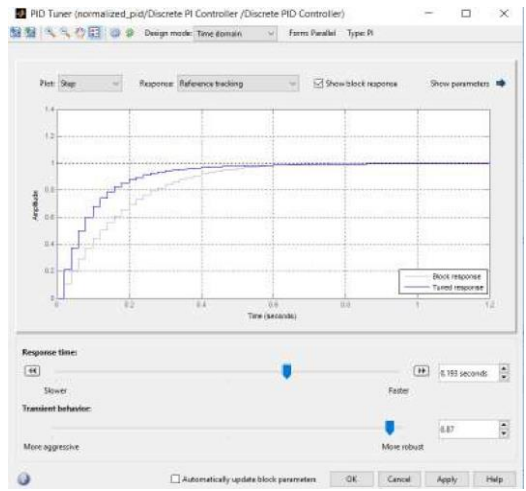

Figure 2.21: Tuning of the discrete PI controller in MATLAB PID tuner

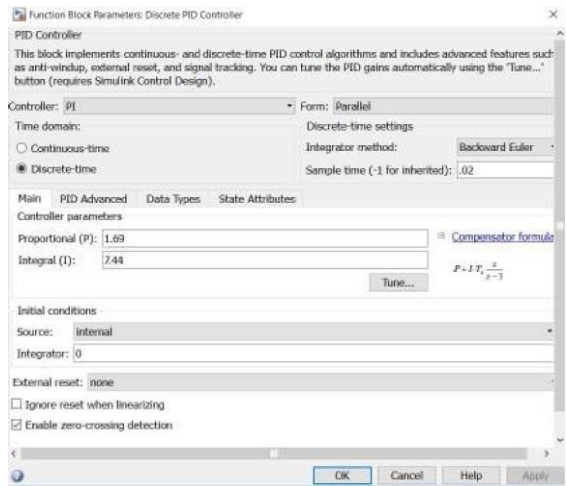

Figure 2.22: The interface of the MATLAB PID tuner

With proper tuning as per the practical requirements, the discrete PI controller parameters are obtained from MATLAB PID Tuner are, $K_p = 1.69$ & $K_I = 7.44$. It should be noted that the $K_I$ is to be multiplied by the sampling time during implementation.

### 2.6.3.2 Root Locus Design of Discrete PI Controller

From figure 2.19, using equation 2.57 and 2.44, the open loop transfer function of the speed control system is,

$$G_c(z)G(z) = K\frac{z - \frac{K_P}{K}}{z - 1}\frac{.0831}{z - 0.92} \tag{2.59}$$

We locate the open loop poles of the DC motor control system in the z-plane as shown in figure 2.23 below. Here, the location of open-loop zero of the PI controller need to be determined.

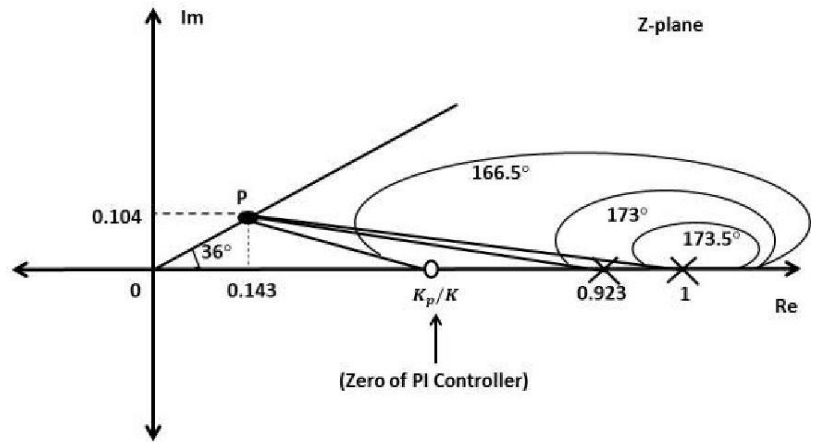

Figure 2.23: The representation of closed loop system in z-domain

We take a dominant closed-loop pole at point $P$ in the upper half of the z-plane.

The other design cosiderations in the design process are, damping coefficient, $\zeta = 0.94$ & $\omega_d/\omega_s = 1/10$ at point $P$ .

The ratio $\omega_d/\omega_s = 1/10$ means that there are 10 samples per cycle of damped sinusoidal frequency which ensures that there is no loss of significant information.

Now, for the closed loop system,

the magnitude of line joining origin & the point $P$ ,

$$|z| = exp(-\sqrt{\frac{2\pi\zeta}{1 - \zeta^2}}\frac{\omega_d}{\omega_s}) = 0.177 \tag{2.60}$$

the angle of the line at which point $P$ is located is,

$$\angle z = 2\pi\frac{\omega_d}{\omega_s} = 36° \tag{2.61}$$

Therefore, the location of closed loop pole at point $P$ is,

$$z = |z| \angle z = 0.177 \angle 36° = 0.143 + j0.104 \qquad (2.62)$$

Using angle criterion, the angle deficiency for point $P$ to be a closed loop pole is,

$$-173.5° - 173° + 180° = -166.5°$$

The PI controller zero at $z = K_P/K$ must contribute an angle compensation of $+166.5°$

Therefore, by drawing the required angle compensation, from 2.23, we get the PI controller zero at $z = 0.54$

$$\frac{K_P}{K} = 0.54 \qquad (2.63)$$

Now from equation 2.59, using magnitude criterion,

$$K \left| \frac{z - 0.54}{z - 1} \frac{0.0831}{z - 0.92} \right|_{z=0.143+j0.104} = 1 \qquad (2.64)$$

$$\Rightarrow K = 19.83 \qquad (2.65)$$

Now, from equation 2.63, $K_P = 10.7$

Again, $K = K_P + K_I = \Rightarrow K_I = 9.13$

Therefore, z-domain representation of the designed discrete PI controller is given as,

$$G_c(z) = \frac{U(z)}{E(z)} = 10.7 + 9.13 \frac{1}{1 - z^{-1}} \qquad (2.66)$$

Therefore, from the root locus design method, the PI controller parameters obtained are $K_P = 10.7$ and $K_I = 9.13$

## 2.7 Practical Implementation of Discrete PI Controller on $\mu C$

For practical implementation we have selected the PI controller parameters obtained through tuning with the help of the MATLAB PID Tuner. This is because the simulated response due to root locus designed PI controller is found to be aggressive (rise time = 20 ms) and that of the MATLAB PID tuner designed PI controller is little less (rise

time = 180 ms) than the time constant of the DC motor plant (257.7 ms). Also the settling time is quite larger in case of the root locus designed discrete PI controller. Figure 2.24 below shows the simulated responses of the speed control system with discrete PI controller designed by both root locus method and MATLAB PID Tuner.

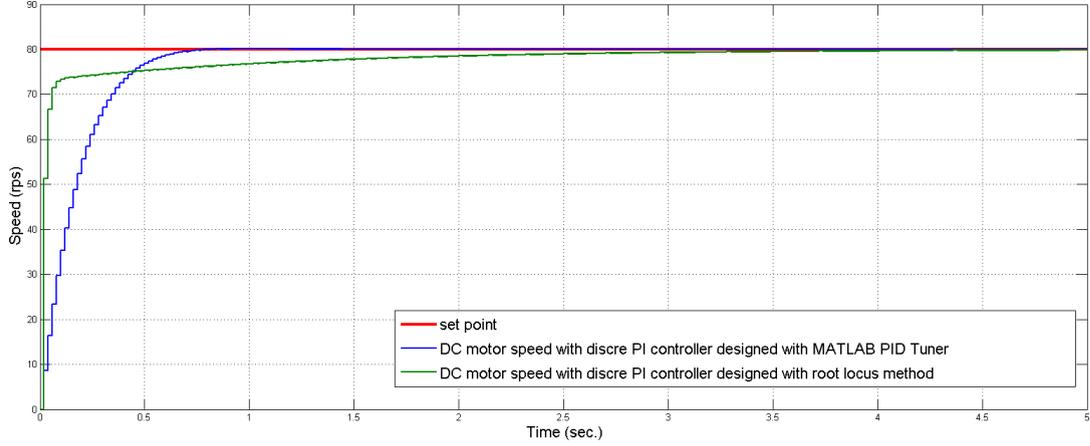

Figure 2.24: Comparison of system responses with different design strategy for discrete PI controller in SIMULINK

For practical implementation, the position algorithm of discrete PI controller is selected. The position algorithm is shown below in equation 2.67[6]. Here sampling time $T_s = 0.02 sec.$.

$$u(k) = K_P \left[ e(k) + \frac{T_s}{T_i} \sum_{k=1}^{k} e(k) \right] \tag{2.67}$$

Now, from the above equation,

$$K_I = \frac{K_P}{T_i} = 7.44 \tag{2.68}$$

$$K_P = 1.69 \tag{2.69}$$

Therefore the representation of the implemented PI controller in discrete time domain is,

$$u(k) = K_P e(k) + K_I * T_s \sum_{k=0}^{k=1} e(k) = 1.69 e(k) + 0.1488 \sum_{k=1}^{k} e(k) \tag{2.70}$$

The main theoretical problem with position algorithm of discrete PI controller is that we need to store all the past error values. This problem is avoided on the Arduino $\mu C$ by defining an integer variable for the summation of error signal part. An integer in the $\mu C$ is of 32-*bit* size & can give a value as high as $(2^{32} - 1)$ which is sufficiently large enough to sum up all the past errors. The discrete PI controller algorithm implemented
in micro controller is shown below.

1: $K_P$ , $K_I$, $T$ (sampling time), integral threshold
2: integral_sum = 0
3: $error$ = setpoint - output
4: integral_sum = integral sum + error
5: proportional = $K_P * error$
6:  integral part = $K_I * T_s *$ integral_sum
7: controller output = proportional + integral part
8: if (controller output > actuator_maximum) then (controller_output = actuator maximum
&& integral sum = integral_ threshold)
9: goto:step 3 if (setpoint available && output available)

The practical set up for implementation of DC motor speed control system is repre-
sented in figure2.25. Figure2.26, illustrates the response of the DC motor speed control
system in the wired setup with the discrete PI controller implemented in Arduino Due
board.

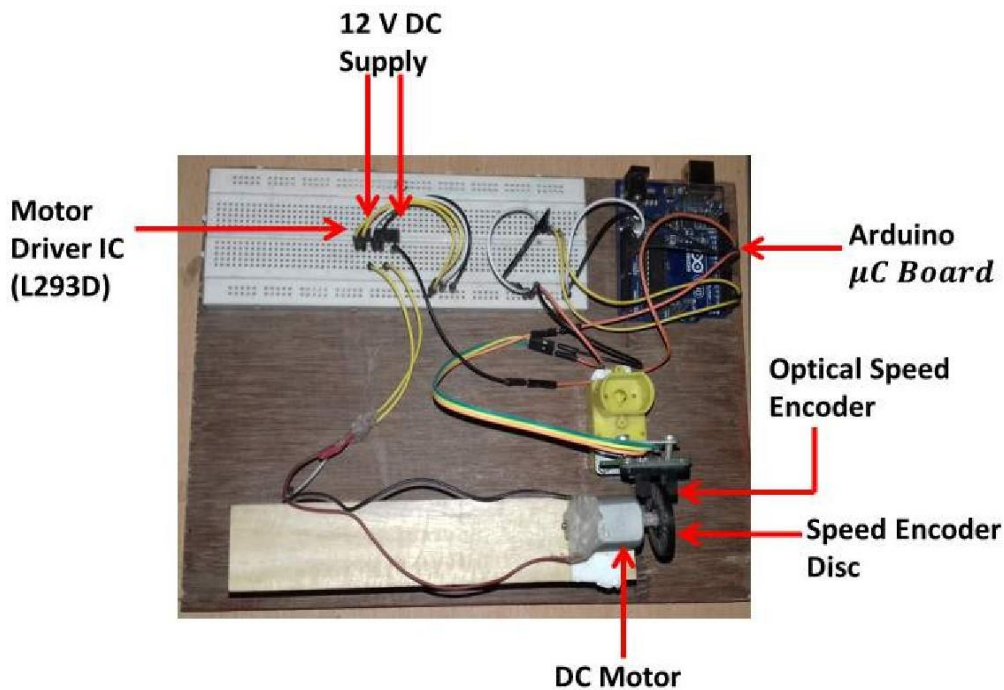

Figure 2.25: The practical setup for speed control of DC motor (wired fashion)

Figure2.27, illustrates the response of the DC motor speed control system in the wired
setup with the discrete PI controller implemented in Arduino Uno board.

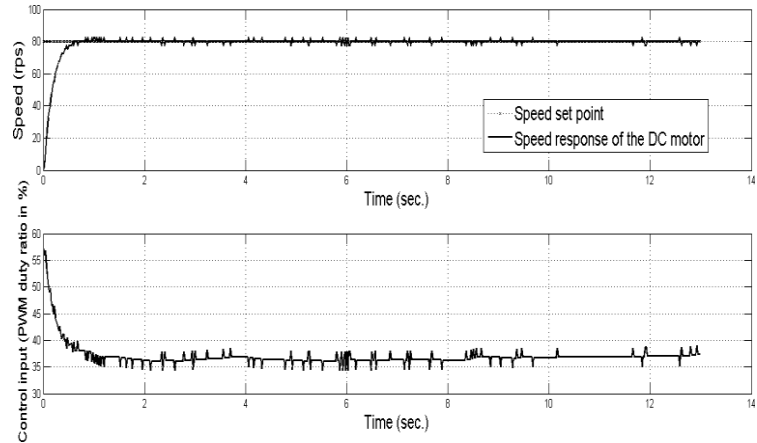

Figure 2.26: Wave forms of the wired speed control setup with discrete PI controller (with Arduino Due board)

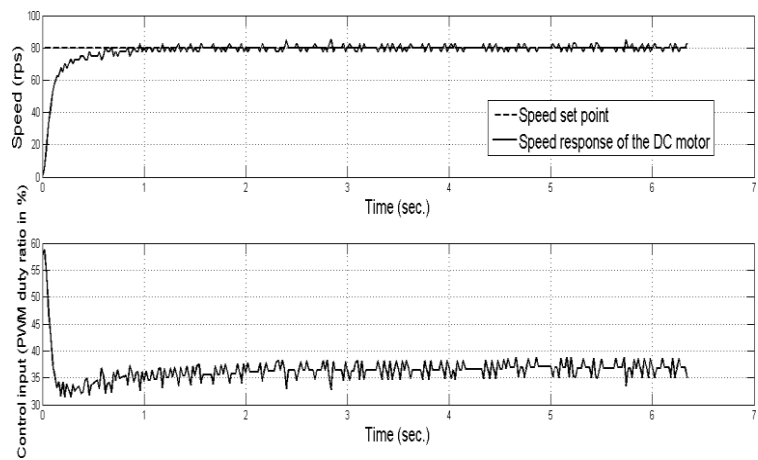

Figure 2.27: Wave forms of the wired speed control setup with discrete PI controller (with Arduino Uno board)

There are some jitters in the response of the practical closed loop system and the reasons to this may be as follows,

- The encoder disc attached to the shaft of the DC motor has 20 slots and for 20 ms sampling the resolution is 2.5 rps value. Hence speed encoder reading may fluctuate some times.

- The control input calculated by the discrete PI logic is quantized into an 8 bit number discarding the floating point numbers. This is because in Arduino embedded environment, PWM duty ratio is an 8 bit number (0-255) in discrete steps of 1.

- The resolution of PWM duty ratio of Arduino board is (100/255)% = 0.392%.

Also from figure 2.26 and 2.27, it is clear that Arduino Due performs better speed control due to its better numerical computational capability and quantization capability.

## 2.8 Chapter Summary

In this chapter we have discussed brief introduction to the DC motor speed control method adopted. Then components and devices used for practical set up has been discussed. Modeling of the DC motor plant using System Identification procedure has been performed and the transfer function of the plant has been obtained. Then we have tried discrete PI controller design based on the model of DC motor obtained from system identification using both Root Locus method and MATLAB PID tuner. MATLAB PID Tuner tuned PI controller is seen to be performing better in simulation and hence adopted for implementation in hardware. Then discrete PI algorithm implemented in micro controller has been given. Then we observed the response of the speed control system with the discrete PI controller implemented in Arduino embedded environment.

# Chapter 3

# Bluetooth Technology & The Wireless Networked Control System

## 3.1 Introduction

Bluetooth is the most popular *personal-area network* (PAN) system [2]. A *personal-area network (PAN)* is a very small wireless network that is created informally or on an ad hoc basis [2]. Bluetooth industry standard started developing from the late winter of 1998 when Ericsson, IBM, Intel, Nokia, and Toshiba formed the Bluetooth Special Industry Group (SIG) to develop and promote a global solution for short range wireless communication [3]. The IEEE standardized Bluetooth as IEEE 802.15.1, but standard is no longer maintained by IEEE [4]. The Bluetooth SIG supervises the development of the specification and manages the qualification program, and protects the trademarks [4].

The primary aim of Bluetooth wireless technology is to serve as the universal low cost, user friendly, wireless interface replaces the variety of proprietary cables that need to be carried to connect personal devices such as notebook computers, cellular phones, personal digital assistants (PDAs), digital cameras etc. [3]. Personal devices usually employ RS-232 serial port protocol. Therefore, It impossible to use the same set of cables to interconnect devices from different manufactures, and sometimes even from the same manufacturer due to need various types of proprietary connectors and pin arrangements [3]. Bluetooth provides replacement for cables to interconnect, and also to form ad hoc network between various personal devices.

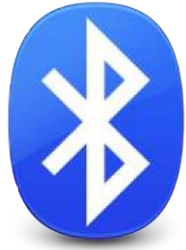

Figure 3.1: The Bluetooth Technology Logo

## 3.2 General Communication Network Architecture

The protocols & components required to satisfy application requirements are defined by architecture of a network [6]. Open System Interconnect (OSI) is a seven-layer reference model for explaining the architecture of a network [6]. OSI is developed by International Standards Organization (ISO). Complete set of network functions are specified by OSI and the OSI reference model is a instrumental model for depicting various standards and interoperability of a wireless network [6].

The different OSI layers and their network functionality are explained below as per [6]:

- **Application layer (layer 7):** Communications among users are established by this layer and the layer also gives basic communication services such as file transfer and e-mail. Examples of application of this layer are: Simple Mail Transfer Protocol (SMTP), HyperText Transfer Protocol (HTTP), File Transfer Protocol (FTP).

- **Presentation layer (layer 6):** This layer does negotiation of data transfer syntax for the application layer and translations between different data formats depending upon necessity.

- **Session layer (layer 5):** Sessions between applications are established, managed and terminated by this layer. In wireless networks, wireless middleware and access controllers provide this type of connectivity.

- **Transport layer (layer 4):** Mechanisms for establishment, maintenance, and orderly terminations of virtual circuits are provided by this layer, while the higher layers are shielded from the network implementation details. Virtual circuits are connections between network applications from one end of the communications circuit to another 9for example; between web browser on a PC to a server). Transmission Control Protocol (TCP) operates at this layer.

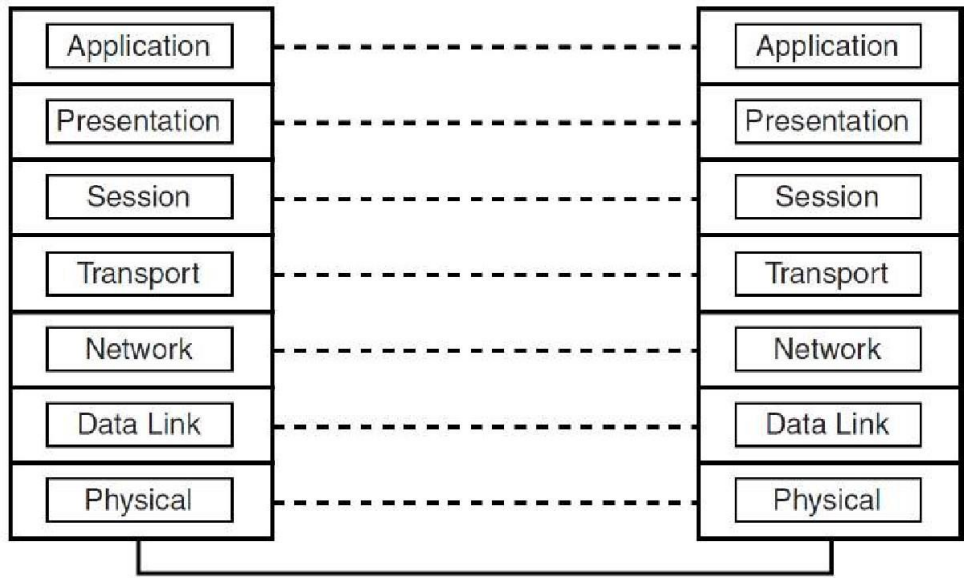

Figure 3.2: OSI Refernce Model Layers [6]

- **Network layer (layer 3):** This layer does routing data packets through the network from source to destination. This routing is done for ensuring that the data packets are sent in a direction leading to the destination. Protocols such as Internet Protocol (IP) operate at this layer.

- **Data link layer (layer 2):** This layer tries to ensure medium access, synchronization and error control between two entities. In case of wireless network, this layer is responsible for the coordination of access to the air medium and minimization of propagation error. Most of the wireless networks have a common technique for performing data link layer operations which is independent of actual way of transmission.

- **Physical layer (layer 1):** Actual transmission of information through the medium is provided by this layer. Radio waves and infrared light are included in this layer.

The combination of the different layers of OSI network architecture depicts the functionality of a wireless network. It is important to notice that wireless networks directly implement only the lower layers of the architecture [6]. it is seen that a wireless *Network Interface Card* (NIC) implements only the data link layer and physical layer functions. Each layer of the OSI architecture provides support to the layers above it. Figure 3.2 shows that each OSI model layer communicate across the network to the respective layer of the peer. But, the actual data transmission takes place only at physical layer. Therefore, in this layering process, a particular layer puts its protocol information into frames which are place within frames of lower layers [6]. The frame sent by the physical layer includes frames from all higher layers. At the destination, each layer sends the respective frames to the higher layers to ease the protocol between peer layers.

## 3.3 The Bluetooth Radio

Bluetooth is a digital radio standard that operates globally in the unlicensed 2.4 *GHz* ISM (Industrial, Scientific and Medical) short-distance radio frequency band [2,4]. Blutooth employs a radio technology called frequency-hopping spread spectrum (FHSS) [2,4]. Bluetooth operates over 79 channels with a frequency band 2.402-2.480 GHz in most of the countries and over 23 channels with frequency a band 2.4465 2.4835 GHz in very few countries [2,7].Each of the 79 or 23 channels has a bandwidth of 1 *MHz* [5]. The hop rate is 1600 hops/second and the dwell time on each frequency is $1/1600 = 625 \mu s$ [2]. Bluetooth divides transmitted data into packets, and transmits each packet on one of 79 or 23 designated Bluetooth channels [4]. There are also guard bands 2 *M Hz* wide at the bottom end and 3.5 *M Hz* wide at the top to prevent interference [4,7].

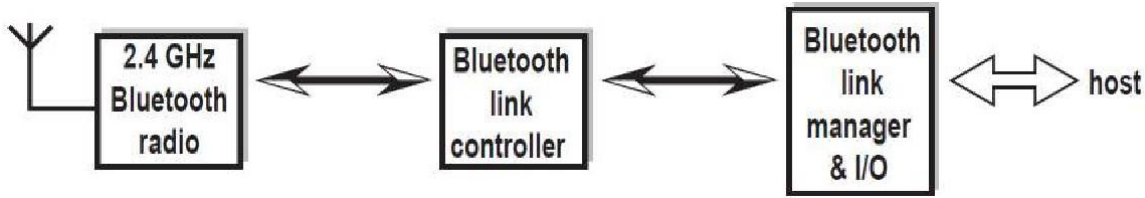

Figure 3.3: Functional Blocks in a Bluetooth System [7]

Since Bluetooth is a radio frequency communications technology, Bluetooth devices do not have to be in visual line of sight of each other [4]. Range of a Bluetooth device is power dependent and the effective range varies with application [4]. There are various classes of Bluetooth radios and their power range along with the range have been provided in Table3.1.

| Class | Max. Permitted Power (mW) | Max. Permitted Power (dBm) | Range (m) |
|-------|---------------------------|----------------------------|-----------|
| 1 | 100 | 20 | 100 |
| 2 | 2.5 | 4 | 10 |
| 3 | 1 | 0 | 1 |

Table 3.1: Different Classes of Bluetooth Radio [4]

# 3.4 Bluetooth as WPAN

A WPAN (Wireless Personal Area Network) is a wireless network for interconnecting devices centered around an individual person's workspace. Bluetooth PAN (Personal Area Network) is used for close-quarters connections between Bluetooth compatible devices such as PCs, laptops, mobile phones, PDAs. In earlier days, special-purpose proprietary cables have been used for interconnection of personal devices [7]. But user experience with these cables are not convenient as they are prone to damage, loss and aslo they bulky, messy to carry. So in order to adress these issues, WPAN technology has evolved. The Bluetooth WPAN technology utilizes a short-range radio communication link optimized for low power, battery operated, light weight and small personal devices [7]. It is created in a ad hoc or temporary manner whenever required.

## 3.4.1 General Requirements of a WPAN

There are various requirements for a WPAN to be accepted widely by people as well as industry and they are as follows [7]

- The WPAN technology should not effect the personal device's weight, power consumption, form factor, cost, convenience of use or other characteristics in a significant way.

- The WPAN technology should fulfill the design and marketing requirements of not only the consumer market but also the business market.

- Users should be able to use the WPAN connectivity to connect the personal device around their vicinity without data services or ethernet connectivity.

- Communicating devices under a WPAN should remain connected without line of sight of each other. This standard leads to use of radio frequency (RF) technologies in WPAN technology and Bluetooth is one of the RF communication technology.

## 3.4.2 Difference of WPAN from WLAN

Fundamentally, WPAN differs from Wireless Local Area Network (WLAN) in three ways [7]:

- **Power Levels and Coverage:** WLAN technology is intended and often optimized for larger coverage. Therefore, the power consumption of the devices for covering larger distances increases. WLAN devices are usually connected to a power plug or they use the wireless network for a relatively short time when unplugged. Therefore, WLAN is usually said to connect "portable" devices which are moved less frequently, powered from wall sockets and have relatively longer periods of connections

On the other hand, WPAN technology trades off coverage for lower power consumption. Therefore, WPAN is said to be oriented towards "mobile" personal devices which operate on batteries and have relatively lower periods of connection.

- **Control of the Medium:** WPAN implements a controlling mechanism that regulates the data transmissions of the devices connected to the network in order to provide required bandwidth guarantees for various connections. A typical WPAN device does not require to have a network-observable and network-controllable state. WPAN is ad hoc in nature i.e. a master must be present for communications to occur.

  On the other hand, WLAN does not usually employ controlling mechanism for regulating data transmissions of the devices as it may not be always desirable considering the large distance covered. Primary objective of WLAN is to provide the data services infrastructure typically available through a LAN to the client devices. Once a client device joins the WLAN network, the device typically remains connected to the WLAN until the device moves away from the network's boundaries. A typical WLAN device require to have a network-observable and network-controllable state being part of a larger infrastructure.

- **Life Span of Network:** WLAN's existence is independent of the constituent devices. Therefore, they are said to be existed uninterrupted. On the other hand, WPAN exists only when master device participates and the network no longer exists when the master does not participate. So the lifespan of WPAN is finite.

## 3.5 The Bluetooth Protocol Stack

The protocol stack of a communication network is nothing but hardware or software realization of the standard protocols which enables the connected devices to communicate with each other [1]. There are certain layers in the protocol stack with different applications.

The lowest layer of the Bluetooth protcol stack is the physical radio, which acts as the radio front and also depicts the attributes of the permissible power levels, channel arrangements, frequency bands and radio receiver's susceptivity [1]. The baseband layer looks after Bluetooth's physical and media access control. The devices to be paired share various control messages for managing and configuring the baseband configurations [1].

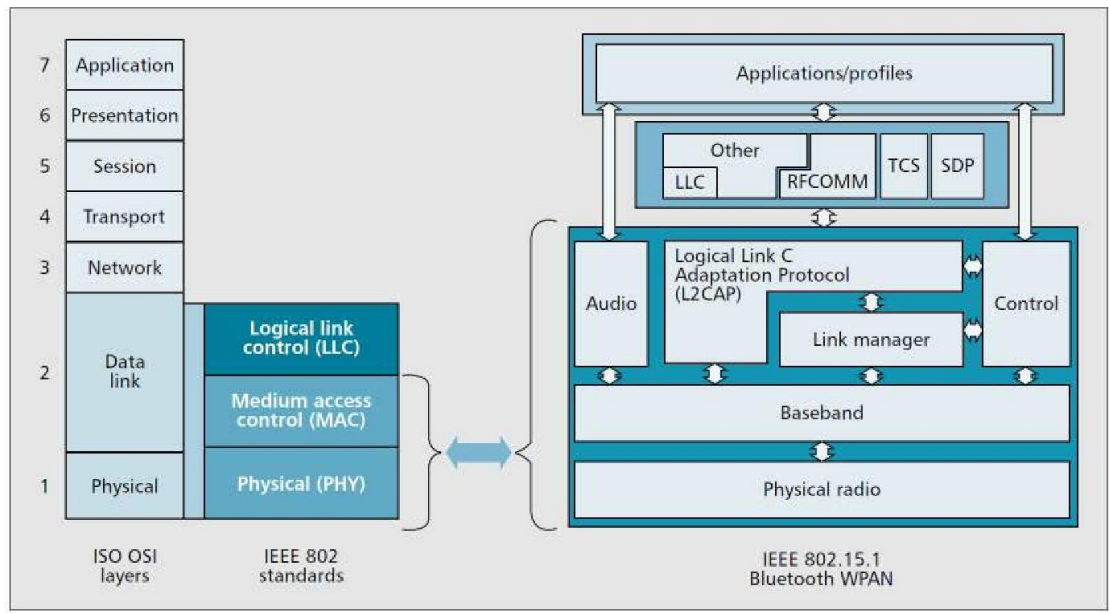

Figure 3.4: The Bluetooth Protocl Stack & Its Mapping with OSI ISO Reference Model[5,7]

The Link manager layer looks after the processing related to link manager protocol [1]. There are two different types of links exist in Bluetooth : *Asynchronous Connection Links* ACLs and *Synchronous Connection-oriented Links* SCOs [5]. An SCO link ensures guaranteed delay and bandwidth apart from the possible interruption caused by high priority link manager protocol messages [5]. SCO links are suitable for average quality music and voice transmission [5]. A slave can have three SCO links with the same master at max, or two SCO links with different masters at most and a master can have up to three SCO links with three different slaves at most [5]. ACLs are useful for non-real time traffic. Only a single ACL link can be there between a given master and a slave [5].

The control layer or the *Host Controller Interface* layer defines a standard interface or independent way of communication with the Bluetooth module [1]. The L2CAP (*Logical Link Control Adaptation Protocol* ) layer provides services to the upper levels off the connection and with the connection on [5]. The functions of L2CAP layer are as follows [5]

- Protocol Multiplexing.

- Segmentation and reassembly of the protocol data units that comes from upper layers.

- Quality of service (QoS) support.

Internet Protocol (IP) can be directly implemented in L2CAP layer. But, it is usually implemented in PPP (*Point-to-Point Protocol* ) layer above RFCOMM [5]. RFCOMM layer mainly does emulation of serial communication port [1]. Due to RFCOMM layer, Bluetooth chip or module can be easily connected to PC or other such devices using USB or UART protocols [1]. SDP (*Service Discovery Protocol* ) layer enables a Bluetooth slave device to find services and its peers [1]. TCS layer is basically provides telephony control specifications.

### 3.5.1 Security in Bluetooth  Protocol

Bluetooth protocol has certain security measures which are explained here as per [5]. Bluetooth protocol security can be divided into three modes as follows:

- **1st Mode:** Nonsecure.
- **2nd Mode** Service level enforced security (after establishment of channel).
- **3rd Mode** Link level enforced security (before establishment of channel).

Link level authentication and encryption are handled with the help of basic four entities:

- 48-bit unique Bluetooth device address assigned to each device.
- A random private authentication  key.
- A random private encryption  key.
- A 128-bit random number dynamically generated by each Bluetooth device. This number changes frequently.

In Bluetooth protocol, there are two security  levels

- **Trusted**.
- **Untrusted**

There are also three security levels defined for services:

- **Open services**.
- **Services with authentication  requirement**.
- **Services with authentication and authorization requirements**.

# 3.6 Bluetooth Communication Topology

Bluetooth communication is established when a master device gets paired with one or more slave devices [1]. A Bluetooth device is capable of operating in both *master* and *slave* role [5]. Therefore Bluetooth devices are capable of forming ad hoc networks. The master devices transmits in even slots and the slave devices transmits in odd slots [5], as shown in Figure3.5.

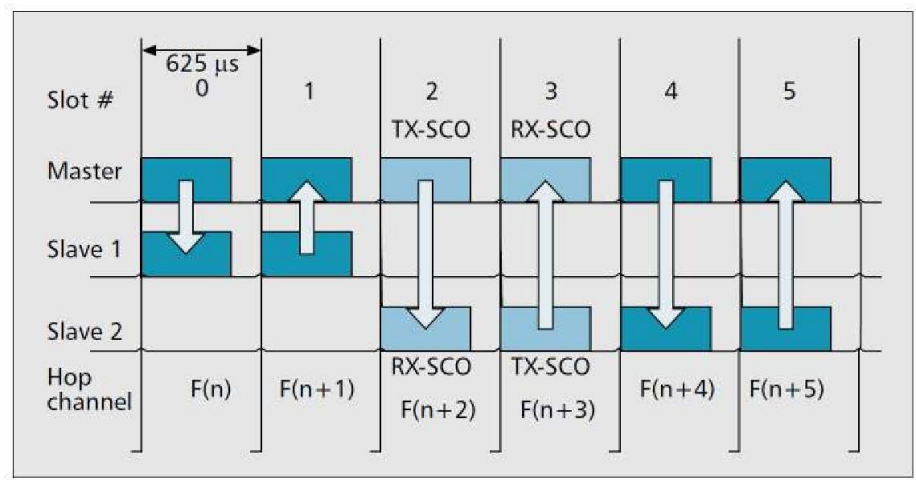

Figure 3.5: Example of data packet exchange in Bluetooth[5]

There are two distinct communication topologies with which devices get connected with each other [7]:

- **Bluetooth Piconet WPAN**

- **Bluetooth Scatternet WPAN**

## 3.6.1 Bluetooth Piconet WPAN

Piconet WPAN is a Bluetooth network formed by a device serving as a master and one or more slave devices connected together [7]. Each piconet is defined by a frequency-hopping channel based on the address of the master device [7]. Slave devices communicate with the master in point to point manner and master device may communicate in either point-to-point or point-to-multipoint manner [7]. In piconet topology, maximum seven slaves can get connected to the master at a time [5].

In piconet configuration, if a master wishes to get connected to more that seven slave devices at a time, then first it must command a slave to go to low power park mode and then request the other parked slave to become active [1]. Reiteration of this procedure allows the master to get paired with more than seven slaves. The master device determines allocation of bandwidth and mode of communication among the slaves [1].

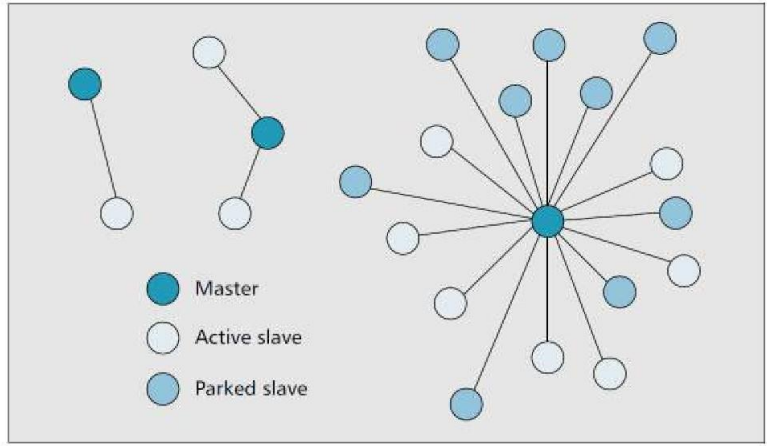

Figure 3.6: The Bluetooth Piconet Topology[5]

## 3.6.2 Bluetooth Scatternet WPAN

Bluetooth *scatternet* topology enables us to create *multihop* WPAN [5]. A multihop network is one where two nodes can communicate with each other with the help of other nodes as relays, without direct connection between them [5]. In scatternet configuration, a number of operational Bluetooth piconets overlap in time and space through common nodes [7]. A node in scatternet topology can be master in only one of the piconets, but can be a slave in several piconets [5,7].

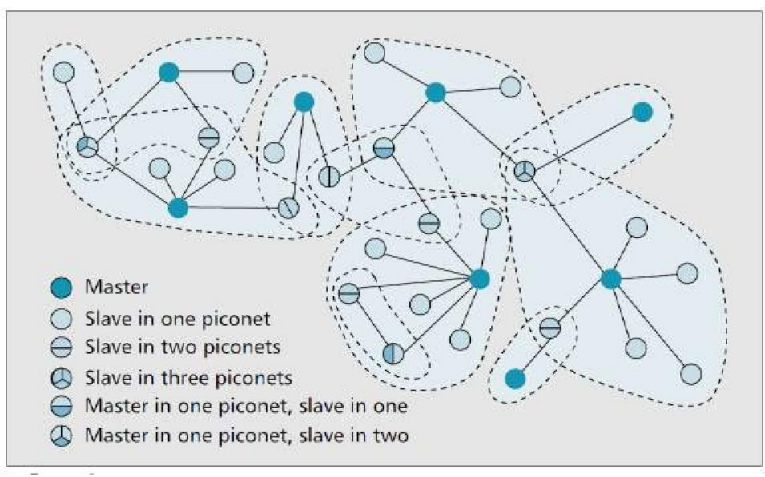

Figure 3.7: The Bluetooth Scatternet Topology[5]

Bridge nodes in overlapping areas of a scatternet network involves with members of different piconets on time sharing basis [1]. Interference between different piconets is not seen as they operate with different frequency hopping sequence [1].

## 3.7 States of a Bluetooth Device

After Bluetooth connection is established, a Bluetooth device can be at any one state among total eight states [1]. The eight states are explained below [1]:

- **Standby State**: When the Bluetooth device is powered on and it has not established any connection to other devices.

- **Inquiry State**: It is the state where a device sends request to discover devices around its coverage zone to pair up.

- **Page Mode:** Paging is the process of forming a connection between two Bluetooth devices. Before this connection can be initiated, each device needs to know the address of the other.

- **Connected State**: In connected state, a device establishes Bluetooth link with one or more other devices and to do this, the device requires an active address.

- **Transmit State**: A slave device is said to be in transmit state when it transmits data to master device.

- **Sniff State:** In sniff state a slave is said to be connected to piconet network virtually with a reduced duty cycle.

- **Park State:** A device in park state surrenders its active address and does not transmit data even if it is linked to the piconet network.

- **Hold State:** Hold mode is a temporary, power-saving mode where a device sleeps for a defined period and then returns back to active mode when that interval has passed. The master can command a slave device to hold.

## 3.8 Different Versions of Bluetooth

Since the inception of Bluetooth, different versions have been evolved. Various versions of Bluetooth are explained below [4]:

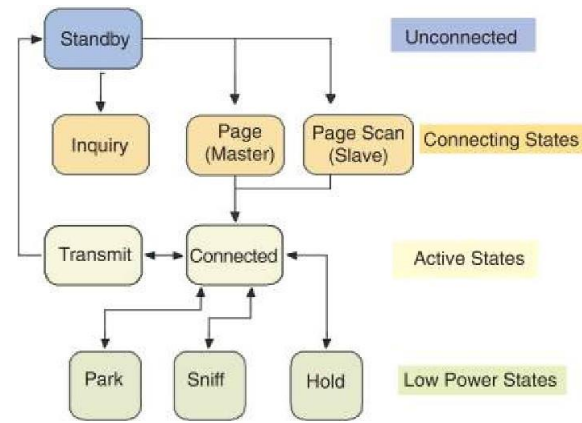

Figure 3.8: Different States of a Bluetooth Device[1]

- **Bluetooth 1.0 and 1.0B:** They are the earlier most versions of Bluetooth. Their connecting process involved mandatory hardware device transmission making anonymity impossible at the protocol level. This was a major setback for some certain services planned for Bluetooth communications. These devices also had problems with interoperability.

- **Bluetooth 1.1:** This version was ratified as *IEEE Standard 802.15.1-2002*. Many faults found in version 1.0B were fixed in this version. Possibility of non-encrypted channel and *Received Signal Strength Indicator* (RSSI) were also incorporated in this version of Bluetooth.

- **Bluetooth 1.2:** Major improvement in this version were faster connection and discovery. *Adaptive Frequency-Hopping* spread spectrum (AFH) has been introduced in this version, which improves resistance to radio frequency interference by avoiding the use of crowded frequencies in the hopping sequence. Extended Synchronous Connections (eSCO) have been introduced, which uplift voice quality of audio links by allowing retransmissions of corrupted packets, and may optionally increase audio latency to provide better concurrent data transfer. This version was retified as Ratified as *IEEE Standard 802.15.12005.*

- **Bluetooth 2.0 + EDR :** This version introduced Enhanced Data Rate (EDR) for faster data transfer. Although EDR is an optional feature. Apart from EDR, there other minor improvements are there in this version.

- **Bluetooth 2.1 + EDR:** This version was adopted by Bluetooth SIG in 2007. The key feature os this version is *Secure Simple Paring*. Other improvements include **Extended Inquiry Response**, which gives more information during inquiry procedure to enable better filtering of devices before connection.

- **Bluetooth 3.0 + HS:** The main feature of this version is *Alternative MAC/-PHY* (AMP) which enables the use of alternative MAC and PHYs to transport

Bluetooth profile data. This allows bluetooth to be in low power mode when in idle and to use faster radio when large data need to be sent. Enhanced power control feature was incorporated in this version.

- **Bluetooth 4.0 + LE:** It was incorporated in 2010 and called as *Bluetooth Smart*. This version includes Classic Bluetooth, Bluetooth high speed and Bluetooth low energy protocols. Bluetooth high speed is based on Wi-Fi, and Classic Bluetooth consists of legacy Bluetooth protocol.

- **Bluetooth 4.1 :** Formal introduction of Bluetooth 4.1 took place in 2013. This specification is an incremental software update to Bluetooth Specification v4.0, and not a hardware update. The update incorporates Bluetooth Core Specification Addenda (CSA 1, 2, 3 4) and adds new features that improve consumer usability.New features include increased co-existence support for LTE, bulk data exchange ratesand aid developer innovation by allowing devices to support multiple roles simultaneously. The other important new features included are *Low Duty Cycle Directed Advertising, L2CAP Connection Oriented and Dedicated Channels with Credit Based Flow Control, Dual Mode and Topology, Limited Discovery Time* etc.

- **Bluetooth 4.2:** It was introduced in 2014. This version introduced features for *Internet of Things*. Other major improvements are as folows:

  - Low Energy Secure Connection with Data Packet Length Extension.
  - Link Layer Privacy with Extended Scanner Filter Policies
  - Internet Protocol Support Profile (IPSP) version 6 ready for Bluetooth Smart things to support connected home.

- **Bluetooth 5:** Bluetooth 5 was unveiled in June, 2016. The feature of this version is mailny focused on emerging *Internet of Things* technology. It also has quadruple the range, double the speed, and provides an eight-fold increase in data broadcasting capacity of low energy Bluetooth transmissions compared to Bluetooth 4.x versions, which could be important for IoT applications where nodes are connected throughout a whole house. It is important to note that park state was removed in this version. The major improvements of this version are as follows:

  - Slot Availability Mask (SAM).
  - 2 Mbit/s PHY for LE.
  - LE Long Range.
  - High Duty Cycle Non-Connectable Advertising.
  - LE Advertising Extensions.

| Bluetooth Version | Data Rate | Range (m) |
|---|---|---|
| 1.2 | 721 kbit/s | 10 |
| 2.x | 2.1 Mbit/s | 10 |
| 3.0 | 25 Mbit/s | 10 |
| 4.x | 25 Mbit/s | 60 |
| 5 | 50 Mbit/s | 100 |

Table 3.2: Different Versions of Bluetooth with Their Speed and Range [4]

# 3.9 HC-05 Bluetooth Module

Bluetooth module is a device where Bluetooth wireless facility is embedded as replacement of serial cables and which can be interfaced with any computer or any other computing device [1]. Bluetooth modules use UART *(Universal Asynchronous Receiver/-Transmitter)* protocol for serial communication. They have one pin for transmission (TX) and another pin (RX) for receiving serial data. Therefore Bluetooth communication is bidirectional or duplex in nature. Different types of Bluetooth modules available in the market are [1]:

1. Industrial Level : HC-03, HC-04

2. Civil Level : HC-05, HC-06

The modules with even number (e.g. HC-04) comes with a default master or slave configuration which can't be changed later [1]. On the other hand modules with odd number (e.g. HC-05) can be configured to act as both master & slave mode by using appropriate AT commands [1].

## 3.9.1 Connection of HC-05 with Arduino Board

The HC-05 module has both transmitter pin (TXD) and receiver pin (RXD) which are used to obtain data from the Arduino board for wireless transmission and to send wirelessly transmitted data to the Arduino board as per UART (Universal Asynchronous Receiver/Transmitter) protocol. Arduino Due board has four pairs of receiver and transmitter pins(TX0-RX0, TX1-RX1, TX2-RX2, TX3-RX3) whereas Arduino Uno board has only single pair of receiver and transmitter pins (TX-RX). The receiver pin of Arduino board needs to be connected to the transmitter pin of HC-05 module and the transmitter pin of Arduino board needs to be connected to the receiver pin of HC-05 module, for e.g. TX3 pin with RXD and RX3 pin with TXD. Power supply to the HC-05 module is given (at VCC pin) from the 5V or 3.3 V pin of the Arduino board along with a ground or reference (GND pin) connection. In our case, we have used the 5V supply. Figure3.9shows the connection of an HC-05 module with Arduino Due board. Once the connections are done as per figure3.9and the Bluetooth module gets connected to any other module, we can transmit or receive data using *Serial.write*(*variable*) and *Serial.read*() commands in Arduino board respectively. As there are four pairs of transmitter and receiver pins in Arduino Due board, therefore while giving command for serial communication through Bluetooth module, we have to specify the specific pair, e.g. for TX1-RX1 pair we have to give command as *Serial*1.*write*(*variable*) and *Serial*1.*read*(), for TX3-RX3 pair we have to give command as *Serial*3.*write*(*variable*) and *Serial*3.*read*() etc. It is to be noticed that the baud rate of the Arduino board and the baud rate of the Bluetooth module should be same for serial communication.

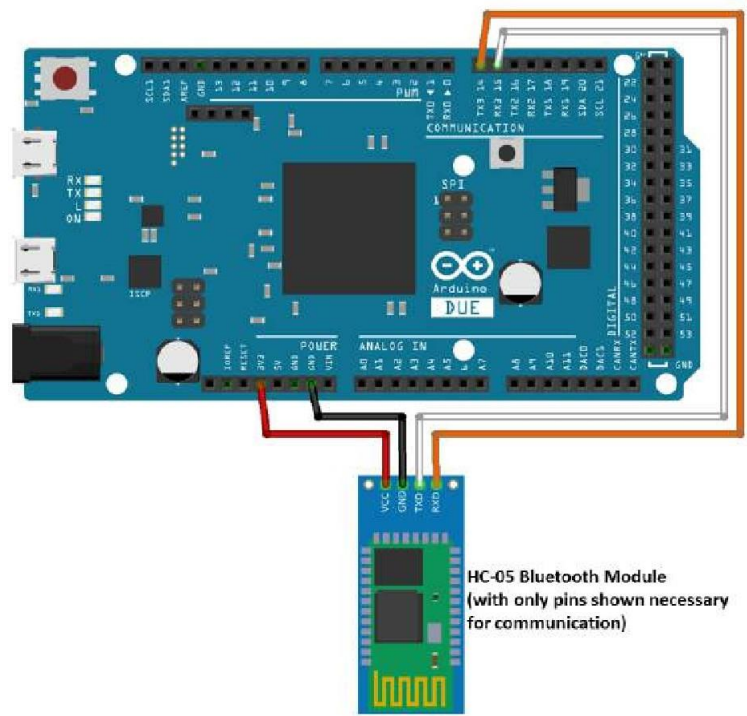

Figure 3.9: Connection of an HC-05 module with Arduino Due board

### 3.9.1.1 Baud Rate

In communication engineering, Baud or Baud rate is the number of distinct symbol changes (signaling events) made to the transmission medium per second in a digitally modulated signal or a line code [8]. Baud rate is related to gross bit rate expressed as bits per second [8]. In digital systems with binary code, 1 Baud (Bd) = 1 bit/sec [8]. Therefore in the serial port context, "X baud rate" means that the serial port is capable of transferring a maximum of X bits per second.

### 3.9.1.2 Software Serial Transmitter and Receiver in Arduino Board

Any two digital I/O pin of an Arduino board can be configured as a pair of serial transmitter and receiver pins by using a default software serial library available in Arduino IDE software. Software serial configuration is needed when we want serial communication for an Arduino board with all the serial transmitter and receiver hardware pins engaged. The software serial library is imported with the code *#include < Sof twareSerial.h >* followed by *Sof twareSerial BT serial*(*Rx, T x*); (Where *BT serial* is the name given to the software defined serial communication port and *Rx, T x* are the numbers of the digital pins to be used as receiver and transmitter respectively). In this case the baud rate for serial communication is defined with the code *BT serial.begin*(*baud_rate*).

Use of sofware serial port is mainly needed in case of Arduino Uno board as it has single dedicated serial communication port. If we connect the Arduino Uno board to the PC via USB cable for data acquisition while Bluetooth module connected to the TX-RX pins, then proper data acquisition is not possible. This is because serial communication with the PC and the Bluetooth module uses the same dedicated hardware. Hence to perform both data acquisition and Bluetooth communication simultaneously using Arduino Uno, we can use software defined serial communication port for connecting the Bluetooth module.

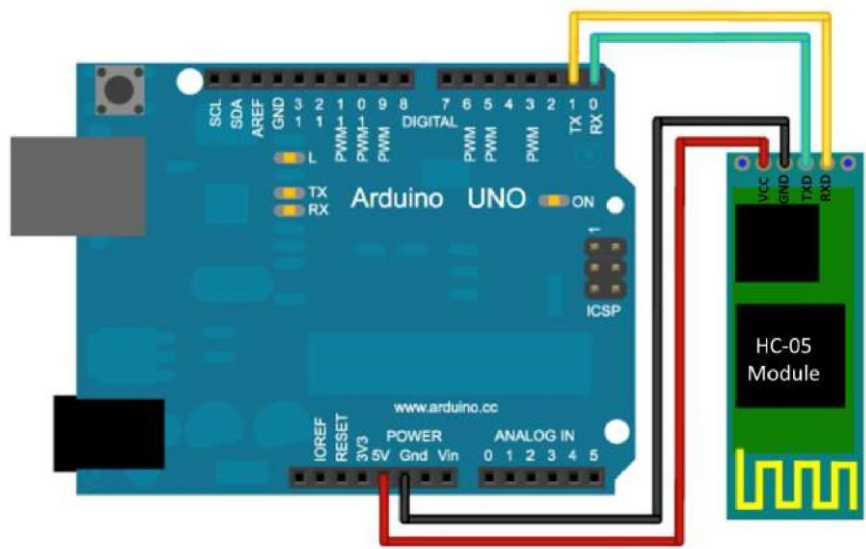

Figure 3.10: Connection of an HC-05 module with Arduino Uno board

## 3.9.2 Configuration of HC-05 Using AT Commands

HC-05 Bluetooth module can be configured using AT commands by connecting it to an Arduino board connected to a PC via USB cable. The steps necessary to drive an

HC-05 module to AT command mode are given below,

- TX pin of Arduino should be connected to TXD pin of HC-05 and RX pin of Arduino should be connected to RXD pin of HC-05.

- Arduino board should be connected to the PC via USB cable.

- A blank sketch, as shown in figure3.11, should be uploaded to the Arduino board from Arduino IDE software.

- 5V power supply should be given to the enable (EN) pin of HC-05 from the Arduino board (with the ground connected).

- The reset button of the HC-05 module is pressed. The button should be kept pressed.

- The 5v power supply should be provided at the VCC pin of HC-05 from arduino board.

- The HC-05 module goes to AT command mode and the reset button should be released.

- Arduino IDE console should be opened up and the baud rate of the console should be set to 38400.

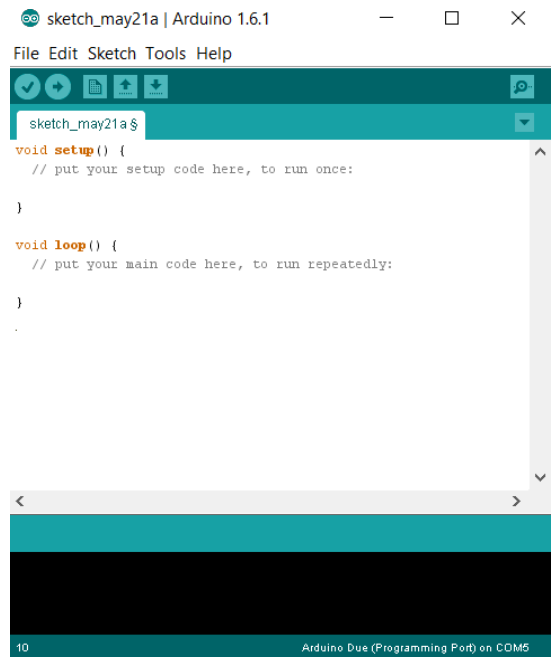

Figure 3.11: Blank sketch uploaded to Arduino board

Once HC-05 module goes to AT command mode, we can send AT commands to it from the Arduino IDE console. There are many types of AT commands for the HC-05 Bluetooth. We are going to highlight the commands that we have used for our purpose in table 3.3 below.

| Command | Response | Parameter |
|---|---|---|
| AT | OK | - |
| AT+VERSION | +VERSION:<Param><br>OK | Param: firmware version |
| AT+ADDR? | +ADDR:<Param><br>OK | Param: address of Bluetooth module |
| AT+NAME? | +NAME:<Param> | Param: Bluetooth module name |
| AT+NAME = <Param> | OK | Param: Bluetooth module name |
| AT+ROLE? | +ROLE:<Param> | Param: 1-master , 0-slave |
| AT+ROLE = <Param> | OK | Param: 1-master , 0-slave |
| AT+PSWD? | +PSWD:<Param><br>OK | Param: PIN Code<br>(Default 1234) |
| AT+PSWD = <Param> | OK | Param: PIN Code |
| AT+ UART? | +UART=<Param1>,<Param2>,<br><Param3><br>OK | Param1: Baud rate<br>Param2: Stop bit<br>Param3: Parity |
| AT+UART==<Param1>,<br><Param2>, <Param3> | OK | Param1: Baud rate<br>Param2: Stop bit<br>Param3: Parity<br>(Set Param2 as 1 and Param3 as 2) |
| AT+CMODE? | + CMODE:<Param><br>OK | Param: 0 – connect fixed address, 1- connect any address |
| AT+CMODE=<Param> | OK | Param: 0 – connect fixed address, 1- connect any address |
| AT+BIND? | + BIND:<Param><br>OK | Param: Fixed address<br>(Default 00:00:00:00:00:00) |
| AT+BIND=<Param> | OK | Param: Fixed address |

Table 3.3: Different AT commands used for HC-05 module

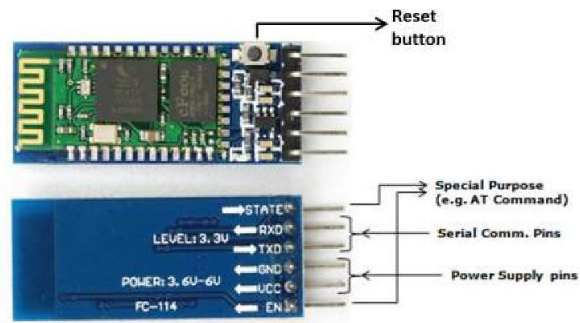

Figure 3.12: Different pins of HC-05 module

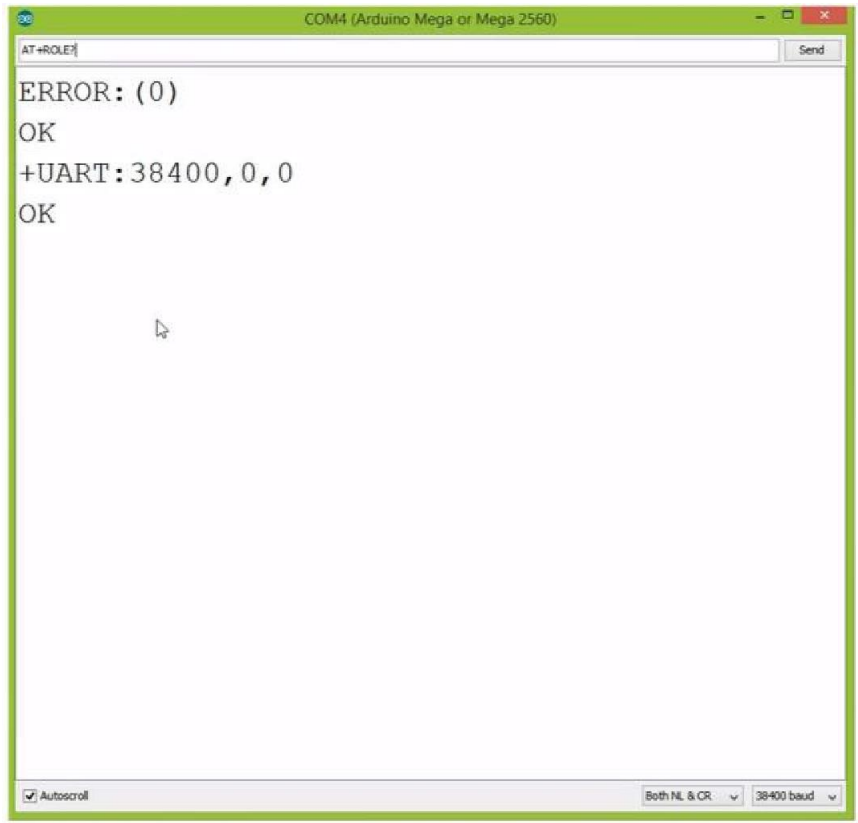

Figure 3.13: Arduino IDE console while giving AT commands to HC-05

### 3.9.3 Bluetooth Connection Between Two Arduino Boards

Bluetooth communication between two Arduino boards is done by connecting HC-05 Bluetooth module with both the boards as shown in figure 3.10 & 3.9. Few moments after power is given to both the Arduino boards, the two HC-05 modules get connected with each other provided both of them are in connection mode (CMODE) 1 or both are bound together if they are in connection mode (CMODE) 0 (refer table 3.3). Once connection is established between the two HC-05 modules, data can be sent from one Arduino board to the other through the Bluetooth link between the two HC-05 modules. It is also important to note that any one of the two HC-05 should be configured as *master* and the other one as *slave* using AT commands. Bluetooth communication takes place only between one master and one slave module. Althouth Bluetooth protocol has multiple device connectivity, that feature is not available in case of HC-05 module. One HC-05 master module can connect with only one slave HC-05 module. Once communication between two HC-05 modules is established, both of them become undiscoverable for other Bluetooth devices.

## 3.10 The Practical Wireless Networked Control System

The functional block diagram of the wireless networked control system for speed control of DC motor is shown below in figure3.14.

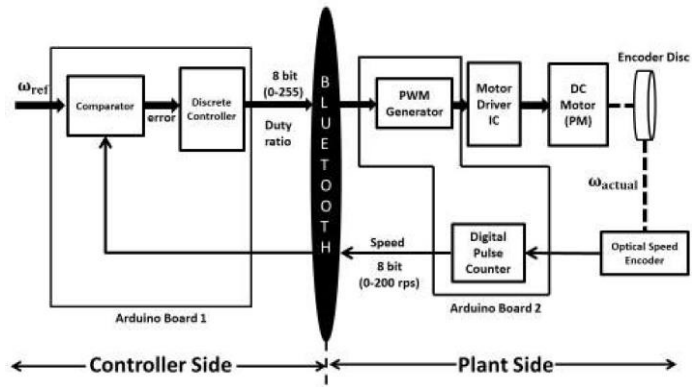

Figure 3.14: The functional block diagram of the wireless networked control system for DC motor speed control

As per the functional block diagram of figure3.14, the practical wireless networked system for speed control of DC motor has been developed as shown in figure3.15. Arduino Due board can also be used in the plant side. We have shown Arduino Uno board in the plant side here.

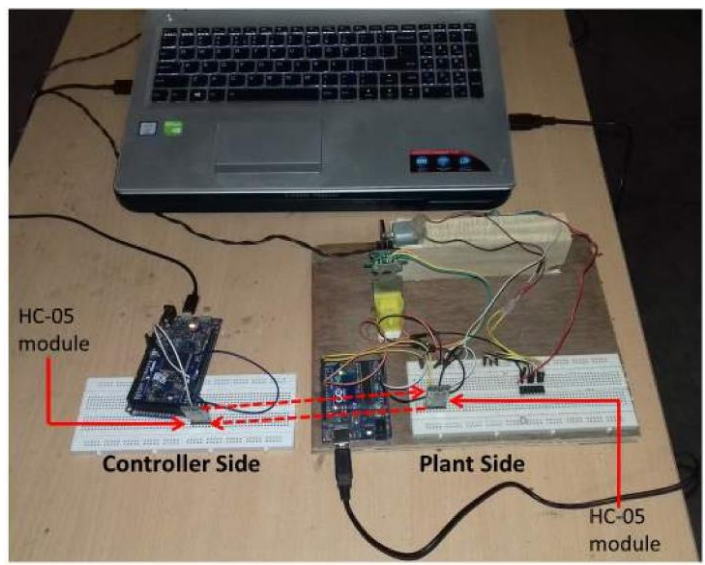

Figure 3.15: The practical wireless networked system for speed control of DC motor (point-to-point configuration

Here, control input calculated by the discrete PI controller is quantized and converted into an *unsigned byte* variable (8 bit) between 0 to 255. This is because Arduino micro controller represents PWM duty ratio with an 8 bit number from 0-255 with steps of 1. The speed of the DC motor varies between 0 to 200 rps and therefore it is also calculate in the form of an *unsigned byte* variable (8 bit). Now the bit rate of a digital system in *bit/s* is given by,

$$bit\_rate = sample\ rate * bit\ depth * channels \tag{3.1}$$

In our system, *sample rate* $= 1/T_s = 50Hz$, *bit depth* $= 8bit$ and *channels* $= 2$ (TX-RX). Hence, the bit rate in the wireless system is, *bit rate* $= 50 * 8 * 2 = 800$ *bit/s*.

The baud rate of the two Arduino boards and HC-05 modules are configured at 115200. That means our system can handle a maximum of 115200 bit/s for Bluetooth communication. Lower baud can also be selected. We have just kept high enough margin of baud for Bluetooth communication to ensure reliable operation.

It is important to note that the controller side node is event driven i.e. it sends data only when it receives a signal from the plant side through the Bluetooth network. On the other hand, the plant side node is clock driven i.e. it is sampled at an interval of 20 ms (50 Hz) using Arduino timer functionality.

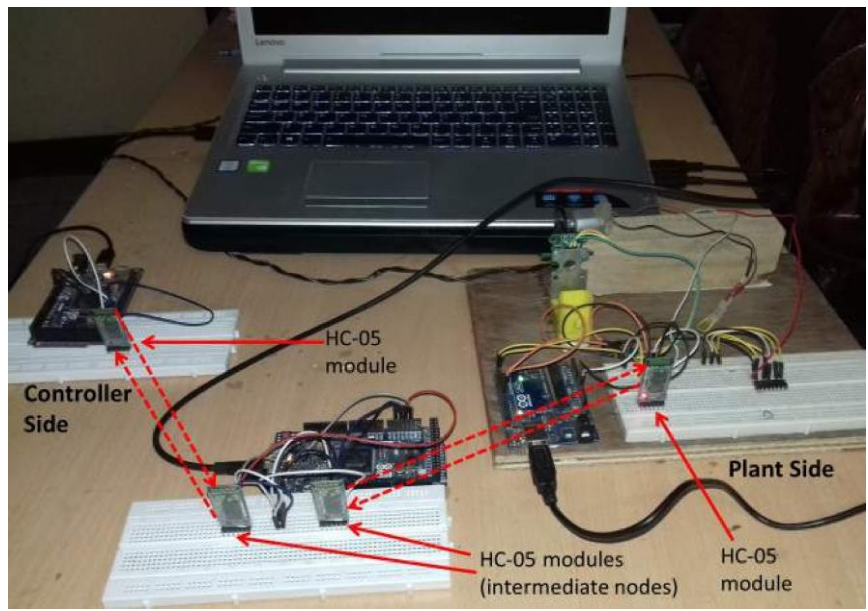

Figure 3.16: The practical wireless networked system for speed control of DC motor (intermediate node configuration)

### 3.10.1 Testing of the Wireless Network To Used in Control Loop

The Bluetooth wireless network which is to be used in the control loop need to be tested to know its the behaviour. The block diagram for testing the network behaviour is shown in figure 3.17. Here, we have sent integer numbers from 1 to 150 in the form of *unsigned byte* (8 bit number) from node 1 to the node 2. At node 2, the data is received and again sent back to node 1. Node 1 is kept clock driven (20 ms sampling) and node 2 is kept event driven. The sent and received information at node 1 for point-to-point configuration is plotted as shown in figure 3.18.

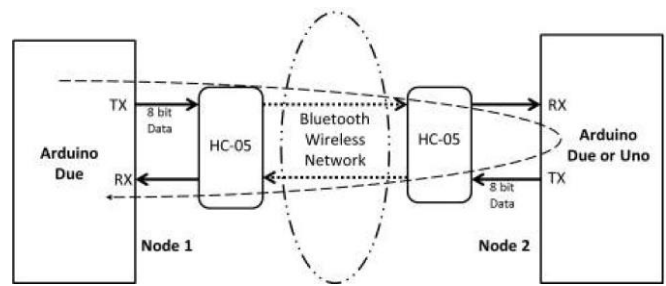

Figure 3.17: The block diagram for testing of Bluetooth network used for control loop

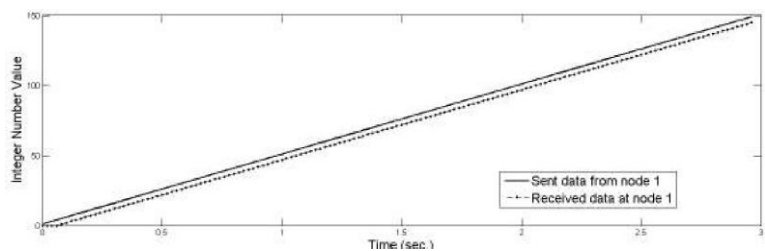

Figure 3.18: The sent and received information at node 1 (point-to-point configuration)

From figure 3.18, we observe that in point-to-point configuration there is a time delay (Round Trip Time) in the wireless network and there is no data loss or packet frop out in the network. The time delay is found to be 80 ms most of the times. Hence we can say that there is a fixed time delay in the point-to-point configuration. These facts are confirmed by repeated observations several times.

In intermediate node configuration, the time delay of the network is not fixed (typically found between 160 ms to 400 ms) and there is no data loss or packet drop out too.

Hence we can draw the following inferences for our wireless networked control strategy,

- There is fixed or varying time delay in the Bluetooth network depending upon type of configuration (point-to-point or intermediate node configuration).

- The Bluetooth network does not suffer from bit error, data loss or packet drop out.

## 3.11 Practical Wave-forms of the wireless Setup with PI Controller

The controller side and plant side of the DC motor speed control system separated and connected through a Bluetooth wireless network as shown in figure 3.14. The discrete PI controller implemented here is the same one as discussed for wired speed control set up in Chapter 2. The wireless setup was run and different responses were recorded for different values of delay. The time delays occurring in intermediate node configuration are much higher compared to point-to-point configuration of the wireless networked system. The graphs obtained from practical experimentation are shown in Figures 3.19-3.22.

Therefore, from figures 3.19-3.22, it is evident that, with increase in delay, the response of the system becomes oscillatory. Thus, the response of the system degrades with increase in delay. In other words, the stability of the system decreases with increase in delay. At $\tau_d = 80ms$, when 3.19 is compared with 2.26, it can be observed that the system response degrades . Small overshoot and jittery response in seen in case of 3.19. At $\tau_d = 240ms$, the overshoot further increases and the system response becomes more oscillatory. At $\tau_d = 300ms$, further degradation in system response is observed .As delay increases to $\tau_d = 400ms$ *(almost twice the time constant of the plant)*, the system is at the verge of instability. Practically at $\tau_d = 400ms$, the system is unstable. Therefore, it can be concluded that a conventional discrete PI controller in not sufficient to compensate for large delays occurring in the system. Therefore, there is a necessity to adopt a different compensation scheme for delay compensation in a network.

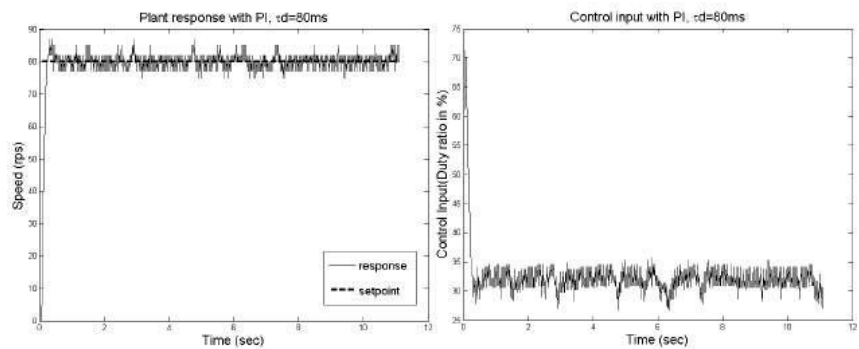

Figure 3.19: Response of the wireless setup with discrete PI controller when $\tau_d = 80ms$ (point-to-point configuration)

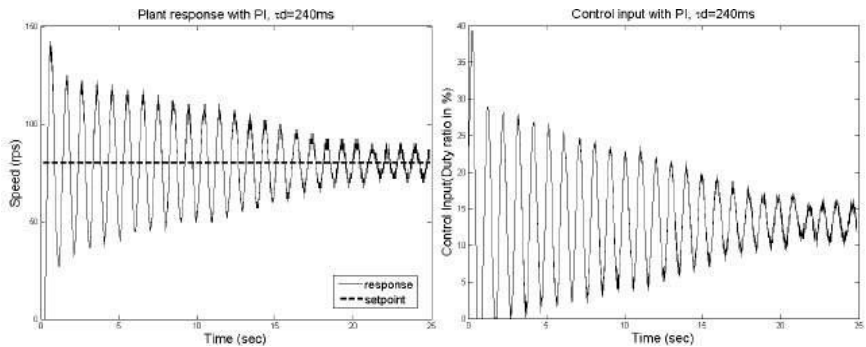

Figure 3.20: Response of the wireless setup with discrete PI controller when $\tau_d = 240ms$ (intermediate node configuration)

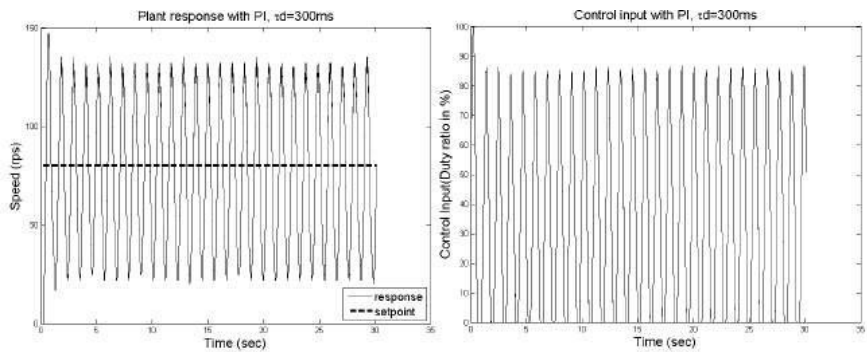

Figure 3.21: Response of the wireless setup with discrete PI controller when $\tau_d = 300ms$ (intermediate node configuration)

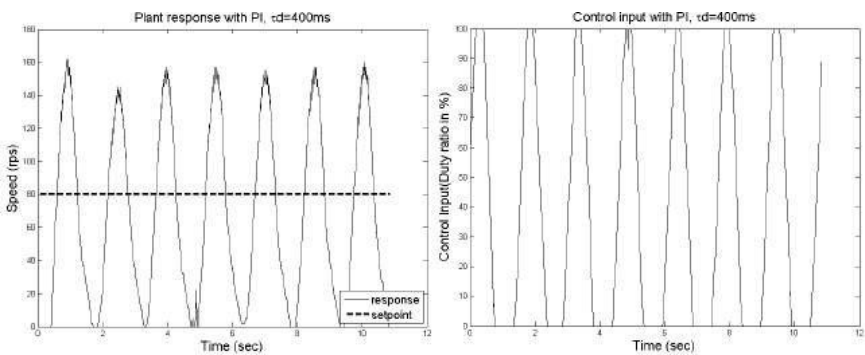

Figure 3.22: Response of the wireless setup with discrete PI controller when $\tau_d = 400ms$ (intermediate node configuration)

Higher values of time delays ($\tau_d$) can also be generated in point-to-point configuration using Arduino timer functionality. This is a software measure and can be implemented to observer the system response with even higher values of time delays than the intermediate node configuration.

### 3.11.1 WNCS With Software Defined Serial Port

In this case, both the boards used in controller and plant side are of Arduino Uno type. Hence for data acquisition purpose we have to use software defined serial transmitter and receiver pins. Both side of the WNCS (Wireless Networked Control System) uses software defined serial communication ports in the Arduino Uno boards.The practical graphs for this type of strategy in point-to-point configuration is shown in figure 3.23.

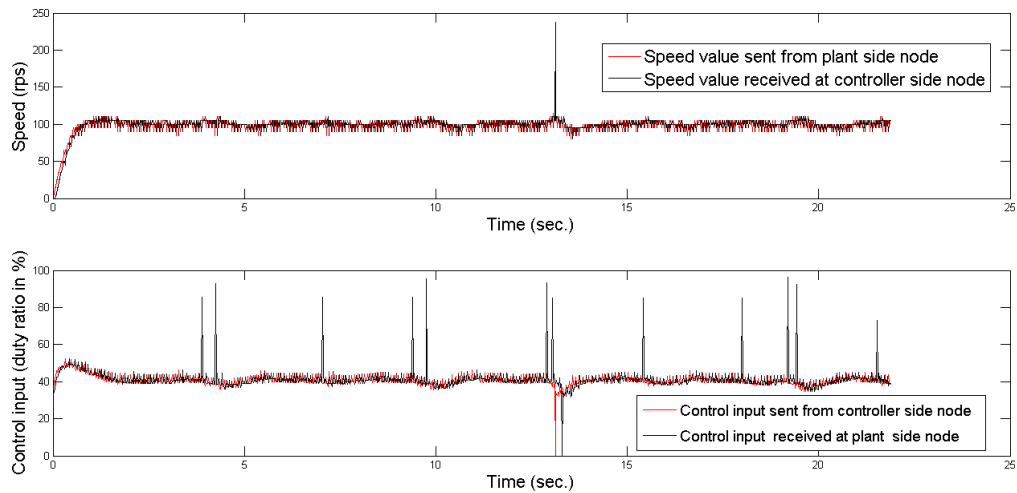

Figure 3.23: Graphs for WNCS With Software Defined Serial Ports (point-to-point configuration)

From figure 3.23, it has been observed that there are several bit errors both in control input and speed value of the WNCS. This is because the dedicated serial communication hardware of the Arduino Uno boards are not used here and the software defined serial ports are meant for digital I/O. Therefore we can observe that bit errors and loss of information due to wireless communication may also occur due to various system components other than the wireless communication modules. This information is provided just for understanding and we are not discussing any remedy or further analysis in thesis.

## 3.12 Chapter Summary

In this chapter basic idea about Bluetooth wireless technology has been provided. The protocol stack, specifications and communication methods of Bluetooth has been discussed. The details about HC-05 Bluetooth module, its configuration using AT commands, its connection with Arduino boards and its connection with another similar module has been discussed. The wireless network behaviour is examined in practical. The practical set up for transforming the conventional embedded speed control system, discussed in Chapter 2, into wireless networked control system has been discussed. It has been observed that the conventional discrete PI controller fails to compensate for larger time delays introduced by wireless network and its performance deteriorates with small time delay introduced by the wireless network.

# Chapter 4

# Time Delay Measurement, Estimation, Aprroximation & Smith Predictor Scheme for Delay Compensation

## 4.1 Time Delay in Network Control System

For the past few decades, NCSs (Networked Control Systems) have been extensively studied. NCSs in which control loops are closed via a communication network has gained huge attention over recent years [6]. With the inclusion of a communication channel in control structure, we invite various challenging problems like packet loss and time delays which is detrimental to the performance of the system, may lead to instability [1] - [6]. Discussion on architecture of NCSs, the effect of packet loss and time delays have been depicted in [4,5] and references therein, where, stability of NCS and controller design to mitigate the effects of packet loss and time delays have been discussed.

Broadly, delay can be categorised as,

- *Static delay:* Processing of data in the protocol and data rate limitations cause such delays and the nature of such delays are always constant [3]. Generally, static delays are small in magnitude and its effect on control performance is the least.

- *Stochastic delay:* The nature of such kind of delay is not constant but varies with time. Mainly caused due to retransmission of wrongly received data [3]. The variations is related to protocol being used [3]. The stochastic nature of the network time delay makes it difficult to apply linear delay-time system analysis [1]. Network induced delays are generally stochastic in nature.

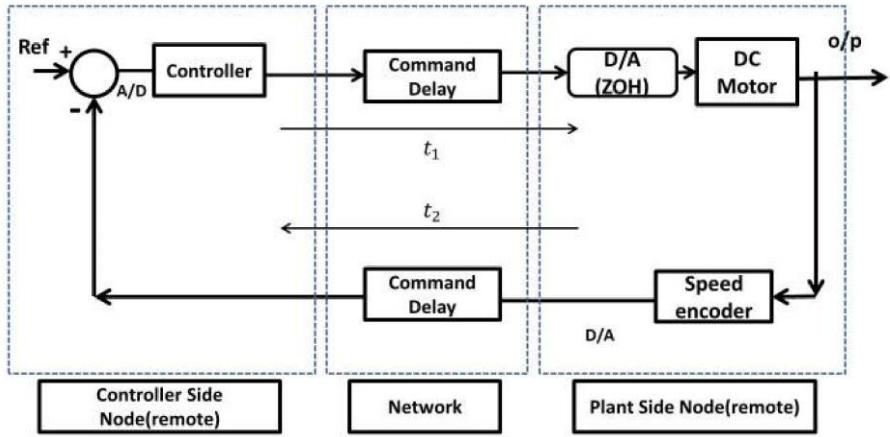

Figure 4.1: General block diagram of NCS

Figure 4.1 depicts the generalised structure of NCS closed-loop model, where the delay in time induced in the network for controller-to-plant+actuator direction is given by $t_1$ and the delay in time induced in the network for sensor-to-controller direction is given by $t_2$. Delay induced in the network depends on various factors, like, the number of nodes in the network, scheduling policies and different protocols used [1]. Design and modelling of NCS with time-varying properties is more challenging. Use of networks introduces uncertain time delays and thus, degrades the performance of a NCS [2] and references therein. The total time-delay can be categorized in the following ways [1],

- *Delay at the server node:* The time-delay at server node is basically the processing time, which includes computation time, waiting time, queuing time, encoding time, and blocking time [1].

- *Delay in the network channel:* The network time-delay includes transmission time of the message and its propagation time [1].

- *Delay at the client node :* The time delay at the client node is basically the post-processing time [1].

The processing and post-processing time-delay is comparative smaller then transmission time-delay. The total delay is stochastic in nature and its variation is called jitter [3] and references therein.

## 4.2 Time Delay Measurement and Estimation in Wireless Network Control System

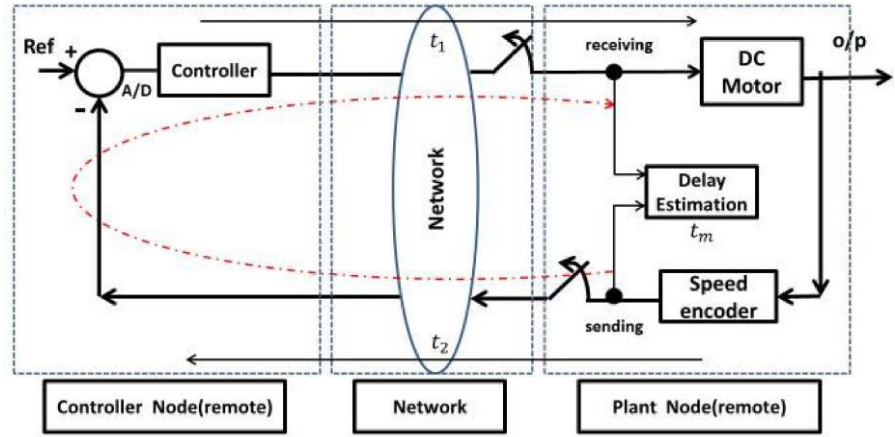

Figure 4.2: Structure of proposed RRT measurement in [1]

In [1]-[6] and references therein, wired connections exist in either feedback or feedforward or both, which means that network is either wired or partially wireless. The purpose of our work is to establish a control scheme where no wired connection exist between the plant node(*server* ) and the controller node(*client* ), i.e, a complete wireless network, creating a Wireless Network Control System (*WNCS* ). Many wireless technologies exist which are capable of reliable transmission of of data, such as WLAN TCP/IP, WLAN UDP/IP, Bluetooth,  and Zigbee [3] and references therein. All protocols guarantee correct arrival of data at the receiving side, by means of packet retransmission except for WLAN UPD/IP [3].

In similarity to Figure4.1, the delay in time induced in the network for controller-to-plant+sensor direction is given by $t_1$ and the delay in time induced in the network for sensor-to-controller direction is given by $t_2$. The total time can be expressed as $t_p = t_1 + t_2$, where $t_p$ is termed as the total time or the round trip time (*RTT* ) [1]. The round-trip delay time (RTD) or round-trip time (RTT) is defined as the amount of time taken by a signal to be sent plus the amount of time it takes for an acknowledgement of that signal to be received  [7].

To illustrate the estimation of time delay induced by the network, a strategy as depicted in [1] has been adopted.  Figure4.3shows pictorially how  the time delay estimation can be made. Referring to Figure4.2, at the beginning of the sampling period, the clock driven plant node transmits the feedback data (rpm) to the controller node. Assuming the plant node-to-controller node delay to be $t_2$, the event driven controller node uses this information to compute the control signal (*control input to the plant*) and then transmits it to the plant node. Assuming the controller node-to-plant node

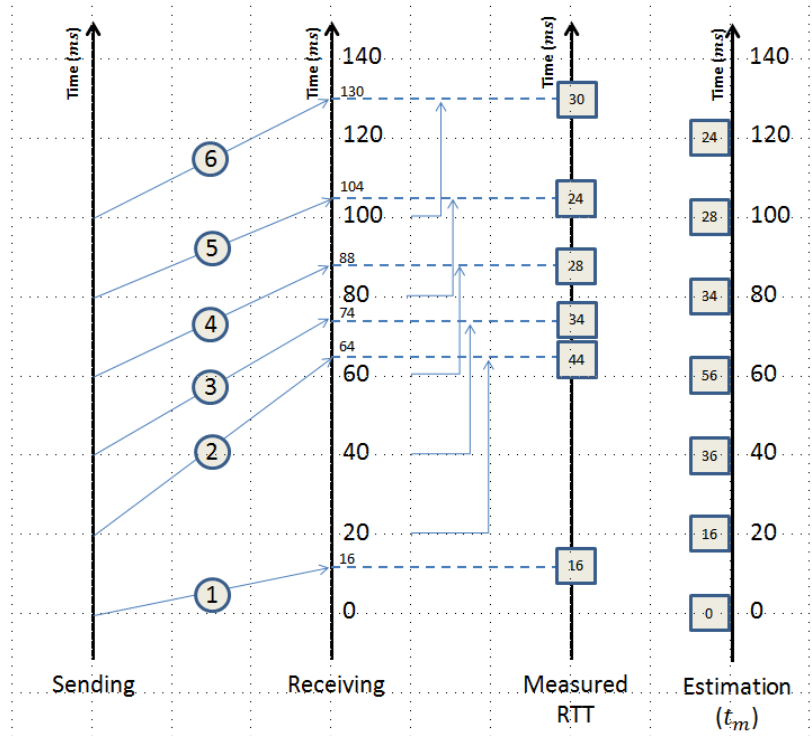

Figure 4.3: RTT measurement and delay estimation

delay to be $t_1$, the plant node uses this information to drive the plant.  Due to ease of implementation and no requirement of clock synchronisation, measurement of RTT has been adopted. Also for accurate periodic assessment of delay, RTT measurement is crucial [1].  Through Figure 4.3, a pictorial example has been presented on measurement of RTT by considering a sampling period of $20ms$, the details are as follows,

- The effect of delay in control performance for a time delay less than one sampling period is one sample delay and the first frame is in normal transmission [1].

- In the second frame, the data which was sent $20ms$ later is received at $64ms$. Accordingly, the corresponding RTT is $44ms$. No data was received at $40ms$ & $60ms$ sampling times. This phenomenon is termed as vacant sampling [1] and reference therein.

- Two data messages (2 & 3) arrived in the same sampling period, and only the most recent data is accepted and all previous data will be discarded. This phenomenon is termed as message rejection [1] and references therein.

- For messages 4-6, all data arrived sequentially at each sampling period with slight variation in receiving time. This phenomenon is called delayed  transmisstion.

Thus, the rules for delay estimation ($t_m$ ) are given as follows ,

- *Normal transmission:* If the time-delay is less than one sampling period, the effect of this delay on control performance is negligible and the measured RTT is directly adopted as estimated delay ($t_m$) [1].

- *Vacant sampling:* If data sample has not arrived before the arrival of next sampling period, the the RTT measured before added with one sampling period is considered as estimated delay ($t_m$) [1].

- *Message Rejection:* If more than two data samples arrive at the same sampling period, only the most recently arrived data is considered and the corresponding RTT is considered as estimated delay ($t_m$) [1] and all previous measured data are discarded.

- *Delayed transmission:* The RTT measured continuously is considered as estimated delay ($t_m$)  [1].

# 4.3 Practical Implementation of The Proposed RTT Measurement in [1] With Necessary Modifications

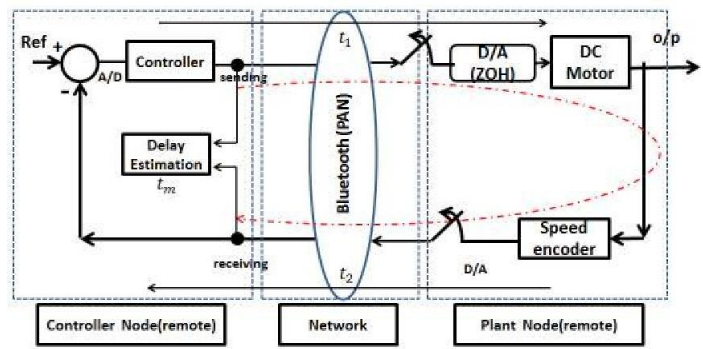

Figure 4.4: Structure of RRT measurement for the experimented setup

Figure 4.4 shows the structure of the control scheme established by us, where, the controller is placed in the controller node and plant is placed in the plant node and both the nodes are allowed to communicate over a wireless network created by bluetooth (*point-point* ). The plant node consists of four parts, namely *µc* (*Arduino DUE/UNO* ), bluetooth module, speed encoder *(sensor)* and the DC motor *(plant)* . The controller node consists of *µc* (*Arduino DUE/UNO* ), bluetooth module and the controller logic implemented on the *µc*. As shown in Figure 4.4 and [1], For the purpose of illustration of RTT measurement, we refer to Figure 4.5 which presents an illustration on how delay has been estimated practically. It should be noted that in Figure 4.2, the delay was

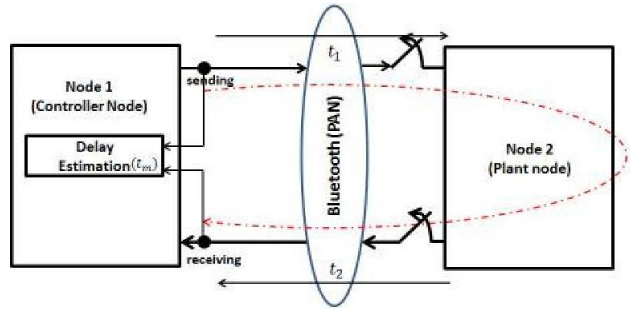

Figure 4.5: Measurement and estimation of delay on practical setup

estimated on the plant node but for the setup under consideration, as in Figure 4.4, we have estimated delay on the controller node. If RTT is measured on the plant side, for compensation of the estimated delay, the information has to be sent to the controller node. The transfer of which may suffer from delay. Therefore the control action required for compensation of the delay for that very sample period may ultimately get delayed and the overall performance of the system may not be satisfactory. In order to avoid such a situation and as a matter of our convenience, we have measured RTT on the controller node itself. For the system considered in Figure 4.4, a point-to-point bluetooth communication between node 1 *(Controller node)* & node 2 *(Plant node)* has been established. A sequence of byte data were sent from node1 and received back. The sending time, the byte data sent and the receiving time, the byte data received were recorded and accordingly RTT was measured. Based on the measured RTT, delay was estimated keeping in mind the rules discussed in the former section. It should be noted that for the system under consideration and for the system considered in [1], Figure 4.2, **the sensor node is clock driven and the controller node is event driven**. Figure 4.6, an illustration has been presented on how the RTT measurement and delay estimation has been carried out on the practical setup. As stated earlier, the sensor node is clock driven but the controller node is event driven. The illustration is explained as follows,

- At the beginning of the sampling period, a byte value of 50 was sent from the controller node.

- The controller node established on a $\mu$c keeps advancing in time. At time $23ms$, 0 appears, i.e, no value is received. Thus, according to the rules, it is considered as a vacant sample. On appearance of 0, the event driven controller node sends another byte value of 60 at $23ms$.

- The same phenomenon continues for the following two sample time ($40ms$ - $60ms$). Thus, there is occurrence of vacant samples till $60ms$. On appearance of zero at $45ms$, the event driven controller node sends a byte value of 70.

- Thus for sample time $20ms$, $40ms$ & $60ms$; the estimated delays are $20ms$, $40ms$ & $60ms$. *(see rule for delay estimation on appearance of vancant samples.)*

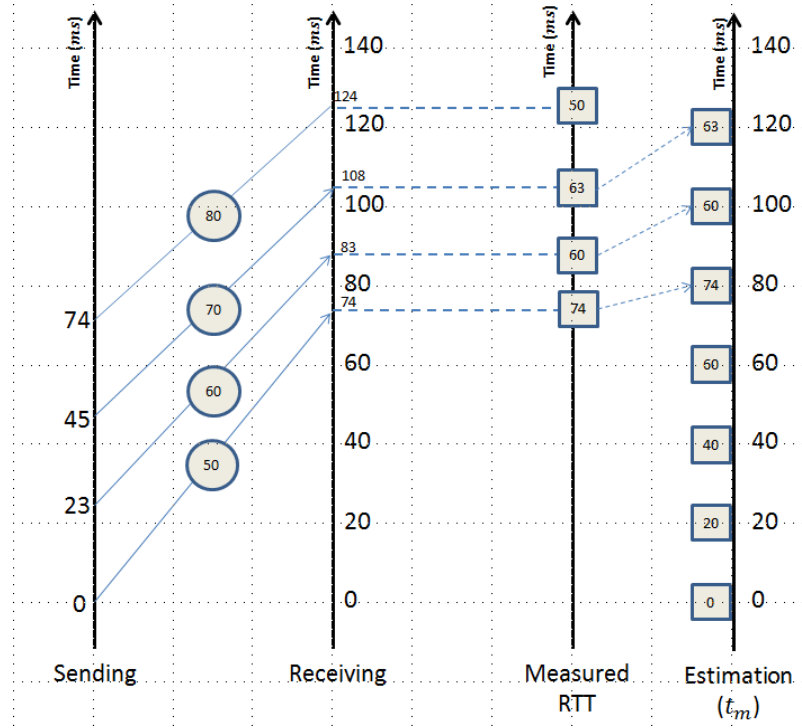

Figure 4.6: RTT measurement and delay estimation on practical setup

- At time 74$ms$, the data sent at 0$ms$, i.e, 50 appears. Thus the measured RTT is 74$ms$(74$ms$ - 0$ms$). Again at 83$ms$, the data sent at 23$ms$, i.e, 60 appears. Thus, the measured RTT is 60$ms$(83$ms$-23$ms$). This phenomenon continues thereafter. It is called *delayed transmission.(see rule for delay estimation on delayed transmission.)*

- The effect of these delays, are seen at their following sample periods as illustrated in Figure4.6.

Now, we present a table depicting the above illustration, where, $t_1 = time - stamp$ *when data is sent*, $t_2 = time$–*stamp when data is received*, $t = t_2$ $t_1$, $t_m = estimated$ *delay*

Once the theoretical illustration of the concept of RTT measurement and delay estimation is clear,the next task is to implement the concept on a $\mu$c which involves designing an algorithm/code for the purpose of delay estimation. Referring to Table4.1; considering the estimated delay at sampling time 100$ms$, which is 60$ms$, i.e., (83$ms$-23$ms$) **(in case of any confusion please refer to Figure4.6& [1])** . When the $\mu$c is at 100$ms$, the most recent time-stamp data available are 74$ms$ & 83$ms$, all previous time stamps are overwritten unless stored . This means the time-stamp data 23$ms$ is overwritten by 74$ms$. Therefore there is a necessity to store the past values. In relation to this, we introduce a new variable $t$ which is the difference between $t_2$ & $t_1$.

| Sent data | $t_1$ | $t_2$ | $t$ | Received data | $t_m$ | Sample time |
|-----------|-------|-------|-----|---------------|-------|-------------|
| 50  | 0   | 23  | 23 | 0   | 0  | 0   |
| 60  | 23  | 45  | 22 | 0   | 20 | 20  |
| 70  | 45  | 74  | 29 | 50  | 40 | 40  |
| 80  | 74  | 83  | 9  | 60  | 60 | 60  |
| 90  | 83  | 108 | 25 | 70  | 74 | 80  |
| 100 | 108 | 124 | 16 | 80  | 60 | 100 |
| 110 | 124 | 143 | 19 | 90  | 63 | 120 |
| 120 | 143 | 167 | 24 | 100 | 50 | 140 |
| 130 | 167 | 184 | 17 | 110 | 60 | 160 |

Table 4.1: Tabular representation of delay estimation

It is interesting to note that the delay estimated at a sampling time is equal to the summation of the difference between the time stamps ($t_1$&$t_2$) summed for the number of times vacant samples occur after $0ms$. In connection to this observation, we propose an empirical formula for estimation of delay,

$$t_m = \sum_{i=0} t_i \qquad (4.1)$$

Where,

$t_m$ = estimated delay,

$t = t_2 - t_1$,

$n$ = number of vacant samples

which can also be represented as,

$$t_m = t_{2pr} - t_{1pr} + \sum_{i=1} t_{ipa} \qquad (4.2)$$

where,

$t_m$ = estimated delay,

$t_{1pr}$, $t_{2pr}$ = present time-stamps,

$t_{ipa}$ = difference of past time-stamps,

$n$ = number of vacant samples,

$t_{2pr} - t_{1pr} = t_0$ *(difference between present time stamps)*

The proposed formulas for estimation of delay provides an easy way to implement the
logic of delay estimation as per theory. One just needs to implement the formula in form
of an algorithm to estimate the delay. In our case, we have implemented equation 6.2.
Whereas, any of the two formulas can be conveniently implemented as both equations
6.1 & 6.2 are alike.

Figure 4.7 shows online estimated time delay versus time graph for the WNCS with
intermediate node configuration (for 3 different readings). It is clear from the graphs
that the time delay is not fixed in case of intermediate node configuration.

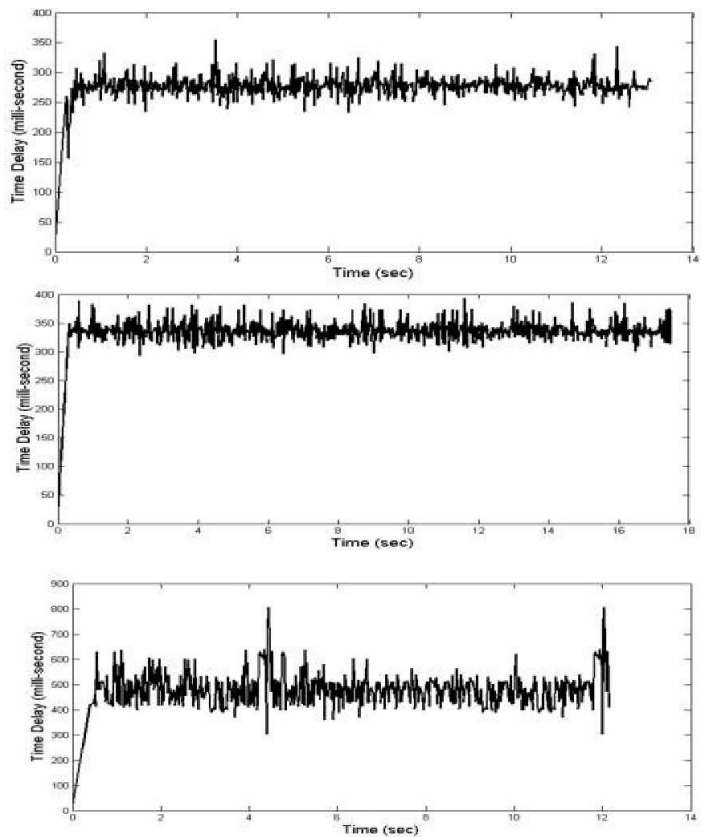

Figure 4.7: On line estimated time delays of the practical WNCS system (interme-
diate node configuration

For point-to-point configuration, time delay is found to be fixed (with very small fluc-
tuations) around 80 ms.

## 4.4 Time-Delay  Approximation

Time delays are also known as transport delay/transport lag, dead time or time lag.
Time delays prohibits the flow of a signal inside a system. They are common in almost
all physical systems, like, chemical, biological as well as economic systems. Their
occurrence is also found common in measurement and computation processes[9]. This
delay in time can be approximated by  use of a polynomial series which will allow
one to analyze a time-delayed system in quite the same manner as in case of a non
time-delayed system [9].

A pure time delay, $e^{-\tau_d s}$ has infinite number of roots, i.e, a pure time delay is infinite
dimensional [8]. By approximating it to a polynomial series, we are actually attempting
to visualize it in finite dimensions in order to,

- analyze and design a system using standard methods [8].

- use standard software for simulation [8].

- implement the delay transfer function on an embedded platform (*if implementa-
  tion demands*).

- also we never work over infinite bandwidth. Hence, we need to approximate $e^{-\tau_d s}$
  in finite frequency range [8].

The usage of computers and embedded hardware requires the delay to be approximated
using polynomial series [9], as these finite word length machines. The occurrence of
error in approximation is inevitable.    The error also depends on the type and order
of the series being considered for approximation [9].  In some literature and in [9], it
is stated that if high accuracy is not a requirement or if the delay is quite large in
comparison to the system time constant, using a lower order series is often sufficient.
The formulation of commonly used series to approximate time-delay is depicted in
Table 4.2.

To evaluate the performance of the series, we have  compared the response of each series
to an unit step with the response of MATLAB time-delay block at different delays as
depicted in Figures 4.8 & 4.9. The performance criteria was chosen as Integral  Squared
Error given as,

$$ISE = \int_{t_0}^{t} |e|^2 dt \qquad (4.3)$$

where,

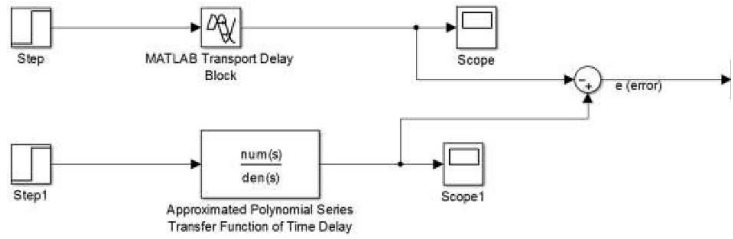

Figure 4.8: SIMULINK block for comparison of response between the MATLAB time-delay block & the polynomial series considered

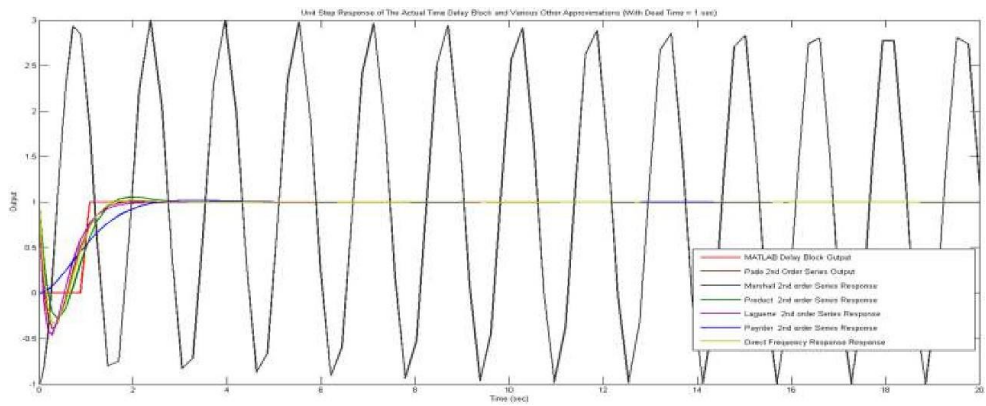

Figure 4.9: Step input responses of the MATLAB delay block & different polynomial series

| Series | Formulation |
|---|---|
| Pade | $\frac{1-0.5s\tau+0.0833s^2\tau^2}{1+0.5s\tau+0.0833s^2\tau^2}$ |
| Marshall | $\frac{1-0.0625s^2\tau^2}{1+0.0625s^2\tau^2}$ |
| Product | $\frac{1-0.5s\tau+0.125s^2\tau^2}{1+0.5s\tau+0.125s^2\tau^2}$ |
| Laguerre | $\frac{1-0.5s\tau+0.0625s^2\tau^2}{1+0.5s\tau+0.0625s^2\tau^2}$ |
| Paynter | $\frac{1}{1+s\tau+0.405s^2\tau^2}$ |
| Direct Frequency Response(DFR) | $\frac{1-0.49s\tau+0.0954s^2\tau^2}{1+0.49s\tau+0.0954s^2\tau^2}$ |

Table 4.2: Tabular representation of different delay approximation series

$t_0$ = initial time, $t$ = final time , $e$ = error between the responses of MATLAB time-delay block and the series under consideration.

Table 4.3, shows the approximation errors. It is evident from Table 4.3, that the DFR

| *Series* | $\tau = 0.04s$ | $\tau = 0.120s$ | $\tau = 0.240s$ | $\tau = 0.300s$ | $\tau = 1s$ | *AverageISE* |
|----------|----------------|-----------------|-----------------|-----------------|-------------|--------------|
| Pade     | 0.0057         | 0.0172          | 0.0345          | 00.0431         | 0.1437      | 0.0592       |
| Marshal  | 43.9888        | 41.6051         | 41.1387         | 40.9687         | 41.8798     | 41.321       |
| Product  | 0.0054         | 0.0161          | 0.0321          | 0.0402          | 0.1339      | 0.05516      |
| Laguare  | 0.0068         | 0.0203          | 0.0406          | 0.0507          | 0.169       | 0.06964      |
| Paynter  | 0.0058         | 0.0174          | 0.0347          | 0.0434          | 0.1447      | 0.05962      |
| DFR      | 0.0053         | 0.016           | 0.0319          | 0.0399          | 0.1331      | 0.05482      |

Table 4.3: Delay approximation errors

series gives the minimum average error followed by the Product series and the Pade series. Therefore, the DRF series has the best performance among the six series. The difference between the average errors of the Pade series and the Product series is not significant. Therefore, for the purpose practical implementation, we choose the DFR series and Pade series. Though Product series yields less average error in comparison to Pade, we still consider it as it is one of the widely used polynomial series for delay approximation.

## 4.5 Implementation of the Approximated Time-Delay

The matter of our concern is implementation. For the kind of solution we sought after for the given problem on delay mitigation, it is required to implement the delay polynomial on the embedded platform. This is the reason behind approximating the delay in form of a polynomial series which can be easily implemented on the embedded device. The formulation of the solution includes conjunction of delay transfer function and a compensation scheme which has been discussed in a later chapter. In this section, we illustrate how the delay polynomial series can be implemented on the embedded device. Different polynomial series has been depicted in Table 4.2. However, as discussed before, we choose Pade series and the DFR series which are in continuous/analog domain. To implement them on the hardware, one has to obtain the discrete version of them. More specifically, a difference equation form from the continuous form of the series has to be obtained which can be achieved through a proper transform operation. One such transform is Bilinear Transform, also known as Tustin's method or Mobius Transformation.

In areas of digital signal processing and discrete-time domain control theory, the bilinear transform is widely used for transformation of continuous-time domain system representation to discrete-time and vice-versa [10]. The bilinear transform is a first-order approximation of the natural logarithm function that is an exact mapping of the *z*-plane to the *s*-plane [10]. For a discrete-time domain signal which has elements of the discrete-time sequence in conjugation with a delayed unit impulse, the *Laplace*

*T transform* of such a sequence results in the *Z-transform* of the discrete-time domain sequence under consideration with the substitution of [10],

$$
\begin{aligned}
Z &= e^{\tau_d s} \\
&= \frac{e^{\tau_d s/2}}{e^{-\tau_d s/2}} \\
&\approx \frac{1 + \tau_d s/2}{1 - \tau_d s/2}
\end{aligned}
\tag{4.4}
$$

Where,

$\tau_d$ = sampling period.

Now, 4.4 can be solved for $s$ on a similar consideration for $s = \frac{1}{\tau_d} ln(z)$. Therefore, the first-order bilinear approximation is,

$$
\begin{aligned}
s &= \frac{1}{\tau_d} ln(z) \\
&= \frac{2}{\tau_d} \frac{z-1}{z+1} + \frac{1}{3} \left[ \frac{z-1}{z+1} \right]^3 + \frac{1}{5} \left[ \frac{z-1}{z+1} \right]^5 + \frac{1}{7} \left[ \frac{z-1}{z+1} \right]^7 + \dots \\
&\approx \frac{2}{\tau_d} \frac{z-1}{z+1} \\
&= \frac{2}{\tau_d} \frac{1-z^{-1}}{1+z^{-1}}
\end{aligned}
\tag{4.5}
$$

Therefore, given a transfer function, $G(s)$, its bilinear transform is given by,

$$
s \leftarrow \frac{2}{\tau_d} \frac{z-1}{z+1}
\tag{4.6}
$$

which results in,

$$
\begin{aligned}
G(z) &= G(s)\big|_{s = \frac{2}{\tau_d} \frac{z-1}{z+1}} \\
&= G\left[ \frac{2}{\tau_d} \frac{z-1}{z+1} \right]
\end{aligned}
\tag{4.7}
$$

Here, we present an example considering Pade first order approximation given as,

$$e^{\tau_d s} \approx \frac{2 - \tau_d s}{2 + \tau_d s}$$

$$G_d(s) = \frac{4}{2 + \tau_d s} - 1 \tag{4.8}$$

Now,

$$s \leftarrow \frac{2}{\tau_d} \frac{z - 1}{z + 1} \tag{4.9}$$

Substituting $\tau_d = 0.02s$ in (4.9), we obtain,

$$s = \frac{2}{0.02} \frac{z - 1}{z + 1}$$
$$= 100 \frac{z - 1}{z + 1} \tag{4.10}$$

Now, substituting (4.10) in (4.8), we obtain,

$$G_d(z) = \frac{4}{2 + 100\tau_d \frac{z-1}{z+1}} - 1 \tag{4.11}$$

Through algebraic manipulation we can obtain,

$$G_d(z) = \frac{(2 - 100\tau_d)z + (2 + 100\tau_d)}{(2 + 100\tau_d)z + (2 - 100\tau_d)} \tag{4.12}$$

similarly, the  $Z-$ *domain* transformed model of the other series under consideration can be obtained.

Once the $Z$ *domain* model is obtained, for formulation of the solution, it will be used in conjugation with the adopted compensation scheme to mitigate the effects of delay. The overall formulation and the compensation scheme has been discussed in a later section.

# 4.6 Introduction to Smith Predictor

A time delay is defined as the time difference between the start of an event at one point of a system and its resulting action at another point of the system [14]. Time delays are also called as transport lags or dead times [14]. Time delays may arise in physical, electric, pneumatic, hydraulic networks, chemical processes, long transmission lines, robotics, biological and economic systems, and also in measurement and computation [12,14]. Time delay occurs due to some phenomena as given below [13]:

- The time required to transport mass, energy or information.

- The accumulation of time lags in a significant numbers of low order systems connected in series.

- The required processing time for sensors.

- Time needed by controllers to implement a complicated control algorithms or process.

PID controllers are the most widely used controllers in process industries [14]. They comprise 90-95% of the total controllers used in process industries [14]. However, PID controllers are less effective when there is dominant time delay in the process dynamics [14]. Predictive control is necessary if high performance is needed for a process with significant time delay [13].Smith predictor structure was first proposed by O.J. Smith to compensate systems with time delay in 1957. For servo applications, a PID controller with smith predictor serves up to 30% better than a appropriately tuned PID controller without smith predictor [14]. Smith predictor is a method which does reduction in the dominance of the time delay term in the closed loop characteristic equation thereby allowing us to design a controller for the nondominant time delay process [14]. For an optimally tuned system, an introduction in time delay requires reduction in system gain to ensure stability [17]. The Smith predictor algorithm helps in avoiding this reduction in gain and hence the resulting poorer performance [17].

# 4.7 The Smith Predictor Scheme

The block diagram of a typical Smith predictor scheme is show in Figure4.10as per [14,17] without considering any.external disturbances The Smith predictor scheme can be rearranged as shown in Figure4.11. Here, $\hat{G}(s)$ represents the nominal plant model, $\tau_m$ represents measured or modeled plant delay, $G_c(s)$ represents the controller along with actual plant & time delay denoted as $G(s)$ & $\tau_d$ respectively.

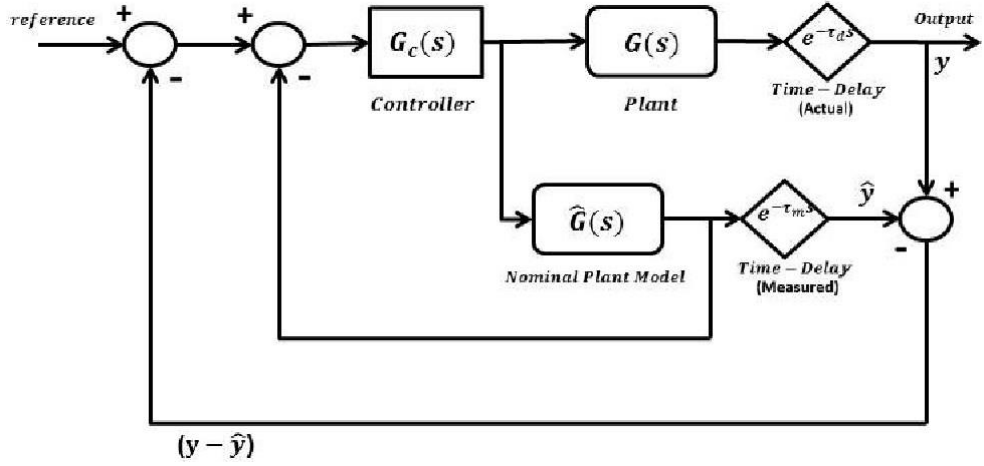

Figure 4.10:  Block diagram of a typical Smith predictor scheme

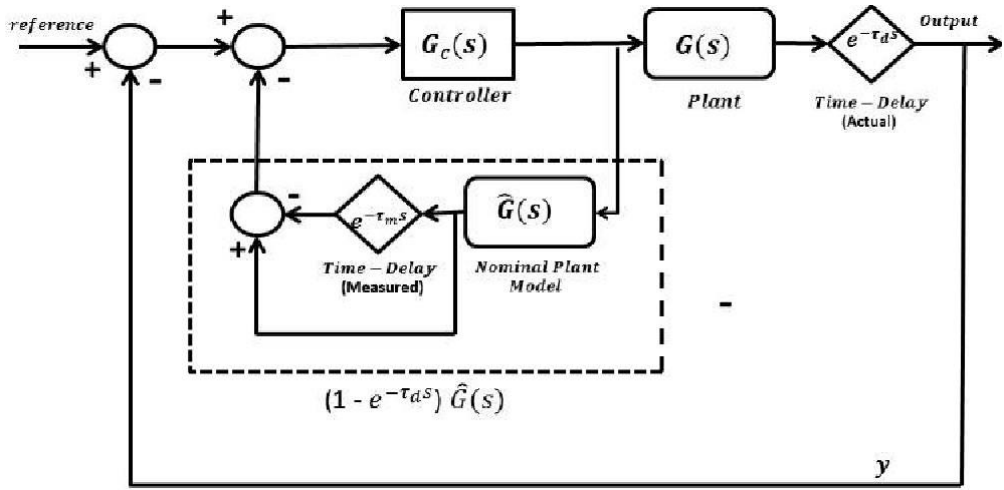

Figure 4.11: Rearranged form of the typical Smith predictor scheme

Now, let the controller with the inner loop as represented in Figure 4.11 is represented as $G_s$(s).  Then,

$$G_s(s) = \frac{G_c(s)}{1 + (1 - e^{-\tau_m s})\hat{G}(s)G_c(s)} \qquad (4.13)$$

Then the closed loop transfer function becomes of the plant with smith predictor scheme becomes,

$$\frac{Y(s)}{R(s)} = \frac{G_s(s)G(s)e^{-\tau_d s}}{1 + G_s(s)G(s)e^{-\tau_d s}} \qquad (4.14)$$

Now, using equation 4.13 in equation 4.14,

$$\frac{Y(s)}{R(s)} = \frac{G_c(s)G(s)e^{-\tau_d s}}{1 + \hat{G}(s)G_c(s) - \hat{G}(s)G_c(s)e^{-\tau_m s} + G(s)G_c(s)e^{-\tau_d s}} \qquad (4.15)$$

When plant is as accurately modelled and the time delay is estimated accurately, $\hat{G}(s) = G(s)$ and $\tau_d = \tau_m$. In this case, equation 4.15 becomes,

$$\frac{Y(s)}{R(s)} = \frac{G_c(s)G(s)}{1 + G_c(s)G(s)}e^{-\tau_d s} \qquad (4.16)$$

Equation 4.16 shows that the transfer function of the whole system represented in Figure 4.11 is simply the transfer function of the closed loop system without time delay cascaded with a pure time delay.

Difficulties in implementation of classical smith predictor structure are,

- A nominal plant model may not always accurately represent a plant as process parameter may change with operating conditions and there may exist unmodeled dynamics in the model [14]. In this case perfect delay compensation may not be possible by the smith predictor.

- From equation 4.15, it is clear that if the delay is not correctly estimated then also perfect delay compensation is not achievable through classical smith predictor. Therefore the time delay should be known, fixed and accurately , measured to obtain a perfect delay compensation.

### 4.7.1 Adaptive Smith Predictor

Adaptive smith predictor is required to take care of the process and model mismatch in dead time compensation. This problem can be mitigated by using self-tuning regulator approach where the plant model parameters are estimated online & updated [19]. The model parameter estimation is done online by Recursive Least Squares (RLS) method. By updating the model parameters we can prevent the process model mismatch in dead time compensation by smith predictor scheme. The adaptive parameter estimation process for a smith predictor scheme is shown in Figure 4.12 below.

Classical smith predictor structure can be implemented when the time delay in the system is fixed and predetermined. But in actual practice, the process time delay may vary due to which actual delay differs from the pre estimated value. To tackle this issue, on line time delay estimation is used for better performance of the smith predictor.

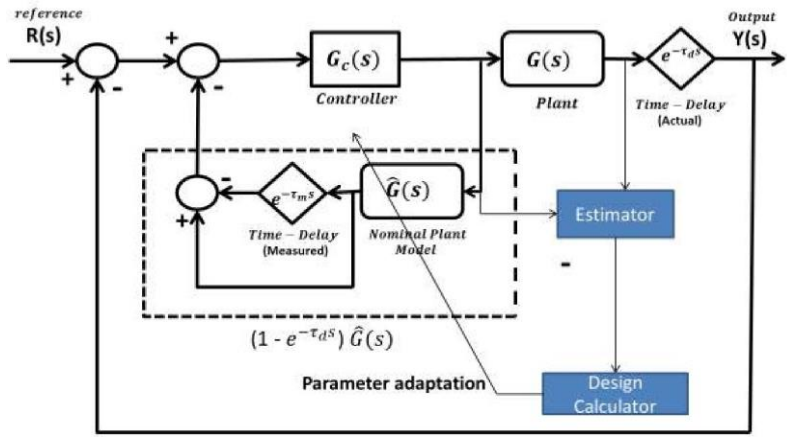

Figure 4.12: Adaptive self-tuning of a smith predictor scheme for process model mismatch

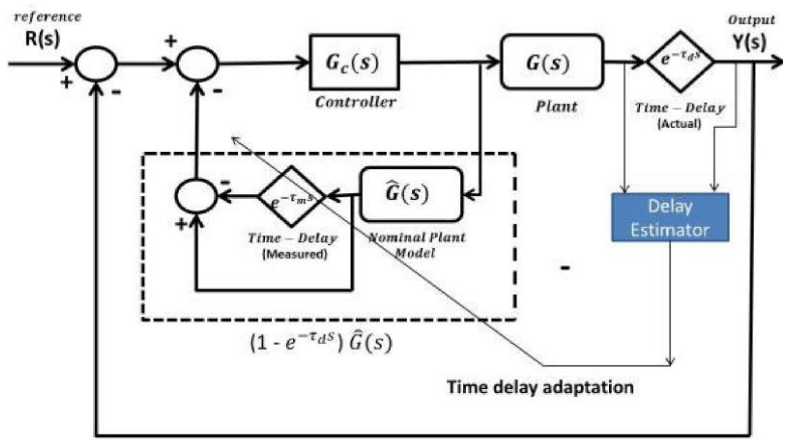

Figure 4.13: Adaptive time delay estimation for smith predictor

Both the adaptation technique reprensted in figure4.12nad figure4.13can also be implemented simultaneously for better  performance.

## 4.7.2 Digital Smith  Predictor

Until 1980, smith predictor schemes developed were mostly analog & hence were not used in the industry due complexity of analog technique. In 1980, a digital control scheme for dead time compensation.    The smith predictor scheme is also sensitive to time delay variation. For smooth operation, the time delay should be accurately determined and for this purpose online delay estimator can be used. The modeled or estimated delay used should match with the actual delay and the use of delay estimator can take care of this thing.  The function of the digital version of the smith predictor is similar to the classical version with similar functionality of the different blocks in discrete domain. The basic digital smith predictor structure is shown in figure **??**. Here, all the blocks are represented in discrete domain similar like in the analog type. It is important to note that the actual delay is shown as $G_d(z)$ (discrete transformation of $e^{-\tau_d s}$) and the estimated delay is shown as $G_d m(z)$ (discrete transformation of $e^{-\tau_m s}$).

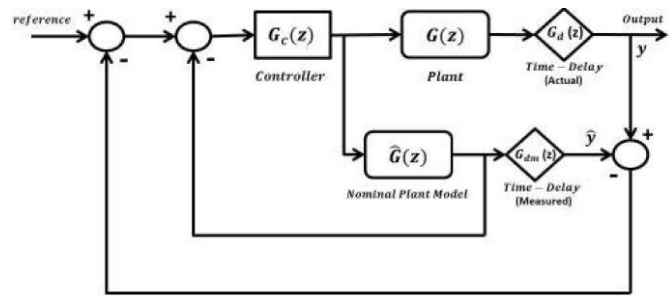

Figure 4.14: Digital smith predictor structure [19]

The scheme represented in figure4.14is be rearranged and shown in figure4.16.

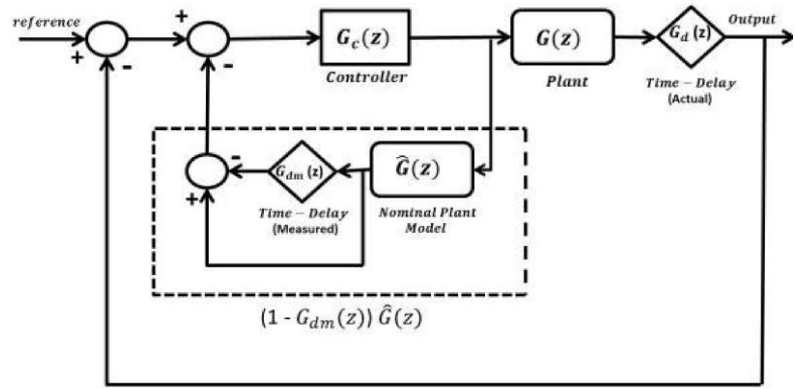

Figure 4.15: Rearranged form of the digital smith predictor represented in figure 4.14

## 4.8 Implementation of Digital Smith Predictor in WNCS

In a earlier chapter we have discussed estimation of time delay in WNCS using round trip time (RTT) concept. The network induced time delay in point to point configuration of two Bluetooth node is found to be 0.06 second most of the time. Hence, we have taken 0.06 second as the dead time for implementation of smith predictor structure as per the scheme represented in figure 4.16.

Here the nominal plant model is taken as the discretized transfer function of the DC motor model cascaded with the ZOH block. The nominal plant model used in smith predictor block is given as,

$$\hat{G}(z^{-1}) = \frac{0.0832z^{-1}}{1 - 0.92z^{-1}} \tag{4.17}$$

Now for implementation in embedded platform, the delay block need to be represented as a discrete transfer function. For this purpose, we have done system identification of the MATLAB transport delay block as shown in figure 4.16.

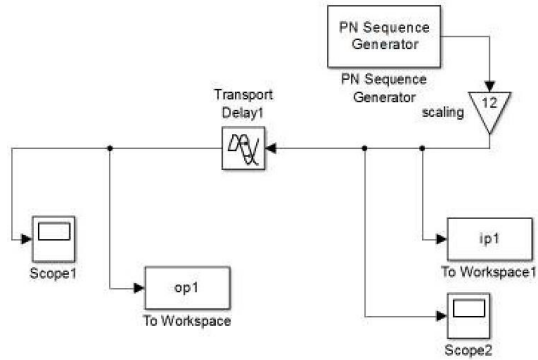

Figure 4.16: Data collection in SIMULINK for identification of delay block transfer function

Now the identified delay block discrete transfer function with best fitting (in feasible order) is obtained from MATLAB system identification toolbox as,

$$G_d(z^{-1}) = \frac{0.0006313z^{-1} + 0.000636z^{-2} + 0.9971z^{-3}}{1 - 0.0006345z^{-1} - 0.000633z^{-2} + 0.00007223z^{-3}} \tag{4.18}$$

Now using the $(1-G_d(z^{-1}))\hat{G}(z^{-1})$ block as diifference equation in embedded micro controller as shown in figure 4.17, we have implemented the smith predictor scheme.

The practical response of the plant with discrete PI controller without smith predictor scheme for the point-to-point network configuration has been shown in figure 4.18.

Now the practical response of the plant with digital smith predictor scheme has been shown in figure 4.19.

It has been observed that with the implementation of digital smith predictor scheme in wireless point-to-point congiguration of the servo system, good performance is obtained compared to conventional PI controller when there is a fixed network induced time delay of 0.06 second.

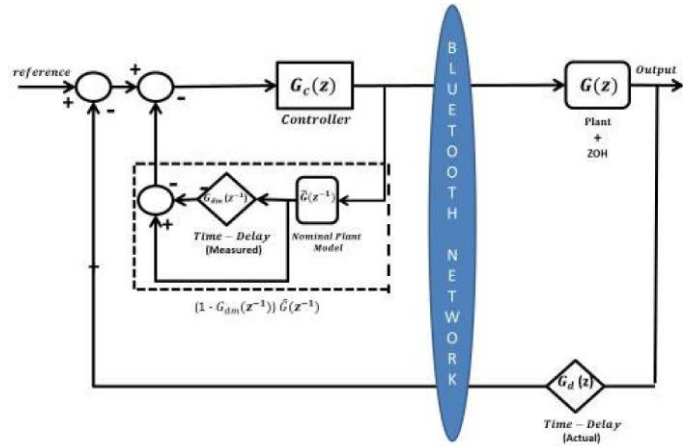

Figure 4.17: Implemented digital smith predictor scheme for the wireless point-to-point configuration

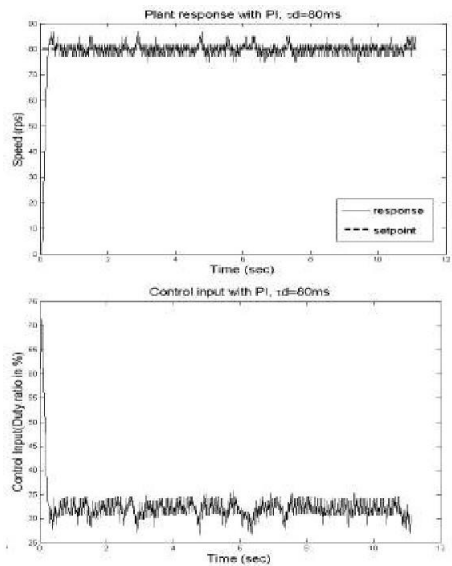

Figure 4.18: Practical response and control signal without smith predictor scheme in wireless point-to-point configuration

## 4.8.1 Digital Smith Predictor With Intermediate Communication Nodes

We can see that the digital smith predictor scheme implemented for point-to-point Bluetooth networked servo system works fine as the delay is fixed. But when intermediate two nodes are inserted between the controller node and plant side node, the time delay no longer remains fixed and it varies from 0.06 second delay in the point to point case. In this network configuration we observe the response of the servo system for two time with the digital smith predictor.

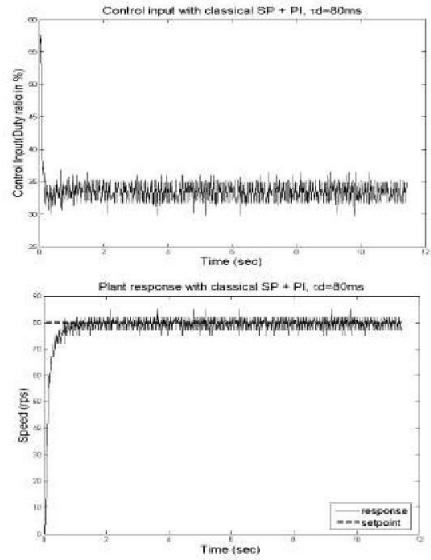

Figure 4.19: Practical response and control signal with smith predictor scheme in wireless point-to-point configuration

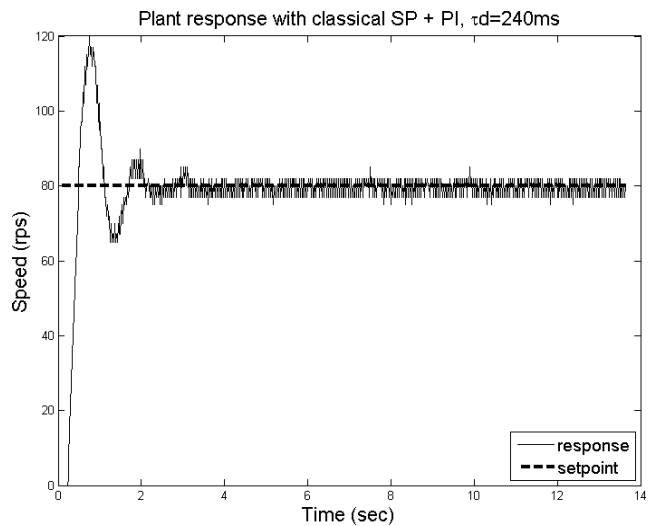

Figure 4.20: Practical response with smith predictor scheme in network having intermediate nodes (1st time)

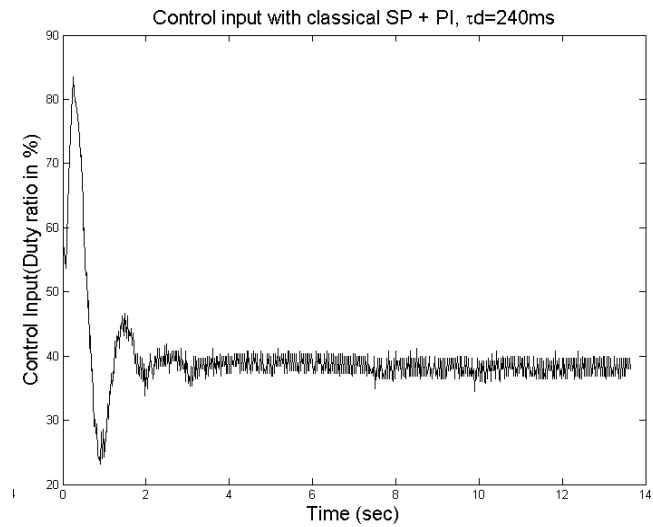

Figure 4.21: Control input with smith predictor scheme in network having intermediate nodes (1st time)

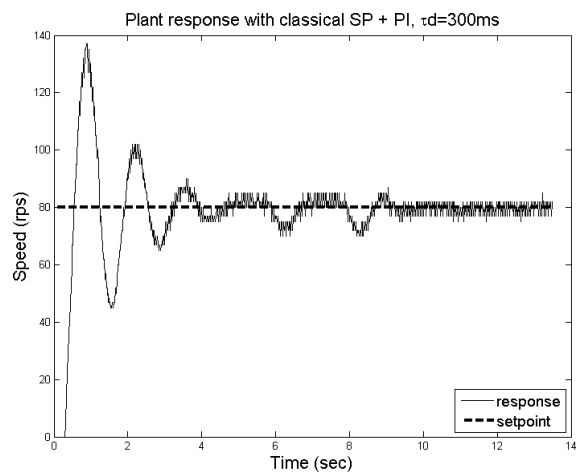

Figure 4.22: Practical response with smith predictor scheme in network having intermediate nodes (2nd time)

It has been clearly observed that the digital smith predictor implemented with fixed estimated time 0.06 sec. can not provide satisfactory performance when used in the wireless networked control system having intermediate two nodes between the controller side node and plant side node. This is due to the fact that the delay introduced by the intermediate nodes is not fixed and it varies.

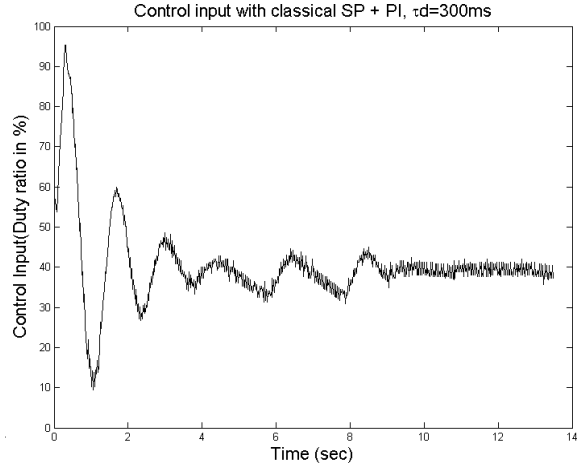

Figure 4.23: Control input with smith predictor scheme in network having inter-
mediate nodes (2nd time)

## 4.8.2 Adaptive Digital Smith  Predictor

As the performance of the digital smith predictor scheme with fixed estimated time de-
lay can not provide satisfactory response when dead time is varying, therefore adaptive
smith predictor scheme is needed. The performance of smith predictor significantly
degrades (may become unstable) due to modeling or estimation errors of both delay
and process [16]. Specially dead time estimation error is very dangerous as it can vary
widely depending on working conditions [16]. Therefore we are not using parameter
adaptation of the process model while implementing the adaptive digital smith predic-
tor. The adaptive smith predictor proposed here simply updates the process time delay
as measured and estimated on line by the method discussed in an earlier chapter. The
adaptive digital smith predictor proposed is given in figure4.24.

Here, $G_d(z)$ is the discrete delay model implemented in the embedded computer. Here
we are using both discrete 2nd order Pade series and Direct Frequency Response (DFR)
series approximation for the estimated time delay. The discretization of the polynomial
series has been discussed in a earlier chapter. Therefore the 2nd order discrete DFR
series approximation for an estimated time delay $\tau_m$ is given as,

$$G_d(df\,r)(z) = \frac{cz^2 + dz + e}{ez^2 + dz + c} \qquad (4.19)$$

Where,

c = 1 + 100 $\tau_m(9.54\tau_m - 0.49)$

d = 2 - 100 $\tau_m(9.54\tau_m + 0.49) - 100\tau_m(9.54\tau_m - 0.49)$

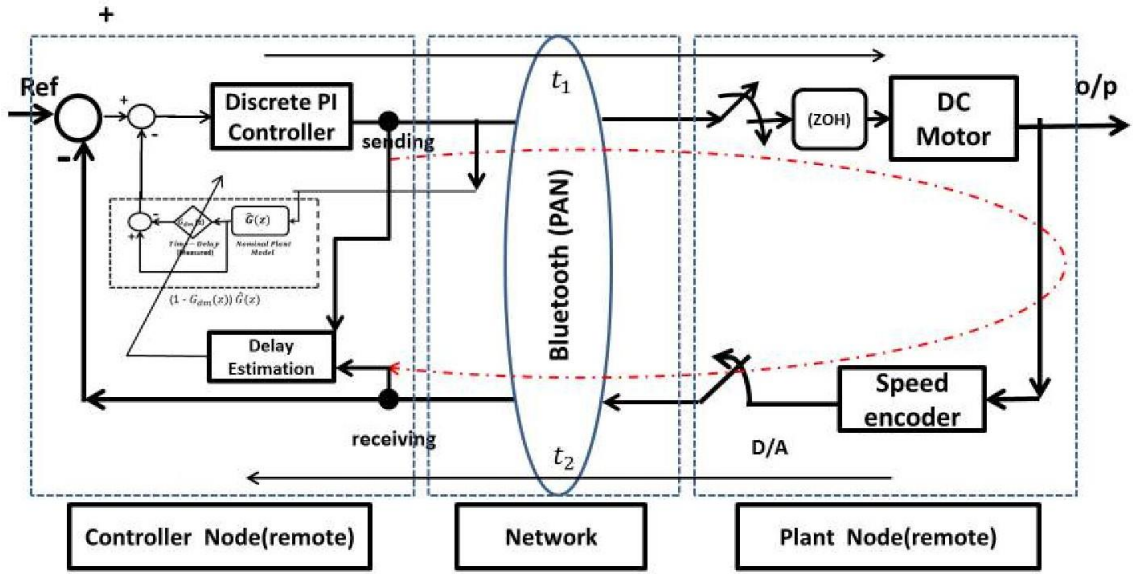

Figure 4.24: The adaptive digital smith predictor scheme for the wireless network with intermediate nodes

e = 1 + 100 $\tau_m$(9.54$\tau_m$ + 0.49)

The 2nd order discrete Pade series approximation of the estimated time delay $\tau_m$ is given as,

$$G_d(pade)(z) = \frac{az^2 + bz - c}{dz^2 + ez + f} \qquad (4.20)$$

Where,

A = $100\tau_m$

a = $12 - 6A + A^2$

b = $24 + 2A$

c = $6 + A$

d = $12 + 6A + A^2$

e = $24 - 2A$

f = $A - 6$

Both the Pade and DFR approximations are used in the smith predictor block's $(1 - G_d(z))\hat{G}(z)$ part and implemented as difference equation.

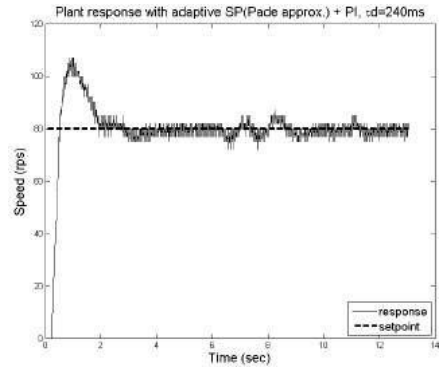

Figure 4.25: Response with Pade series delay approximation of the adaptive digital smith predictor in wireless network with intermediate nodes

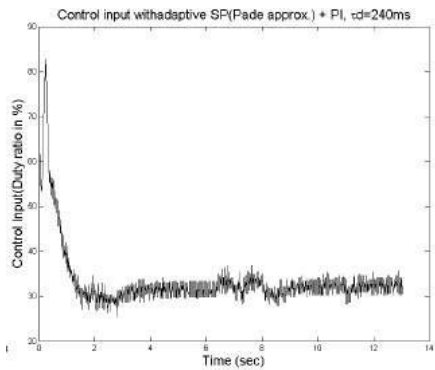

Figure 4.26: Control input with Pade series delay approximation of the adaptive digital smith predictor in wireless network with intermediate nodes

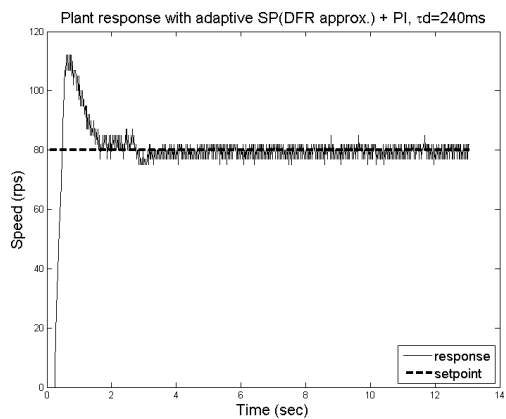

Figure 4.27: Response with DFR series delay approximation of the adaptive digital smith predictor in wireless network with intermediate nodes

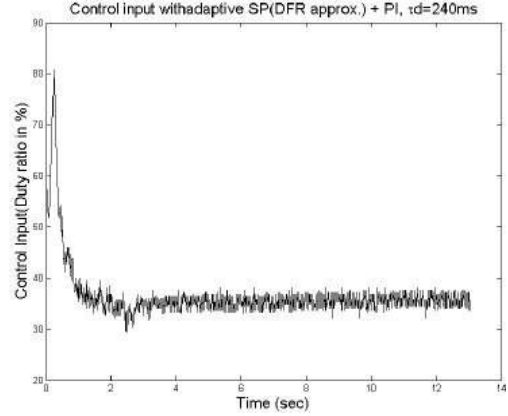

Figure 4.28: Control input with DFR series delay approximation of the adaptive digital smith predictor in wireless network with intermediate nodes

### 4.8.3 Implementation of Discrete Transfer Function in Embedded Platform

The Pade and DFR discrete domain approximation series are often very long, hence we are showing a simplified example transfer function showing the implementation. Let us consider a simple discrete transfer function,,

$$G(z) = \frac{2z}{z^2 - z + 1} \tag{4.21}$$

The we can also represent the transfer function as,

$$G(z^{-1}) = \frac{2z^{-1}}{1 - z^{-1} + z^{-2}} \tag{4.22}$$

Now we can represent, the transfer function in terms of discrete input X(k) & output Y(k) as,

$$\frac{Y(k)}{X(k)} = G(z^{-1}) = \frac{2z^{-1}}{1 - z^{-1} + z^{-2}} \tag{4.23}$$

Now,

$$\Rightarrow 2z^{-2}X(k) = Y(k)(1 - z^{-1} + z^{-2}) \tag{4.24}$$

$$\Rightarrow 2X(k-1) = Y(k) - Y(k-1) + Y(k-2) \tag{4.25}$$

$$\Rightarrow Y(k) = 2X(k-1) + Y(k-1) - Y(k-2) \tag{4.26}$$

Equation 4.26 provides the difference equation of the discrete transfer function G(z) which can be implemented in embedded controller. In this way the delay approximated series and the nominal plant model combined transfer function in discrete domain $((1-\hat{G}_dm(z))G(z))$ has been implemented in embedded controller to get smith predictor action.

## 4.9 Practical Waveforms Obtained from Extensive Experimentation of the Digital Smith Predictor

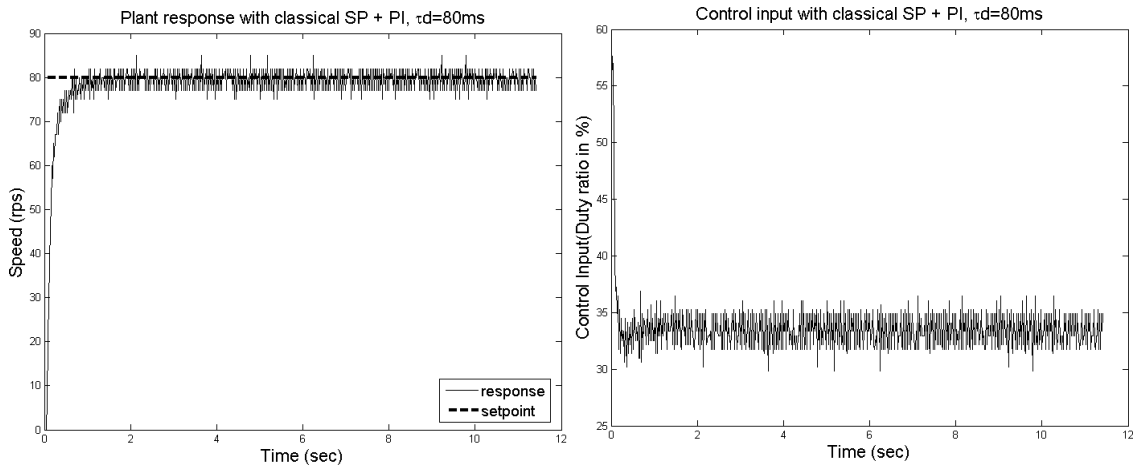

Figure 4.29: Response of Classical Digital Smith Predictor + PI controller when time delay = 80 ms

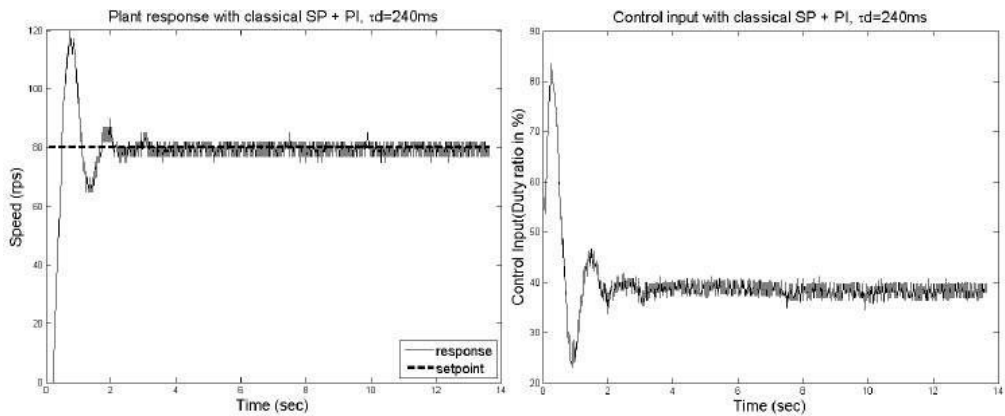

Figure 4.30: Response of Classical Digital Smith Predictor + PI controller when time delay = 240 ms

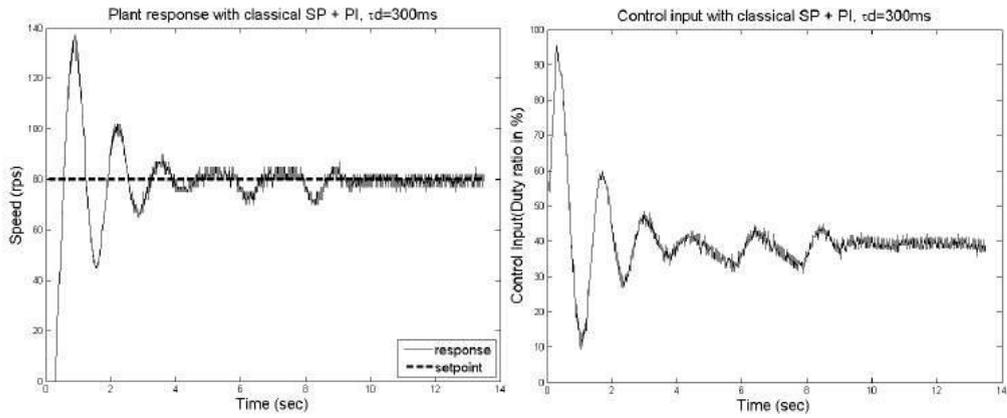

Figure 4.31: Response of Classical Digital Smith Predictor + PI controller when time delay = 300 ms

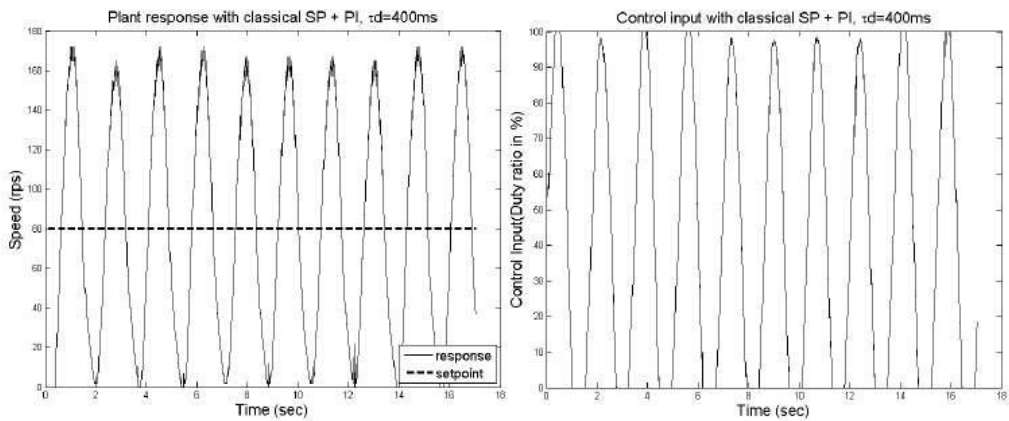

Figure 4.32: Response of Classical Digital Smith Predictor + PI controller when time delay = 400 ms

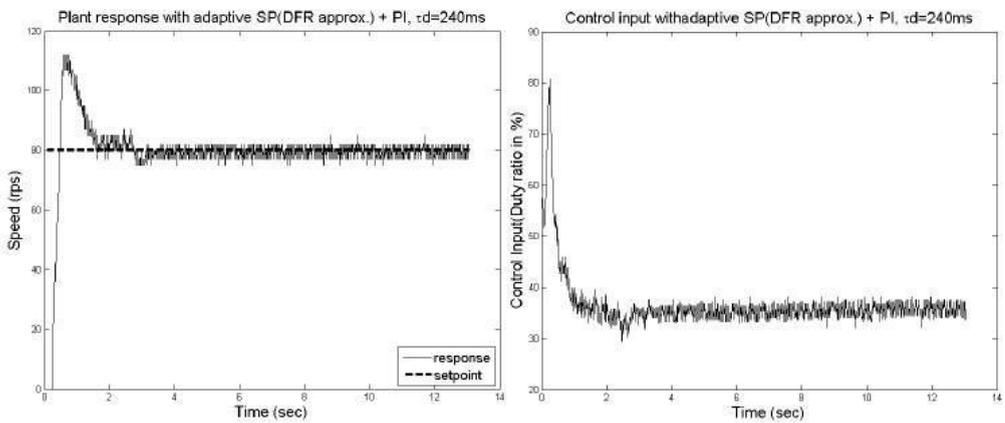

Figure 4.33: Response of Adaptive Digital Smith Predictor( DFR approx.) + PI controller when time delay = 240 ms

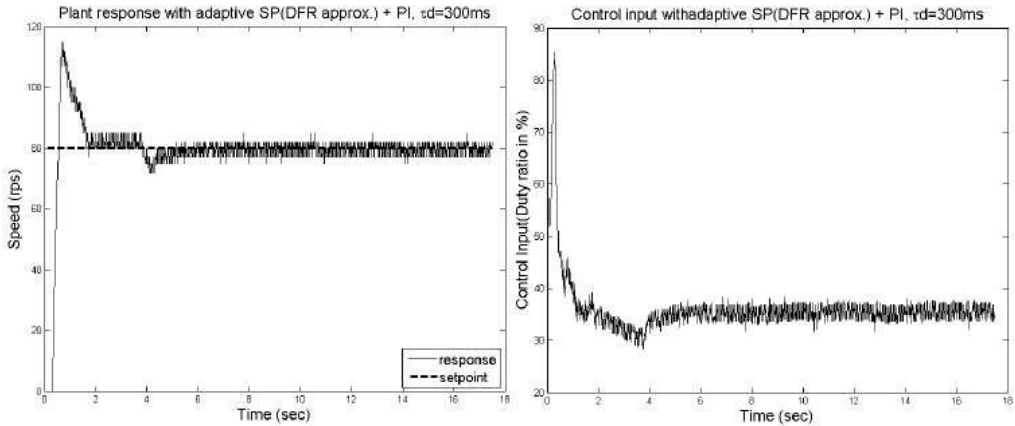

Figure 4.34: Response of Adaptive Digital Smith Predictor( DFR approx.) + PI
controller when time delay = 300 ms

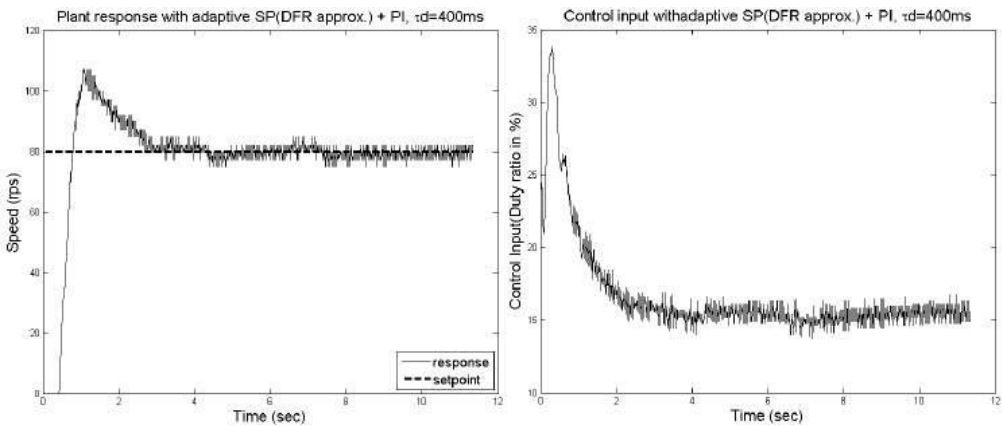

Figure 4.35: Response of Adaptive Digital Smith Predictor( DFR approx.) + PI
controller when time delay = 400 ms

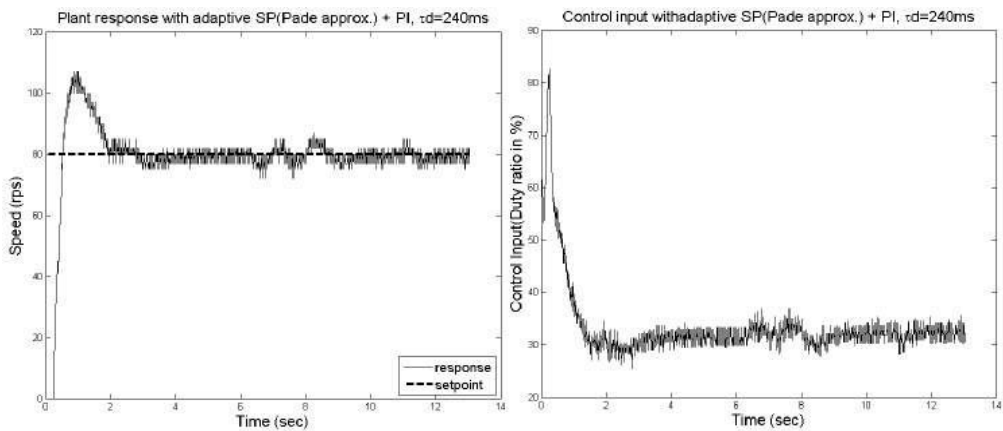

Figure 4.36: Response of Adaptive Digital Smith Predictor( Pade approx.) + PI
controller when time delay = 240 ms

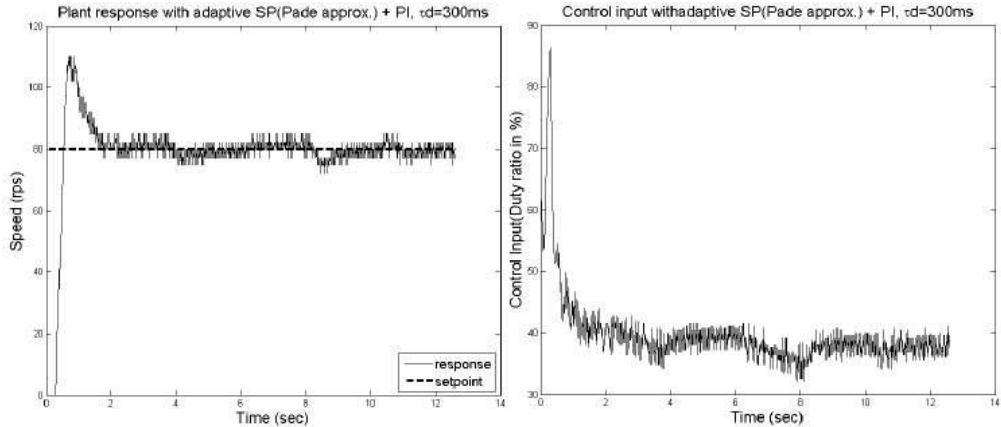

Figure 4.37: Response of Adaptive Digital Smith Predictor( Pade approx.) + PI controller when time delay = 300 ms

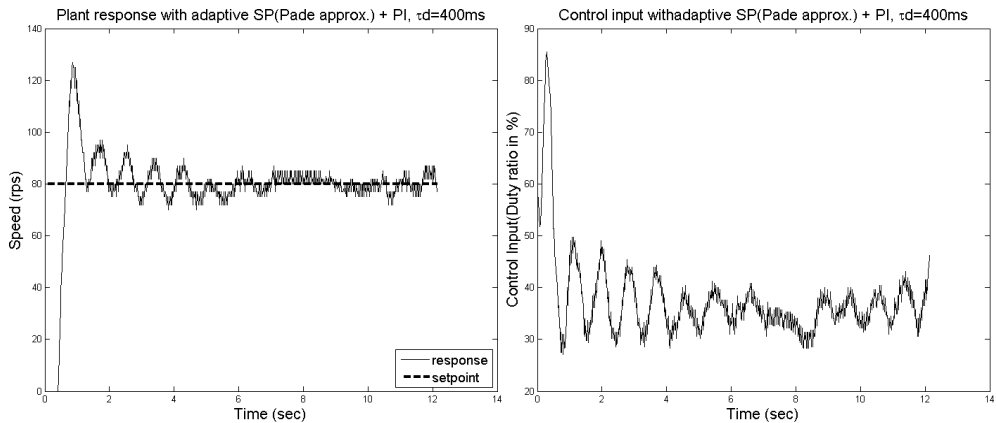

Figure 4.38: Response of Adaptive Digital Smith Predictor( Pade approx.) + PI controller when time delay = 400 ms

## 4.10 Chapter Summary

In this chapter, the measurement of RTT and delay estimation adopted for the project/research has been discussed. RTT measurement and delay estimation is the heart of the delay mitigation scheme adopted in this project/research. Next, the approximation of the time-delay model and the requirement of it from the view of implementation has been discussed where different delay approximation polynomials were considered and the best was chosen for implementation. Next, the strategy of implementing the approximated delay model on embedded platform has been discussed. Next, we introduce the Smith predictor scheme where a generic discussion on classical Smith predictor was done followed by its adaptive version and then the digital Smith predictor scheme was introduced. It was then followed by the implementation of digital Smith predictor on embedded hardware where a constant delay model scheme for Smith predictor was discussed. It was observed that the Smith predictor as a constant delay model did not perform well for a significant amount of delay (*by introduction of intermediate communication node*). Therefore, an adaptive scheme for the Smith predictor was introduced where delay was considered to be varying and of significant magnitude. It was here, we used the approximated delay model in conjugation with the Smith predictor to make it adaptive, the estimated delay was continuously fed to the Smith predictor block (*as shown in4.24* ) for the required delay compensation. Next, we illustrate the practical waveforms as obtained from experimentation considering different Smith predictor schemes. It was observed that, the DFR series approximation of delay gave better result when compared to Pade approximation polynimial of delay.

# Chapter 5

# Stability Analysis of the DC Motor System

## 5.1 Introduction

In this chapter, we illustrate a qualitative analysis on the stability of the system under consideration with different time delays. Within a system, Dead-time / time delay is the time interval between an event initiated at a point in the system and the output of that event at another point in the same system [1]. The stability of a minimum phase system (*systems that do not have poles or zeros in the right-hand side of s-plane or do not have other delay component* ) is always found to be reduced with inclusion of delay in the system [1]. Therefore, under the presence of delay, the stability analysis of a system is vital.

## 5.2 Nyquist Stability Criterion for System with Time Delays

In system engineering, stability of closed-loop system is a basic concept as only this type of system has practical use [1]. A system is said to be stable if it generates bounded output for bounded inputs [2]. The benefits of analyzing the stability of a system in frequency domain, according to [3], are :

- Simple stability tests and improved accuracy with sinusoidal signal generator.

- Experimentally, frequency response tests can provide complex transfer functions.

- Design & analysis process are extendable to non-linear systems by neglecting noise effects.

The Nyquist stability criterion provides us the stability of a closed-loop system from its corresponding open-loop frequency response and open-loop poles [2]. Both absolute and relative stability analysis can be done with Nyquist stability criterion [1]. The presence of dead time in the system does not alter the criterion,  and therefore it can be utilized for analysis of stability of systems with time-delay [1].

Let us first consider a closed loop system free from time delay,

$$G(s) = \frac{G_p(s)}{1 + G_p(s)H(s)} \tag{5.1}$$

Then, the Nyquist stability criterion is given by,

$Z = N + P$

where,

$Z=$ the number of zeros of 1+G(s)H(s) on the right hand part of the s-plane,

$N =$ the number of locus of G(j)H(j) that encircles point $-1 + j0$ clockwise,

$P =$ denotes the number of poles of G(s)H(s) on the right hand part of the s-plane

Now, if delay is introduced, the characteristic equation can be written as,

$$X(s) = 1 + G_p(s)H(s)e^{-\tau_d s} = 0 \tag{5.2}$$

Now, let,

$$d(s) = X(s) - 1 = G_p(s)H(s)e^{-\tau_d s} = 0 \tag{5.3}$$

Cauchy theorem states that contour integral of d(s) along closed path on the s-plane equals to zero if D(s) is analytic both inside and along the path. Therefore,

$$\oint_C d(s)ds = 0 \tag{5.4}$$

with the integral evaluated in clockwise direction. Now, expressing d(s) in terms of its poles and zeroes, we arrive at a decomposed form of d(s) given by,

$$d(s) = \frac{(s+z_1)^{l_1}(s+z_2)^{l_2}\ldots}{(s+p_1)^{o_1}(s+p_2)^{o_2}\ldots}e^{-\tau_d s} \tag{5.5}$$

Now, the ratio $\frac{d'(s)}{d(s)}$ is written as,

$$\frac{d'(s)}{d(s)} = \overbrace{\frac{l_1}{s+z_1}+\frac{l_2}{s+z_2}+\ldots}^{\Sigma} - \overbrace{\frac{o_1}{s+p_1}+\frac{m_2}{s+p_2}+\ldots}^{\Sigma} - \tau_d \tag{5.6}$$

Using residual theorem,

$$\oint_C \frac{d'(s)}{d(s)} = 2\pi j[(l_1+l_1+\ldots)-(o_1+o_2+\ldots)]$$
$$= -2\pi j(Z-P) \tag{5.7}$$

where,

$Z$ = the number of zeros of d(s) inside the path,

$P$ = the number of and poles of d(s) inside the path

Since d(s) is a complex variable, it can be expressed as,

$$d(s) = |d(s)|e^{j\theta}$$
$$\implies ln(d(s)) = ln|d(s)| + j\theta \tag{5.8}$$

Therefore,

$$\frac{d'(s)}{d(s)} = \frac{dlnd(s)}{ds}$$
$$= \frac{dln|d(s)|}{ds} + j\theta \tag{5.9}$$

hence,

$$
\int_C \frac{d'(s)}{d(s)} = \int_C dln|d(s)| + j \int_C d\theta
$$
$$
= 2\pi(P - Z)
$$

(5.10)

where,

$$
\int_{\substack{C \\ C}} dln|d(s)| = 0
$$

(5.11)

as $ln|d(s)|$ has an equal value on the initial and the end point of the integration. Accordingly,

$$
\frac{\theta_2 - \theta_1}{2\pi} = P - Z
$$

(5.12)

The angular difference between end point and initial point of $\theta$ equals to the total change in the phase angle of d'(s)/d(s). As N denotes the number of closed paths on d(s) plane that encircle the original point in clockwise manner, and $\theta_2$-$\theta_1$ = $2\pi l$, $l$ = 0,1,...., then

$$
\frac{\theta_2 - \theta_1}{2\pi} = -N
$$
$$
\Longrightarrow \ N = Z - P
$$

(5.13)

Therefore, the existence of delay does not alter the Nyquist Stability criterion. Thus, it can be used for systems with time-delays.

Analysis of stability of a linear system with Nyquist stability criterion leads to the following possibilities [[1],[2],

- System is stable if there is no encirclement of the point -1+j0 by the closed path on G(s)H(s) plane, provided there is no pole of G(s)H(s) on the right hand side of the s-plane.

- System is stable if there are one or more counter-clockwise encirclement of the point -1+j0 by the closed path on G(s)H(s) plane, provided the number of counter-clockwise encirclement is equal to the number of poles of G(s)H(s) in the right-half of s-plane. Otherwise, system is unstable.

- System is unstable if there are one or more clockwise encirclement of the point -1+j0 by the closed path on G(s)H(s) plane.

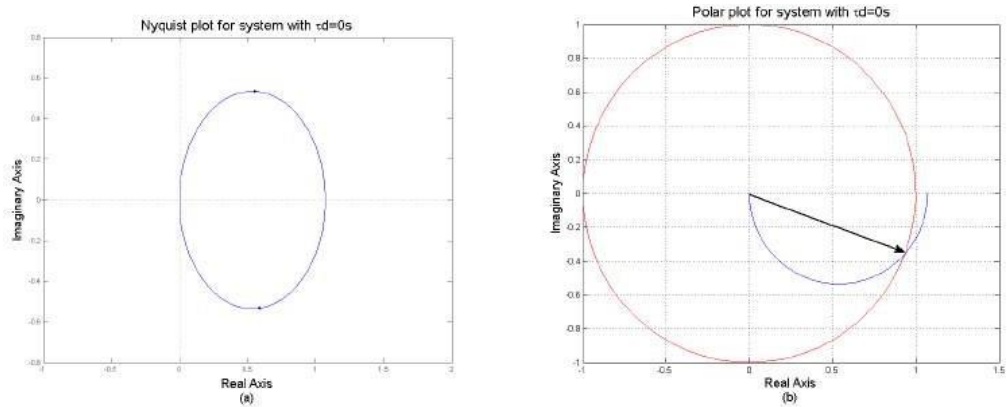

Figure 5.1: Nyquist plot(a) & polar plot(b) for the open-loop transfer function $tt(s)H(s)$ without time-delay

For the system under consideration, the open-loop transfer function is given by,

$$G(s)H(s) = \frac{4.159}{s + 3.888} \tag{5.14}$$

The Nyquist & polar plot of equation 5.14 is shown in figure 5.1. From figure 5.1, it is clear that the open-loop transfer function $G(s)H(s)$ of the plant is absolutely stable in closed-loop configuration as there is neither zero nor pole lies on the right hand side of the s-plane and locus of G(s)H(s) does not encircle the point -1+j0.

An important concept of stability analysis is relative stability which is related with the settling time [1]. Faster settling time of a system implies better stability compared to a system with slower settling time . Degree of stability of a system can be determined with the help of Nyquist plot and polar plot. The relative stability based on the distance of G(j)H(j) locus to -1+j0 point can be stated in the phase margin. The phase margin is defined as the phase lag that needs to be compensated for the system to be at the border of stability. Mathematically, the phase margin can be written as,

$$\varphi M = \gamma = 180^{\circ} + \varphi \tag{5.15}$$

where,

$\gamma$ = denotes phase margin ,

$\varphi$ = the phase of systems OLTF at gain crossover frequency

For time-delay system, the phase margin can be written as,

$$\varphi M = \gamma = 180^\circ + \varphi - \omega_g \tau_d \qquad (5.16)$$

Where,

$\omega_g$ = gain crossover frequency ,

$\tau_d$ = time-delay

The figures 5.2-5.10 illustrates the Nyquist plot & the polar plot for the open-loop transfer function $G(s)H(s)$ for different time-delays.

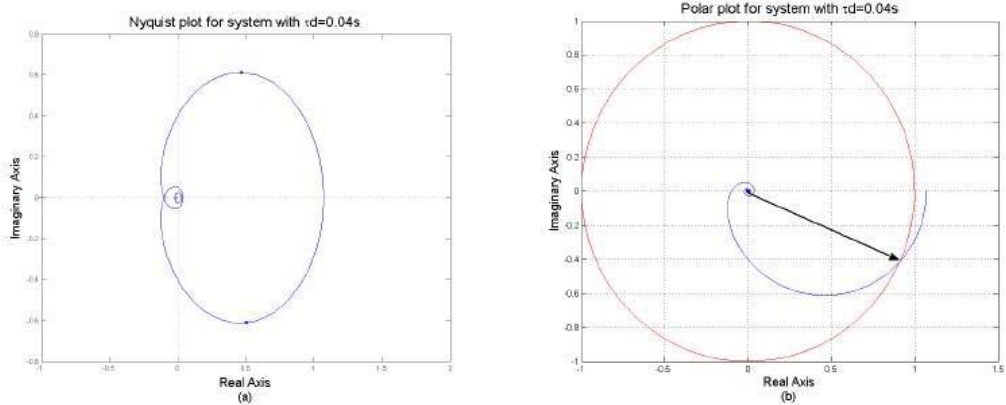

Figure 5.2: Nyquist plot(a) & polar plot(b) for $\tau_d = 0.04s$

From the figures 5.2-5.10, we can draw few conclusions,

- With increase in delay, the locus of $G(j\omega\ H(j\omega)$ tend towards the point -1+j0.

- The Phase Margin($\varphi$M) decreases with increase in delay.

- The relative stability of the system decreases with increase in delay.

- At $\tau_d$=2s, the Phase Margin is $-ve$ ,i.e, the system becomes unstable.

- Time-delay in a system creates larger phase lag, so the delayed system is less stable than the original  system.

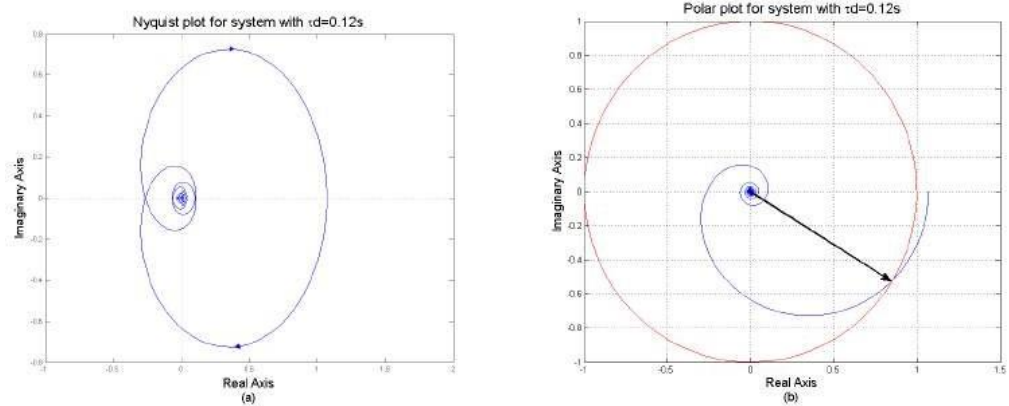

Figure 5.3:  Nyquist plot(a) & polar plot(b) for $\tau_d = 0.120s$

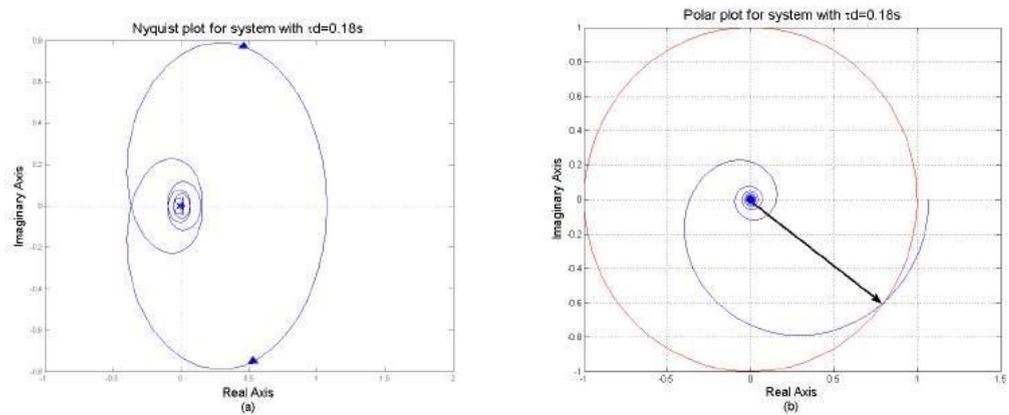

Figure 5.4:  Nyquist plot(a) & polar plot(b) for $\tau_d = 0.180s$

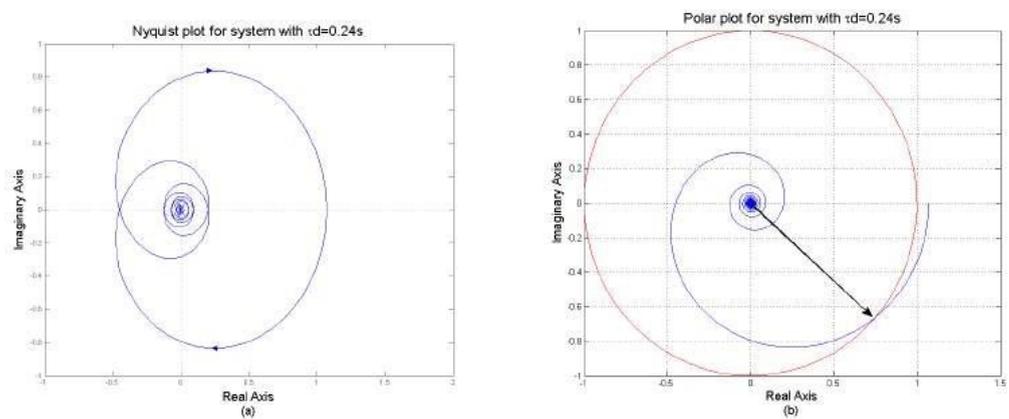

Figure 5.5: Nyquist plot(a) & polar plot(b) for $\tau_d = 0.240s$

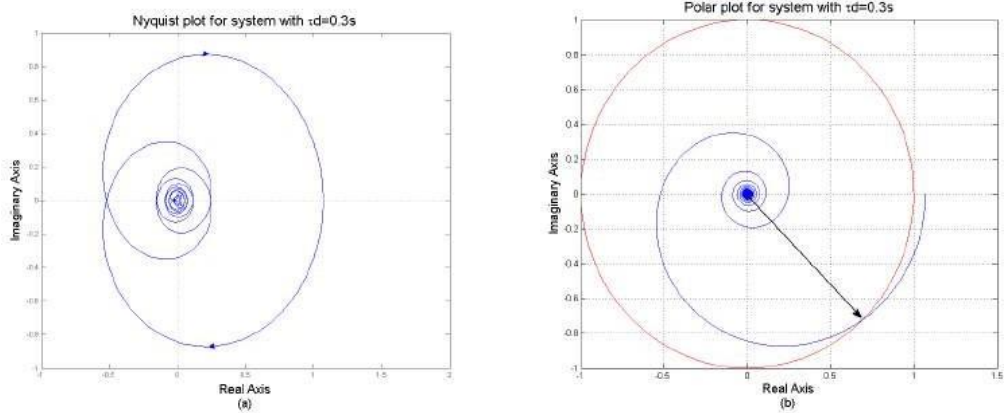

Figure 5.6: Nyquist plot(a) & polar plot(b) for $\tau_d = 0.300s$

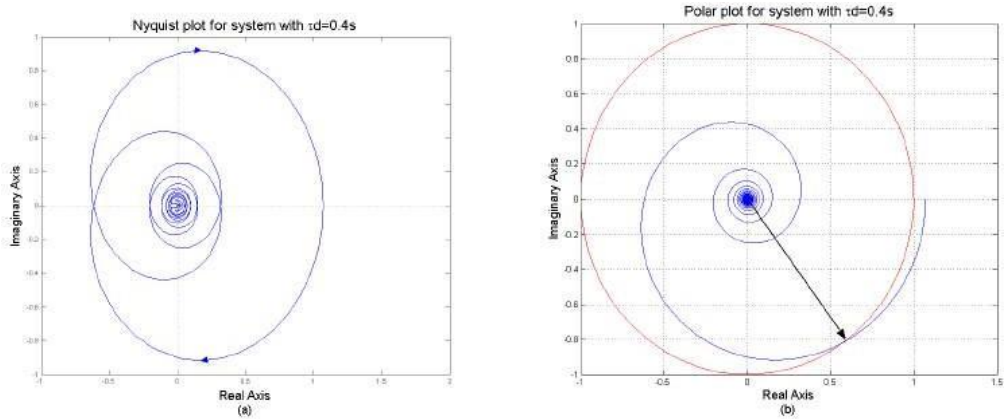

Figure 5.7: Nyquist plot(a) & polar plot(b) for $\tau_d = 0.400s$

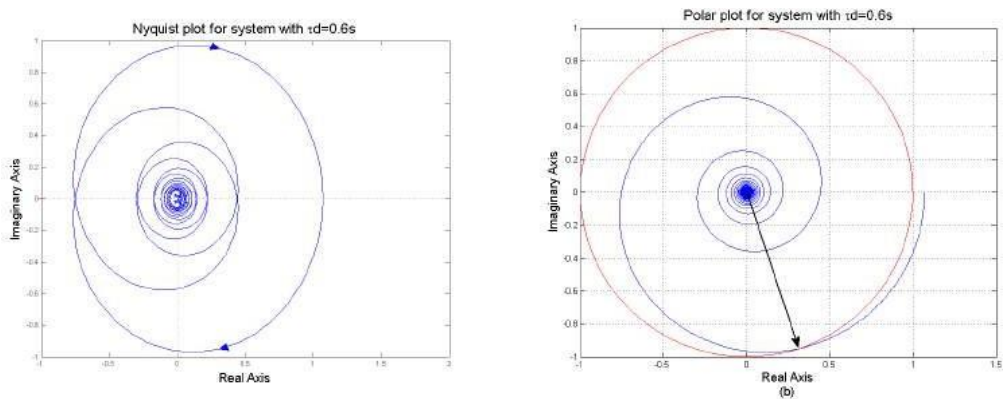

Figure 5.8: Nyquist plot(a) & polar plot(b) for $\tau_d = 0.600s$

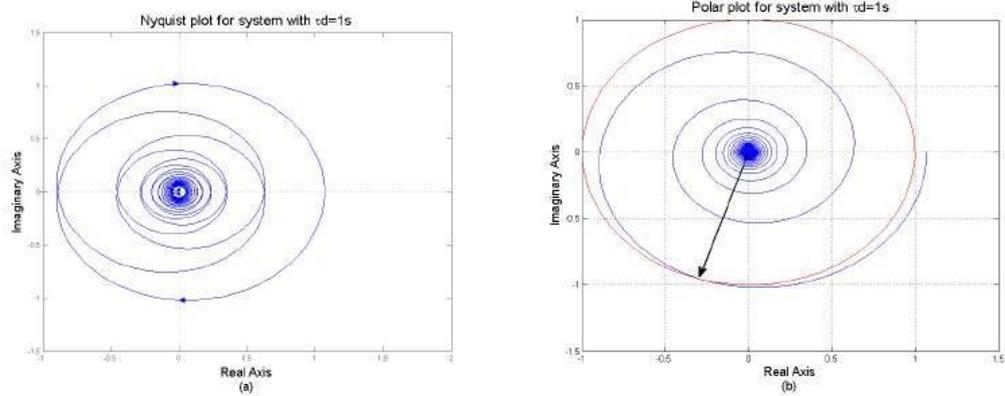

Figure 5.9: Nyquist plot(a) & polar plot(b) for $\tau_d = 1s$

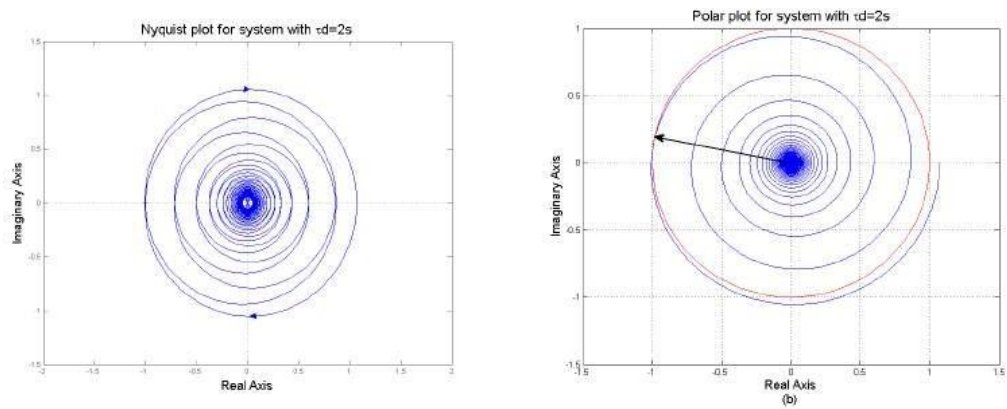

Figure 5.10: Nyquist plot(a) & polar plot(b) for $\tau_d = 2s$

Table 5.1 illustrates the obtained Phase Margin( $\varphi$M) for the system under consideration for different time-delays. The Phase Margins were calculated geometrically.

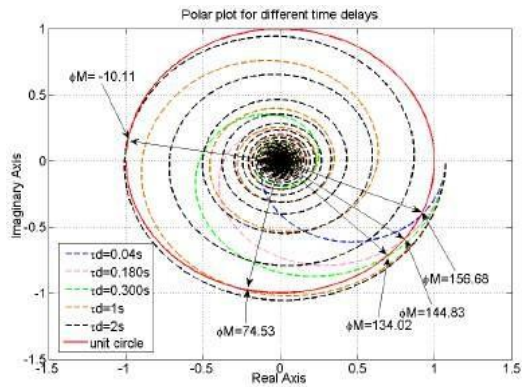

Figure 5.11: Polar plot for different time-delays

| $Time - Delay, \tau_d$ seconds | $Phase\ Margin(\varphi M\ )$ |
|:---:|:---:|
| 0 | 159.19 |
| 0.04 | 156.68 |
| 0.120 | 148.73 |
| 0.180 | 144.83 |
| 0.240 | 139.30 |
| 0.300 | 134.02 |
| 0.400 | 126.31 |
| 0.600 | 106 |
| 1 | 74.53 |
| 2 | -10.11 |

Table 5.1: Variation of $\varphi$M with variation in $\tau_d$

## 5.3 Chapter Summary

In this chapter, a qualitative analysis on the stability of the DC motor system is illustrated. Here, the stability analysis of the plant done using Nyquist criterion of stability. Nyquist and polar plots for different time delays were studied. It was observed that with increase in delay, the phase margin reduced. Therefore, the relative stability of the plant is reduced with increase in delay.

# Chapter 6

# Modifications, Contributions/Proposals of The Research

Though the strategy of estimation of delay was adopted from [1],necessary modification in the existing strategy has been made to adopt it to the system under consideration. The goal was to implement the strategy on the hardware platform (*ARDUINO UNO/- DUE*), hence, modification was necessary for the sake of implementation. Modification pertaining to refitting the existing idea to comply with the system under consideration with due concern on the ease of implementation and etiquette of the strategy were always under consideration.

The modification is given as under,

- *Modification:* In [1], the RTT measurement & estimation of delay was made at the plant node and then the information of delay was sent to the controller node. For the system under consideration, i.e, for our system, the RTT measurement & delay estimation were made at the controller node itself. Necessity for such a modification,

    - If RTT is measured on the plant side, for compensation of the estimated delay, the information has to be sent to the controller node. Therefore, in [1], the delay information at each frame was sent to the controller node at half sampling period and complicated algorithms were designed to achieve this purpose. Whereas, for the system under consideration, the measurement and estimation of delay was made at the controller node itself, hence, such complicated algorithms were avoided. Thus, making the implementation easier and more convenient.

– Measurement of delay at the plant node and then transfer of this informa-
tion to the controller node may itself suffer from delay. If the information of
delay gets delayed itself and does not arrive before the arrival of the feedback
signal, the control action required for the compensation of the delay for that
very sample period may ultimately get delayed and the overall performance
of the system may get degraded. Whereas, for the system under consid-
eration, such an undesirable phenomenon/happening has been avoided by
directly measuring and estimating the delay at the controller node itself.

When it comes down to the level of implementation, especially on a dedicated hardware
platform, two factors are always a matter of concern; namely, **_The Cost of Imple-
mentation_** & **_The Ease of Implementation_**. The implemented idea should be
practical and equally economically viable. A product/idea maintaining a correct ratio
of these two factors is admissible. A growing technology demands for an easy and
cost-effective solution to a given commercial problem. In relation to this, we present
few contributions and proposals put forward by us for an easy and cheap solution to
the problem of delay estimation and mitigation.

The contributions/proposals are listed as under,

- In [1], the system runs on a high performance computer supported with sophis-
ticated hardware and software. The control and estimations algorithms were all
implemented on it. The goals were ultimately achieved but the overall cost of
implementation was high. The objective of our project/research was to find a
cheaper alternative to this. Therefore, in order to reduce the cost of implementa-
tion, the control system as a whole was exclusively implemented on a dedicated
hardware where no high performance PC and other sophisticated software were
involved. The cost of the hardware *(ARDUINO UNO/DUE)* ( 1000 - 1500) is
quite less when compared to a high performance PC. Thus, by bringing down the
implementation on a dedicated hardware gives a compact and economic solution
to the problem that too in form of a device which can be readily used for delay
estimation and compensation.

- In the chapter of delay estimation, it was  discussed that the delay estimated at
a sampling time is equal to the summation of the difference between the time
stamps ($t_1$ & $t_2$) summed for the number of times vacant samples occur after $0ms$.
In connection to this observation, we propose an empirical formula for estimation
of delay,

$$t_m = \sum_{i=0} t_i \qquad (6.1)$$

Where,

$t_m$ = estimated delay,

$t = t_2 - t_1$,

$n$ = number of vacant samples

which can also be represented as,

$$t_m = t_{2pr} - t_{1pr} + \sum_{i=1} t_{ipa} \qquad (6.2)$$

where,

$t_m$ = estimated delay,

$t_{1pr}$, $t_{2pr}$ = present time-stamps,

$t_{ipa}$ = difference of past time-stamps,

$n$ = number of vacant samples,

$t_{2pr} - t_{1pr} = t_0$ *(difference between present time stamps)*

The proposed formulas for estimation of delay provides an easy way to implement the logic of delay estimation as per theory. One just needs to implement the formula in form of an algorithm to estimate the delay. In our case, we have implemented equation6.2. Whereas, any of the two formulas can be conveniently implemented as both equations6.1&6.2are alike. Here we list few key points of the formula,

- The formula gives a convenient way to measure delay. The formula has been proposed by studying the pattern of the delay data. Rigorous experimentation has been conducted by implementing the formula for delay estimation. It was observed that the formula stands good for the occurrence of any scenario as discussed before (*Normal transmission, Vacant sampling, Message Rejection, Delayed transmission*).
- The formula works well even if there are no vacant samples. In this case, it will return only the difference between the present time stamps.
- The formula also stands good in case of message rejection. In this case, the implementer/coder must be aware of considering only the most recently arrived data and rejecting all previous measured data *(as per the rules)*.

• In [1], the control system as a whole is wired. Whereas, in our case, the complete system is exclusively wireless. Thus, it adds flexibility to the device to estimate and mitigate the effects of delay on a plant which is remotely located. Thus, it can be readily used in an industrial environment. Therefore, it is a ***"Wireless Delay Estimation & Compensation Device"***.

- We have also contributed to the development of the Adaptive Smith Predictor scheme using an approximated polynomial series of time-delay. Developing an Adaptive Smith Predictor scheme for delay compensation using such a strategy is new of its kind.

# Chapter 7

# Conclusion & Future Works

## 7.1 Conclusion

The conclusions that have been inferred from the research and experimentation has been listed below,

- A practical set up for DC motor speed control using Arduino embedded platform has been developed. A discrete PI controller has been designed & implemented in embedded platform for speed control. The controller design is based on the DC motor transfer function obtained through system identification technique.

- The embedded DC servo system is distributed with a Bluetooth wireless network having the controller in one side and the actuator, plant, sensor on the other side. For simplicity, the wireless network is said to have connected two sides of the servo system: controller side and plant side. Two configurations of the wireless network link has been developed: point-to-point configuration and intermediate node configuration. The wireless network behavior used in the control loop is investigated and it has been found that the network induces time delay and there is no bit error or data loss in the network. The time delays occurring in intermediate node configuration are found to be varying and much higher in magnitude compared to fixed small time delays in the point-to-point configuration.

- The performance of conventional discrete PI controller deteriorates a little bit in point-to-point configuration compared to the wired set up. The controller fails to compensate for larger time delays in the process in intermediate node configuration of the wireless network. Therefore, a different modification in control strategy need to be adapted for compensation of such time delays.

- The classical digital Smith predictor scheme with discrete PI controller works well in case of small fixed time delay in the system. It performs well when time delay is fixed and small.

- If the time delay is significant and varying, the classical digital Smith predictor designed with fixed time delay fails to compensate for the delay. Therefore the classical digital Smith predictor needs to be modified to adapt it to the significant and varying delay in the system. Thus, an adaptive digital Smith Predictor scheme has been developed.

- It has been observed that the strategy adopted in case of the adaptive digital Smith predictor for significant and varying time delay in the system works well.

- The adaptive digital smith predictor scheme implemented with Direct Frequency Response (DFR) & Pade polynomial series approximation of dead time provided good results with varying time delay. It has been observed that DFR series approximation provides better results.

- Stability analysis of the DC motor plant based on Nyquist and polar plots has been performed for the system under consideration which illustrated that relative stability of the system reduces with increasing delay. This has been validated with experimental results.

- All the conventional PI and predictive control strategies for the DC motor speed control purpose have been implemented for on an embedded platform i.e. Arduino Due micro controller board.

- The results of the research/project itself verifies the modifications done and the contributions made to the research/project.

## 7.2 Future Works

- The performance of developed control strategies with different wireless networks other than Bluetooth can be observed.

- Adaptive Smith predictor along with a nominal delay Smith predictor architecture can be implemented for better performance (overshoot elimination).

- Development of control strategies for networked control systems with both time delay and data packet loss.